\documentclass[3p]{lmcs}
\pdfoutput=1

\usepackage{lastpage}
\lmcsdoi{15}{4}{6}
\lmcsheading{}{\pageref{LastPage}}{}{}%
{Feb.~17,~2018}{Oct.~29,~2019}{}


\usepackage{hyperref}
\theoremstyle{plain} 

 \usepackage{appendix}
\usepackage{stackrel}
\usepackage{rotating}
 \usepackage[T1]{fontenc}
\usepackage{graphicx}
\usepackage{tabularx}
\usepackage{amsfonts}
\usepackage{cmll}
\usepackage{amsmath}
\usepackage{proof}
\usepackage{mathdots}
\usepackage{xcolor}
\usepackage{amssymb}
\usepackage{color}
\usepackage{tikz}
\usepackage{mathrsfs}
\usepackage[all]{xy}
\usepackage{lscape}
\usepackage{stmaryrd}
\usepackage{mathdots}
\usepackage{bbm}
\usepackage{bussproofs}
\usepackage{subcaption}

\EnableBpAbbreviations

\usetikzlibrary{cd}

\theoremstyle{plain}

\newcommand{\B}[1]{\mathbf{#1}}
\newcommand{\TT}[1]{\mathtt{#1}}

\newcommand{\SF}[1]{\mathsf{#1}}

\newcommand{\C}[1]{\mathcal{#1}}
\newcommand{\BB}[1]{\mathbb{#1}}

\newcommand{\OV}[1]{\overline{#1}}
\newcommand{\F}[1]{\mathfrak{#1}}

\newcommand{\To}{\Rightarrow}

\newcommand{\arrow}[2]{\left (#1\To#2\right )}

\newcommand{\sk}[0]{\mathsf{sk}}
\newcommand{\forpiudue}[0]{\forall^{+}}
\newcommand{\formenodue}[0]{\forall^{-}}
\newcommand{\forpiupro}[0]{\forall^{+}_{\TT P}}
\newcommand{\formenopro}[0]{\forall^{-}_{\TT P}}
\newcommand{\Cl}[0]{\mathfrak C}
\newcommand{\CR}[0]{\mathsf{CR}}
\newcommand{\CRuno}[0]{\mathsf{CR}'}
\newcommand{\CRdue}[0]{\mathsf{CR}''}
\newcommand{\Rel}[2]{\langle #2\rangle^{\Cl}_{#1}}
\newcommand{\Relz}[2]{r}

\newcommand{\model}[1]{\llbracket #1 \rrbracket}

\newcommand{\FF}{\mathsf F}

\definecolor{color0}{HTML}{4682B4}

\makeindex

\title[Completeness/parametricity in the realizability semantics of System $\FF$]{On completeness and parametricity\\ in the realizability semantics of System $\FF$}

\author{Paolo Pistone}
\address{Wilhelm Schickard Institut, Eberhard Karls Universit\"at T\"ubingen, Sand 13, D-72076 T\"ubingen}
\email{paolo.pistone@uni-tuebingen.de}

\begin{document}
\maketitle

\begin{abstract}
We investigate completeness and parametricity for a general class of realizability semantics for System $\FF$ defined in terms of closure operators over sets of $\lambda$-terms. 
This class includes most semantics used for normalization theorems, as those arising from Tait's saturated sets and Girard's reducibility candidates. 

We establish a completeness result for positive types which subsumes those existing in the literature, and we show that closed realizers satisfy parametricity conditions expressed either as invariance with respect to logical relations or as dinaturality.
Our results imply that, for positive types,  typability, realizability and parametricity are equivalent properties of closed normal $\lambda$-terms.
\end{abstract}

\tableofcontents

\section{Introduction}\label{sec1}

In this paper, we call realizability semantics for typed
$\lambda$-calculi interpretations of types with well-behaved sets of $\lambda$-terms (or, more generally, with subsets of some \emph{partial combinatory algebra} \cite{Streicher}),
called the realizers of the type.
%
System $\FF$ came equipped since its birth in \cite{Girard72} with two realizability semantics. First, the
 \emph{reducibility candidates} semantics, which constituted the starting point for the now common use of realizability  to prove normalization theorems for higher order type theories (see \cite{Gallier, Krivine}). Second, the semantics of \emph{partial equivalence relations ($\mathsf{PER}$)}, which constituted the starting point for the mathematical investigation of realizability models, leading to striking categorical structures (e.g.\ the \emph{effective topos} \cite{Hyland1988}).

Normalization results for typed $\lambda$-calculi are usually proved by establishing the \emph{soundness} of a given realizability semantics,
that is, by showing that all terms of a given type are realizers of that type. 
The converse property, \emph{completeness}, that is the fact that all realizers of a type provably have that type, has also been established for some realizability semantics \cite{Hindley, Rosolini1989, Nour1998}. 
While soundness is usually established for full System F, 
completeness results are generally restricted to types in which the second order quantifier $\forall$ only occurs in positive position (e.g. algebraic types). Following \cite{Krivine}, we will refer to these types as \emph{positive types}. 

A common feature of realizability models is that universal quantification is interpreted by set-theoretic intersection. Hence the realizers of a universally quantified type $\forall X\sigma$ are $\lambda$-terms which are, at the same time, realizers of all possible instantiations of $\sigma$. 
This property plays an important role in the aforementioned completeness results. Moreover, due to this property, realizability yields one (among several existing) formalisations of Strachey's informal notion of \emph{parametric polymorphism} \cite{Strachey1967}. Strachey notoriously made a distinction between \emph{ad hoc} and \emph{parametric} polymorphic programs, where the latter are, roughly, programs which behave in the same way for all possible type instantiations. A realizer of a universally quantified type is thus parametric in this informal sense, since the same term realizes all instantiations of the type. However, the relationship between realizability and other formalisations of parametricity is not yet completely understood.

\subsection{Contributions}
In this paper we investigate completeness and parametricity for a general class of realizability models defined in terms of closure operators over sets of $\lambda$-terms \cite{Vouillon2004,Riba2008}.
In particular, 
(1) we establish a general completeness theorem for positive types, which subsumes existing ones and applies to other semantics and 
(2) we show that the closed realizers of a positive type satisfy other parametricity conditions given in terms of \emph{logical relations} and \emph{dinaturality}.
(1) and (2) allow to conclude that, for positive types, typability, realizability and parametricity are equivalent properties of closed normal $\lambda$-terms.

The models we consider include the semantics generally used to prove normalization theorems for type theories: \emph{saturated sets}, \emph{reducibility candidates}, \emph{$\beta$ and $\beta\eta$-stable sets} \cite{Tait1967, Girard72, Krivine}.
We do not consider categorical models, but we expect that 
the techniques here presented might scale to a categorical setting (e.g. to the category of \emph{modest sets} of $\lambda$-terms \cite{Streicher}). 

\subsubsection{Completeness}
The completeness problem is the one to show that, for a certain class $\C C$ of types, for any type $\sigma\in \C C$, if $M$ is a closed (normal) realizer of $\sigma$, then $M$ provably has type $\sigma$. 
Several completeness results have been established for simple types and positive types for semantics based on sets of $\lambda$-terms \cite{Hindley, Hindley1982, Hindley1983b, Krivine, Nour1998}. All these results use a similar method which consists in constructing a ``term model'' in which types are interpreted as sets of provably typable terms.
By considering the class of realizability semantics generated by a closure operator over sets of $\lambda$-terms, we provide a general completeness argument based on this method which subsumes the results mentioned. While Girard's original reducibility candidates semantics seems to fall out of this approach, we show that our result applies to two variants of this semantics introduced in \cite{Cousineau}.

\subsubsection{Parametricity}
We compare realizability semantics with the approaches to parametric polymorphism based on \emph{logical relations} \cite{Reynolds1983} and on \emph{dinatural transformations} \cite{Bainbridge1990}.


Logical relations are a standard semantic technique used to prove properties of programs (e.g.\ \emph{program equivalence} or \emph{representation independence}, see \cite{Ahmed2011, Ahmed2017}). The basic result on logical relations is a proof that 
typable terms are \emph{invariant} with respect to logical relations, that is, that they map related objects into related ones. This invariance property can be read as a parametricity condition for polymorphic programs, as it expresses the fact that the programs behave in related ways in related contexts.

We propose a formalisation of logical relations in the realizability semantics generated by closure operators.
Our construction relies on the demand that closure operators be \emph{stable by union} (a property first investigated in \cite{RibaUnion} for reducibility candidates). Under this condition, closed sets form a topology, and closed logical relations can be defined by the product topology.

Our main result is the \emph{invariance theorem} (Theorem \ref{param1}) stating that closed realizers are invariant with respect to closed logical relations (under suitable conditions of the closure operator). 
A consequence is that all realizability semantics used for normalization theorems satisfy the parametricity condition expressed by invariance with respect to $\beta$-stable and $\beta\eta$-stable logical relations.
 The proof is obtained by adapting the ``term model'' technique used for completeness to the case of logical relations.
The invariance theorem generalizes a similar result proved in \cite{FSCD2017} for the $\beta\eta$-stable semantics and restricted to simple types.

For second order types, the invariance property does not coincide with \emph{Reynolds' parametricity} \cite{Reynolds1983}, which is based on a different interpretation of the universal quantifier.
We shortly discuss why, in order to account for Reynolds' parametricity, one has to enrich closed sets of $\lambda$-terms with  equivalence relations extending $\beta\eta$-equivalence (as in parametric $\mathsf{PER}$ models, \cite{Bainbridge1990}). 

%
%

We then consider the approach to parametricity through dinatural transformations \cite{Bainbridge1990, Girard1992}. In categorical semantics, types can be interpreted as functors in which variables may occur both covariantly and contravariantly. Dinatural transformations generalize natural transformations to such functors \cite{MacLane}.

In contrast with logical relations, dinaturality provides a purely equational criterion for parametricity, as it is expressed by a family of commuting diagrams.
We consider a syntactic approach to dinaturality in which commuting diagrams are replaced by $\beta$ and $\beta\eta$-equations.

Syntactic dinaturality can be defined as dinaturality in the syntactic category generated by System $\FF $ terms (as in \cite{Delatail2009}). 
We adopt here a more uniform definition of syntactic dinaturality using indeterminates, already used in \cite{FSCD2017}. 
This definition allows one to express dinaturality by a single equation, making this condition more amenable to syntactic treatment.
While a term satisfying this equational condition yields a dinatural transformation in the syntactic category, we do not know if the converse holds.

We prove that the closed realizers of a positive type are dinatural. This result is obtained by showing that invariant closed terms satisfy the syntactic dinaturality condition. We finally discuss some applications of this result, including a second argument to prove completeness for positive types.

\subsection{Related work}

The two variants $\C S_{\CRuno}$ and $\C S_{\CRdue}$ of reducibility candidates  here investigated come from \cite{Cousineau}, where a completeness problem of a different nature is considered for this semantics.
Moreover, we take from \cite{Cousineau} the remark that the  ``term model'' method to prove completeness relies on a property of closure operators, that we call here \emph{$\FF$-adaptedness} (Definition \ref{fada}).

Our analysis of logical relations relies on stability by union of closure operators. This property is established for reducibility candidates in \cite{RibaUnion, Riba2009}. Here we show that this property extends to the variants $\C S_{\CRuno}$ and $\C S_{\CRdue}$, by adopting a similar proof method.

Connections between realizability and parametricity have been investigated in the literature at a formal level (for formal approaches to parametricity see \cite{Abadi1993}, \cite{Mairson1991} and \cite{Plotkin1993}). 
In \cite{Wadler2007} realizability and parametricity are related as formal translations within a second order intuitionistic predicate calculus $P^{2}$. In particular, the realizers of inductive types are shown to correspond exactly to the parametric terms which satisfy the predicates expressing the so-called ``Reynolds'' translation of those types in $P^{2}$.
As a consequence, binary parametricity, as a formal translation, corresponds to the composition of unary parametricity and realizability. This idea is generalized in \cite{Bernardy2011}, where formal realizability and parametricity are related in the case of Pure Type Systems, and it is shown that $n+1$-ary parametricity corresponds to $n$-ary parametricity composed with realizability. In other words, realizability is seen as a translation which increases arities of parametricity. In the concluding section we suggest that the extension of our treatment of realizability and parametricity to Krivine's functional arithmetic $\C{AF}2$ \cite{Krivine} might allow to establish a similar connection between realizability and parametricity within our approach, as a consequence of the invariance theorem.

\subsection{Structure of the paper}
The paper can be ideally divided in two parts: the first part (sections \ref{sec2}, \ref{sec3}) recalls some syntactic properties of System $\FF$ and introduces our general approach to realizability semantics by means of closure operators; the second part (sections \ref{sec4}, \ref{sec5}, \ref{sec6}) relates realizability with completeness, logical relations and dinaturality.
In Section \ref{sec2}, after recalling System $\FF$ and the simply typed $\lambda$-calculus (Subsection \ref{sec21}), we introduce positive types and prove some syntactic results which relate them with simple types (Subsection \ref{sec22}). In Section \ref{sec3} we introduce the realizability semantics generated by a closure operator and we show how some well-known semantics which fit into our framework. In Section \ref{sec4} we discuss completeness for positive types (Corollary \ref{forallcr3}). In Section \ref{sec5} we formalize $\Cl$-closed logical relations (Subsection \ref{sec51}) and we prove the $\Cl$-invariance theorem (Theorem \ref{param1}) (Subsection \ref{sec52}). Then we shortly discuss the problems to extend our approach to Reynolds' parametricity (Subsection \ref{secRey}).
In Section \ref{sec6} we introduce syntactic dinaturality (Subsection \ref{sec61}), we prove that $\Cl$-invariant closed terms are syntactically dinatural (Subsection \ref{sec62}), and we discuss applications of this result (Subsection \ref{sec63}). In the concluding section (Section \ref{sec7}), we put all results together to state the equivalence of realizability, $\Cl$-invariance, dinaturality and typability for positive types (Theorem \ref{general}) and we suggest some open problems and further directions.

To enhance readability, we enclose an index of defined terms at the end of the paper.

\section{System $\FF$ and positive types}\label{sec2}

We recall System $\FF$ and the simply typed $\lambda$-calculus (in the formulation \emph{\`a la Curry}) and we
discuss the class of positive types, introduced in \cite{Krivine}. 
The main result of this section (Theorem \ref{forpi}) allows to compare simple types and positive types and will be used in several places in the next sections to lift results from simple to positives types.  

\subsection{$\lambda_{\to}$ and System $\FF$}\label{sec21}

\subsubsection{$\lambda$-calculus}\label{lcf}

In the following, by the letters $M,N,P,\dots$ we will indicate elements of the set $\Lambda$ of untyped $\lambda$-terms, subject to usual $\alpha$-equivalence. By the letters  $x,y,z,\dots$ we will indicate \emph{term variables}, i.e. the variables that might occur free or bound in untyped $\lambda$-terms. We let $\TT{TermVar}$ indicate the set of term variables. 
As usual, by the expression $M_{1}M_{2}M_{3}\dots M_{n}$ we will indicate the term $(\dots ((M_{1}M_{2})M_{3})\dots M_{n})$. For all $M\in \Lambda$, $\TT{FV}(M)$ and $\TT{BV}(M)$ indicate, respectively, the set of free and bound term variables occurring in $M$.

Given a reduction relation $R$, we will indicate by $\to_{R}$ one-step $R$-reduction, by $\to_{R}^{*}$ the transitive closure of $\to_{R}$ and by $\simeq_{R}$ the transitive-reflexive-symmetric closure of $\to_{R}$, i.e. the equivalence relation generated by $R$. Finally, by a $R$-normal $\lambda$-term we indicate a term to which no $R$-reduction can be applied.

In this paper we will consider the following reduction relations: $\beta$-reduction, $\eta$-reduction, $\beta\eta$-reduction and $wh$-reduction (weak-head reduction), where the latter is the reduction generated by 
\begin{equation}
(\lambda x.P)QQ_{1}\dots Q_{n} \ \to_{wh} \ P[Q/x]Q_{1}\dots Q_{n}
\end{equation}

We let $\C{SN}$ indicate the set of \emph{$\beta$-strongly normalizing} $\lambda$-terms, $\C N$ indicate the set of \emph{neutral $\lambda$-terms}, i.e. the $\lambda$-terms not beginning with a $\lambda$ and $\C V=\Lambda-\C N$ indicate the set of non-neutral $\lambda$-terms, that we call \emph{values}. We let $\C{N}^*$ be the set of neutral non $\beta$-normal $\lambda$-terms.
\index{$\lambda$-calculus!Strongly normalizing terms $\C{SN}$}
\index{$\lambda$-calculus!Neutral terms $\C N$}
\index{$\lambda$-calculus!Values $\C V$}
\index{$\lambda$-calculus!Neutral non $\beta$-normal terms $\C N^{*}$}
\index{$\lambda$-calculus!$\beta, \beta\eta, wh$, reductions}

Finally, for all $M \in\C{SN}$, we let $\mathsf d(M)$, the \emph{depth of $M$}, be the maximum length of a reduction sequence from $M$ to its $\beta$-normal form.
\index{$\lambda$-calculus!Depth ot a term $\mathsf d(M)$}

\subsubsection{System $\FF$}

We recall the formulation ``{\`a la Curry}'' of System $\FF$:

\begin{defi}[System $\FF$ ``\`a la Curry'']\index{System F!System F ``\`a la Curry''}

Given a countable set $\TT{TypeVar}=\{X_{1},X_{2},X_{3},\dots\}$ (that we will also write as $X,Y,Z,\dots$) of symbols, called \emph{type variables} (or, simply, variables when no ambiguity occurs), the set $\TT {T}$ of types of System $\FF$ is defined by the grammar below:
\begin{align}
\TT T \:= \ X \mid \TT T\to \TT T \mid \forall X\TT T
\end{align}

Given $\sigma\in \TT T$, $\TT{FV}(\sigma)$ and $\TT{BV}(\sigma)$ indicate, respectively, the set of the \emph{free} and \emph{bound} type variables occurring in $\sigma$.

A \emph{type declaration} is an expression of the form $x:\sigma$, where $x$ is a term variable and $\sigma$ is a type. A \emph{context} $\Gamma$ is a finite set of type declarations in which distinct declarations have distinct variables. 
A \emph{judgement} is an expression of the form $\Gamma\vdash M:\sigma$, where $\Gamma$ is a context, $M$ a term and $\sigma$ a type.

The typing derivations of System $\FF$ are generated by the rules in figure \ref{rules}, in which $X$ is bindable in $\Gamma$ if, for any type declaration $x:\sigma\in\Gamma$, $X\notin \TT{FV}(\sigma)$.

\end{defi}

\begin{figure}
$$
\begin{matrix}
\infer[(id)]{\Gamma\vdash x:\sigma}{x:\sigma\in \Gamma}   &   	\\  \ & \ \\
\infer[(\to E)]{\Gamma\vdash MN:\tau}{\Gamma\vdash M:\sigma\to\tau  &  \Gamma\vdash N:\sigma}  & \infer[(\to I)]{\Gamma-\{x:\sigma\}\vdash \lambda x.M:\sigma\to\tau}{\Gamma\vdash M:\tau & x:\sigma\in \Gamma}  \\  \ & \ \\
\infer[(\forall E)]{\Gamma\vdash M:\sigma[\tau/ X]}{\Gamma\vdash M:\forall X\sigma}  &   \infer[(\forall I)]{\Gamma\vdash M:\forall X\sigma}{\Gamma\vdash M:\sigma &  X\text{ bindable in $\Gamma$}}\\
\end{matrix}
$$
\caption{System $\FF$ rules}
\label{rules}
\end{figure}

We indicate by $\TT T_{0}$ the set of \emph{simple types}, i.e. those types $\sigma\in \TT T$ containing no occurrence of the quantifier $\forall$. 
The simply-typed $\lambda$-calculus $\lambda_{\to}$ is the subsystem of $\FF$ in which types are restricted to $\TT T_{0}$ and rules are restricted to $(id), (\to E)$ and $(\to I)$.

The set $S(\sigma)$ of \emph{subtypes} of a type $\sigma\in \TT T$ is defined by induction as follows:
\begin{equation}
\begin{split}
S(X)& :=\{X\}  \\
S(\sigma\to \tau) &:= S(\sigma)\cup S(\tau)\cup \{\sigma\to \tau\} \\
S(\forall X\sigma) & := S(\sigma)\cup \{\forall X\sigma\}
\end{split}
\end{equation}

\begin{conv}\label{convention}
We adopt some conventions about the names of bound variables appearing in types.
We will suppose that in any type $\sigma\in \TT T$, for any variable $X\in \TT{BV}(\sigma)$ there is exactly one subtype of $\sigma$ of the form $\forall X\tau$. Equivalently, that for any two distinct subtypes $\forall X\tau, \forall X'\tau'\in S(\sigma)$, $X\neq X'$. 
We will similarly suppose that in any type judgement $\Gamma \vdash M:\sigma_{n+1}$, where $\Gamma=\{x_{1}:\sigma_{1},\dots, x_{n}:\sigma_{n}\}$, for any variable $X\in \bigcup_{i=1}^{n+1}\TT{BV}(\sigma_{i})$ there is exactly one $1\leq j\leq n+1$ and exactly one subtype of $ S(\sigma_{j})$ of the form $\forall X\tau$.

Given a type $\sigma\in \TT T$ and a finite (possibly empty) list of pairwise distict type variables $\C X=X_{1},\dots, X_{n}$, by $\forall \C X \sigma$ we indicate the type $\forall X_{1}\dots \forall X_{n}\sigma$. We let
$\TT{ListVar}$ indicate the set of finite lists of pairwise distinct type variables.
For $\C X\in \TT{ListVar}$, we indicate by $\C X^{\dag}$ the set of type variables occurring in $\C X$.

\end{conv}

For any $\sigma\in \TT T$, by a position in $\sigma$ we indicate a node in the syntactic tree of $\sigma$. A node is  positive (resp.\ negative) if the unique path from the root to the node passes an even (resp.\ odd) number of times in the lefthand side of a $\to$-bifurcation.

We recall that System $\FF$ \emph{\`a la Curry} enjoys the \emph{subject $\beta$-reduction} property: if $\Gamma\vdash M:\sigma$ and $M\to_{\beta}^{*}M'$, then $\Gamma\vdash M':\sigma$. On the contrary, \emph{subject $\eta$-reduction} fails. For instance $\sigma=(\forall YX\to X)\to (X\to X)$, $ \vdash\lambda x.\lambda y.xy: \sigma$ holds but $\vdash\lambda x.x:\sigma$ does not hold.


\subsubsection{Some useful terms}

We introduce some further notations for some terms typable in System $\FF$. Given terms $M,N$ we let:
\begin{equation}
\index{System F!Terms $M\circ N$ and $\arrow{M}{N}$}
\begin{split}
M\circ N & :=\lambda x.M(Nx)\\
\arrow{M}{N} &:=\lambda x.\lambda y. N(x(My))
\end{split}
\end{equation}

The following properties are easily verified:
\begin{lem}
\begin{enumerate}[label=(\roman*)]

\item If $\Gamma\vdash M:\tau\to \rho$ and $\Gamma\vdash N: \sigma\to \tau$, then $\Gamma\vdash M\circ N:\sigma\to \rho$. 

\item If $\Gamma\vdash M:\sigma\to \tau$ and $\Gamma\vdash N:\sigma'\to \tau'$, then $\Gamma\vdash \arrow{M}{N} : ( \tau\to \sigma'  )\to (\sigma\to \tau'   )$.

\end{enumerate}
\end{lem}

\begin{lem}\label{lemma:comput}
$\arrow{P}{Q} \circ \arrow{P'}{Q'} \ \simeq_{\beta} \ \arrow{P'\circ P}{Q\circ Q'}$.

\end{lem}
\begin{proof}
A short computation shows that both terms $\beta$-reduce to 
$\lambda x.\lambda y. Q(Q'(x(P'(P(y)))))$.
\end{proof}

For any $\sigma\in \TT T$, we define, by induction, a $\beta$-normal term $U_{\sigma}$ with a unique free variable $x$ such that $x:\sigma\vdash U_{\sigma}:\sigma$:
\begin{equation}
\begin{split}U_{X} & := x\\
U_{\sigma_{1}\to \sigma_{2}} & := \lambda z. U_{\sigma_{2}}\left [ xU_{\sigma_{1}}[z/x]/x\right ]\\
U_{\forall X\sigma} & := U_{\sigma}
\end{split}
\end{equation}
\index{System F!Terms $U_{\sigma}$ and $I_{\sigma}$}

\begin{prop}\label{prop:usigma}
For all $\sigma\in \TT T$, $x:\sigma\vdash U_{\sigma}:\sigma$ is derivable and moreover $U_{\sigma}\to_{\eta}^{*}x$.
\end{prop}
\begin{proof}
We argue by induction on $\sigma$. Both claims are obvious for $\sigma=X$. If $\sigma=\sigma_{1}\to \sigma_{2}$, then, by induction hypothesis,
$x:\sigma_{i}\vdash U_{\sigma_{i}}:\sigma_{i}$ for $i=1,2$, so we deduce 
$x:\sigma_{1}\to \sigma_{2}, z: \sigma_{1}\vdash U_{\sigma_{2}}[x U_{\sigma_{1}}[z/x]/x]: \sigma_{2}$, whence
$x:\sigma\vdash U_{\sigma}:\sigma$. Moreover, by the induction hypothesis $U_{\sigma_{i}}\to^{*}_{\eta} x$, for $x=1,2$, whence
$U_{\sigma}\to^{*}_{\eta} \lambda z. xz \to_{\eta} x$. 
If $\sigma=\forall X\sigma'$, by induction hypothesis, 
$x:\sigma'\vdash U_{\sigma'}:\sigma'$; from $x:\sigma\vdash  x:\sigma'$ we deduce $x:\sigma\vdash U_{\sigma'}:\sigma'$, and we conclude $x:\sigma\vdash U_{\sigma'}:\sigma$ since $X$ is not free in $\sigma$. That $U_{\sigma}\to^{*}_{\eta} x$ is just the induction hypothesis.
\end{proof}

For all $\sigma\in \TT T$, we let $I_{\sigma}= \lambda x.U_{\sigma}$. From Proposition \ref{prop:usigma} it follows that 
$\vdash I_{\sigma}:\sigma\to \sigma$ and that $I_{\sigma}\to^{*}_{\eta} \lambda x.x$. Moreover we have the following two lemmas.

\begin{lem}\label{lemma:usigma}
If $M$ is $\beta$-normal and $\Gamma\vdash M:\sigma$, then there exists $M'$ $\beta$-normal such that $I_{\sigma}M\to_{\beta}^{*}M'$ and $M'\to^{*}_{\eta}M$.
\end{lem}
\begin{proof}
  We have $I_{\sigma}M\to_{\eta}^{*} (\lambda x.x)M$. Let $M'$ be the $\beta$-normal form of $I_{\sigma}M$. By the commutation property of $\beta\eta$-reduction \cite[Lemma 3.3.8 p.66]{Baren95}, there exists $P$ such that $M'\to_{\eta}^{*} P$ and $(\lambda x.x)M\to_{\beta}^{*} P$. From $M'\to_{\eta}^{*}P$ we deduce that $P$ is $\beta$-normal  (as $\eta$-reduction does not create $\beta$-redexes), and since $(\lambda x.x)M\to_{\beta}^{*}P$ and $M$ is $\beta$-normal, it must be $P=M$.
\end{proof}

\begin{lem}\label{lemma:usigmaprop}
For all types $\sigma, \tau$:
\begin{enumerate}[label=(\roman*)]
\item $\arrow{I_{\sigma}}{I_{\tau}}\to_{\beta}^{*}I_{\sigma\to \tau}$.
\item $I_{\sigma}\circ I_{\sigma}\to_{\beta}^{*}I_{\sigma}$.
\end{enumerate}
\end{lem}
\begin{proof}
For point (i), we have 
$\arrow{I_{\sigma}}{I_{\tau}}= \lambda x.\lambda z. I_{\tau}(x( I_{\sigma}z))\to_{\beta}^{*} \lambda x.
\lambda z.U_{\tau}[xU_{\sigma}[z/x]/x]= I_{\sigma\to \tau}$.
For point (ii) we argue by induction on $\sigma$. 
If $\sigma=X$, then $I_{\sigma}\circ I_{\sigma}= \lambda z.(\lambda y.y)((\lambda x.x)z)\to_{\beta}^{*}\lambda z.z=I_{\sigma}$.
If $\sigma=\tau\to \rho$, then by point (i) and Lemma \ref{lemma:comput}
$I_{\sigma}\circ I_{\sigma} \simeq_{\beta} \arrow{I_{\tau}}{I_{\rho}}\circ \arrow{I_{\tau}}{I_{\rho}}\simeq \arrow{I_{\tau}\circ I_{\tau}}{I_{\sigma}\circ I_{\sigma}}$, and by the induction hypothesis and point (i) we deduce $\arrow{I_{\tau}\circ I_{\tau}}{I_{\rho}\circ I_{\rho}}\simeq \arrow{I_{\tau}}{I_{\rho}}\simeq_{\beta}I_{\sigma}$.
Hence $I_{\sigma}\circ I_{\sigma}\simeq_{\beta} I_{\sigma}$, and since $I_{\sigma}$ is $\beta$-normal, we conclude
$I_{\sigma}\circ I_{\sigma}\to_{\beta}^{*}I_{\sigma}$.
Finally, for $\sigma=\forall X\tau$, the claim follows from the induction hypothesis as $I_{\sigma}=I_{\tau}$.
\end{proof}

\subsection{Positive types}\label{sec22}

We introduce positive types \cite{Krivine} and we compare typability with respect to such types and with respect to simple types. 
We analyze the role of $\eta$-equivalence by distinguishing between \emph{proper} and \emph{improper} quantification (as in \cite{Nour1998}).
We finally prove that the typability of a closed $\beta$-normal term $M$ at a positive type $\sigma$ is equivalent to the typability of $M$ at a certain simple type, called the \emph{skeleton} of $\sigma$, when all quantifications in $\sigma$ are proper. If $\sigma$ contains some improper quantification, then this equivalence holds only up to $\eta$-equivalence.

\subsubsection{Positive second order types}

Positive second order types (introduced in \cite{Krivine}) are defined by the restriction that the $\forall$ quantifier can only occur in positive positions.

\begin{defi}[positive and negative types, \cite{Krivine}]\label{forall}\index{System F!Classes of types:!$\forpiudue$ and $\formenodue$}
The classes $\forpiudue,\formenodue$ are defined inductively as follows:
\begin{itemize}
\item $X\in \forpiudue, \formenodue$;
\item if $\sigma\in \formenodue$, and $\tau\in \forpiudue$, then $\sigma\to \tau\in \forpiudue$;
\item if $\sigma\in \forpiudue$, and $\tau\in  \formenodue$, then $\sigma\to \tau\in \formenodue$;

\item if $\sigma\in \forpiudue$, then $\forall X\sigma\in \forpiudue$.
\end{itemize}

\end{defi}

It is easily checked that $\forpiudue$ is the class of types in which the $\forall$ quantifier occurs only in positive positions and $\formenodue$ is the class of types in which the $\forall$ quantifier occurs only in negative positions.

\begin{lem}\label{simpletypes}
$ \forpiudue\cap \formenodue= \TT T_{0}$.
\end{lem}
\begin{proof}
That $\TT T_{0}\subseteq \forpiudue\cap \formenodue$ is easily verified by induction on $\sigma\in \TT T_{0}$. 
For the converse direction, if $\sigma\in \forpiudue\cap \formenodue$ then if a quantifier occured in $\sigma$,  it would occur in a position both positive and negative, which is impossible. Therefore we can conclude $\sigma\in \TT T_{0}$.
\end{proof}

\begin{lem}\label{lemma:piumeno}
If $\sigma\in \formenodue$, then $\sigma=\sigma_{1}\to \dots \to \sigma_{p}\to Y$, for some $p\in \BB N$, variable $Y$ and types $\sigma_{1},\dots, \sigma_{p}\in \forpiudue$.
\end{lem}
\begin{proof}
By induction on $\sigma\in \formenodue$: if $\sigma=X$, then the claim obviously holds; otherwise, it must be $\sigma=\sigma_{1}\to \sigma_{2}$, where $\sigma_{1}\in \forpiudue$ and $\sigma_{2}\in \formenodue$, and by induction hypothesis
$\sigma_{2}=\tau_{1}\to \dots \to \tau_{q}\to Y$, for some variable $Y$ and types $\tau_{1},\dots, \tau_{2}\in \forpiudue$, so $\sigma=\sigma_{1}\to \tau_{1}\to \dots \to \tau_{q}\to Y$ satisfies the claim.
\end{proof}

To compare positive types and simple types it is useful to consider a subclass of $\forpiudue$ (as well as a subclass of $\formenodue$) defined as follows:

\begin{defi}[$\B \Pi,\mathsf{co}\text{-}\B\Pi$-types]\label{pi}\index{System F!Classes of types:!$B \Pi$ and $\mathsf{co}\text{-}\B\Pi$}
Let $\sigma$ be a type.
\begin{enumerate}[label=(\roman*)]
\item $\sigma\in \B \Pi$ if for some $\C X\in \TT{ListVar}$ and some $\sigma'\in \TT T_{0}$, $\sigma=\forall \C X\sigma'$;

\item $\sigma\in \mathsf{co}\text{-}\B\Pi$ if for some $\sigma_{1},\dots, \sigma_{p}\in \B \Pi$ and a variable $X$, $\sigma= \sigma_{1}\to \dots \to \sigma_{p}\to X$.

\end{enumerate}
\end{defi}

It is clear that $\B \Pi\subseteq \forpiudue$ and $\mathsf{co}\text{-}\B\Pi\subseteq \formenodue$, but the converse inclusions do not hold.
For instance, the type $ \forall Y((\forall XX\to Y)\to Y)$ is in $\forpiudue$ but is not in $\B \Pi$.

\subsubsection{The skeleton}

Given a type $\sigma$ and a set $S\subseteq \TT{TypeVar}$ of variables not occurring free in $\sigma$, we can define the type $\sigma^{S}$ obtained by 
deleting from $\sigma$ all quantifiers of the form $\forall X$, for all $X$ occurring in $S$. More formally:

\begin{defi}\label{deelt}\index{System F!Types $\sigma^{S}$}
Let $\sigma\in \TT T$ and let $S\subseteq\TT{TypeVar}$ be such that $S\cap \TT{FV}(\sigma)=\emptyset$. We define the type $\sigma^{S}$ as follows:
\begin{equation}
\begin{split}
X^{S} & := X\\
(\tau\to \rho)^{S} &:=  \tau^{S}\to \rho^{S}\\
(\forall X\tau)^{S} &:=
\begin{cases}
\tau^{S}  & \text{ if }X\in S\\
\forall X\tau^{S} & \text{ otherwise}
\end{cases}
\end{split}
\end{equation}
\end{defi}
The \emph{skeleton} $\sk(\sigma)$ of a type $\sigma$ is the simple type obtained from $\sigma$ by deleting all quantifiers.

\begin{defi}[skeleton]\index{System F!Skeleton of a type $\sk(\sigma)$}
For all $\sigma\in \TT T$, we let $ \sk(\sigma):= \sigma^{\TT{BV}(\sigma)}\in \TT T_{0}$.
\end{defi}

From Definition \ref{deelt} it is clear that $ \sk(X)=X$, 
$ \sk(\sigma\to \tau)= \sk(\sigma)\to  \sk(\tau)$ and $\sk(\forall X\sigma)= \sk(\sigma)$.
Given a class of types $\C C$ and a context $\Gamma$, by the expression $\Gamma\in\C  C$ we indicate that, for any type $\sigma$ occurring in some type declaration in $\Gamma$, $\sigma\in\C  C$. Also, given $\Gamma=\{x_{1}:\sigma_{1},\dots, x_{n}:\sigma_{n}\}$ by $ \sk(\Gamma)$ we indicate the context $\{x_{1}: \sk(\sigma_{1}),\dots, x_{n}: \sk(\sigma_{n})\}$.

In the following we will consider derivations $\Gamma\vdash M:\sigma$, where $\Gamma\in \formenodue$ and $\sigma\in \forpiudue$. These derivations satisfy the property below, which is proved in Appendix \ref{app:foralle}

\begin{prop}\label{prop:foralle}
Suppose $M$ is $\beta$-normal and $\Gamma\vdash M:\sigma$ is derivable, where $\Gamma\in \formenodue$ and $\sigma\in \forpiudue$. Then there exists a derivation of $\Gamma\vdash M:\sigma$ which does not employ the rule $\forall$E and such that, for any type judgement $\Gamma'\vdash M':\sigma'$ occurring in it, $\Gamma'\in \formenodue$ and $\sigma'\in \forpiudue$.

\end{prop}

From Proposition \ref{prop:foralle} we deduce the following property of skeletons:

\begin{lem}\label{beta2}
 Let $M$ be a $\beta$-normal $\lambda$-term, such that $\Gamma\vdash M:\sigma$, where $\Gamma\in \formenodue$ and $\sigma \in \forpiudue$. Then $ \sk(\Gamma)\vdash M:  \sk(\sigma)$. 
\end{lem}
\begin{proof}
We argue by induction on a typing derivation which does not use the rule $\forall$E (by using Proposition \ref{prop:foralle}):

\begin{itemize}
\item if the derivation is $x:\sigma\vdash x:\sigma$, then $\sigma\in \formenodue\cap \forpiudue$, hence $\sigma\in \TT T_{0}$ (by Lemma \ref{simpletypes}) and $ \sk(\sigma)=\sigma$, so the claim trivially holds;

\item If the derivation ends by $\AXC{$\Gamma, x:\sigma\vdash M:\tau$}\UIC{$\Gamma\vdash \lambda x.M:\sigma\to \tau$}\DP$ then, since $\sigma\to \tau\in \forpiudue$, $\sigma\in \formenodue$ and $\tau\in \forpiudue$; by the induction hypothesis, then $ \sk(\Gamma),x:  \sk(\sigma)\vdash M:  \sk(\tau)$, whence the claim follows from $ \sk(\sigma\to \tau)= \sk(\sigma)\to  \sk(\tau)$;

\item If the derivation ends by $\Gamma\vdash xM_{1}\dots M_{p}:\sigma$, with $\Gamma\vdash M_{i}:\tau_{i}$ for $i=1,\dots,p$; then $ \tau_{1}\to \dots \to \tau_{p}\to \sigma\in \formenodue$, hence, $\tau_{i}\in \forpiudue$ and $\sigma\in \forpiudue\cap \formenodue$ must be a simple type. By the induction hypothesis then $ \sk(\Gamma)\vdash M_{i}:  \sk(\tau_{i})$, hence the claim follows from $ \sk(\tau_{1}\to \dots \to \tau_{p}\to \sigma)=  \sk(\tau_{1})\to \dots \to  \sk(\tau_{p}) \to \sigma$;
\item If the derivation ends by $\AXC{$\Gamma\vdash M:\sigma$}\UIC{$\Gamma\vdash M:\forall X\sigma$}\DP$, where $X\notin \TT{FV}(\Gamma)$, then by the induction hypothesis $ \sk(\Gamma)\vdash M:  \sk(\sigma)$ and the claim follows from $ \sk(\forall X\sigma)= \sk(\sigma)$. \qedhere
\end{itemize}
\end{proof}


\subsubsection{A translation from $\B \Pi$ to $\forpiudue$}

A first step to compare positive and simple types comes from a comparison between the classes $\forpiudue$ and $\B \Pi$. To this end we introduce translations $[\_]^{+}: \forpiudue\to \B \Pi$ and $[\_]^{-}: \formenodue\to \mathsf{co}\text{-}\B\Pi$ as follows:

\begin{defi}[positive and negative translations]\index{System F!Types $\sigma^{+}$ and $\sigma^{-}$}
If $\sigma\in \forpiudue$, we let 
$$\sigma^{+}:= \forall \C X\sk(\sigma)\in \B \Pi$$
where $\C X$ is any list such that $\C X^{\dag}=\TT{BV}(\sigma)$.
If $\sigma\in \formenodue$, we let
$$\sigma^{-}:= \forall \C X_{1}\sk(\sigma_{1})\to \dots \to \forall \C X_{p}\sk(\sigma_{p}) \to Y\in \mathsf{co}\text{-}\B\Pi$$ 
where 
$\sigma =\sigma_{1}\to \dots \to \sigma_{p}\to Y$ (by Lemma \ref{lemma:piumeno}) and,
for $i=1,\dots,p$, $\C X_{i}$ is any list such that $\C X_{i}^{\dag}=\TT{BV}(\sigma_{i})$.
\end{defi}

The following lemma provides a useful characterization of the types $\sigma^{+}$ and $\sigma^{-}$:

\begin{lem}\label{foral}
For any type $\sigma\in \TT T$, the following hold:
\begin{itemize}
\item if $\sigma= X$, then $X^{+}=X$ (resp.\ $X^{-}=X$);
\item if $\sigma=\tau\to \rho\in \formenodue$, then $\sigma^{-}=\tau^{+}\to \rho^{-}$;

\item if $\sigma=\tau \to \rho\in \forpiudue$, then
$$(\tau\to \rho)^{+}= \forall \C X((\sk(\tau_{1})\to \dots \to \sk(\tau_{p})\to Y)\to \sk(\rho))$$
where $\tau=\tau_{1}\to \dots \to \tau_{p}\to Y$ and $\C X^{\dag}= \TT{BV}(\tau_{1})\cup \dots \cup \TT{BV}(\tau_{n})\cup \TT{BV}(\rho)$.
%

\item if $\sigma=\forall X\tau$, then $\sigma^{+}=\forall X\tau^{+}$.

\end{itemize}

\end{lem}
\begin{proof}
We argue by induction on $\sigma$.
The first case is clear. Suppose $\sigma=\tau\to \rho\in \formenodue$, hence $\tau\in \forpiudue$ and $\rho\in \formenodue$. By
Lemma \ref{lemma:piumeno} $\rho=\rho_{1}\to \dots \to \rho_{p}\to Y$, where $\rho_{1},\dots, \rho_{p}\in \forpiudue$. 
Then $\sigma^{-}=\forall \C X\sk(\tau)\to \forall \C X_{1}\sk(\rho_{1}) \to \dots \to \forall \C X_{p}\sk(\rho_{p})\to Y$, where $\C X^{\dag}=\TT{BV}(\tau)$ and $\C X_{i}^{\dag}=\TT{BV}(\rho_{i})$, that is, $\sigma^{-}= \tau^{+}\to \rho^{-}$.
Suppose now $\sigma=\tau\to \rho \in \forpiudue$, hence $\tau\in \formenodue$ and $\rho\in \forpiudue$. By Lemma \ref{lemma:piumeno}, $\tau= \tau_{1}\to \dots \to \tau_{p}\to Y$, where $\tau_{1},\dots,\tau_{p}\in \forpiudue$. Then
$\sigma^{+} =\forall \C X\sk(\sigma)= \forall \C X((\sk(\tau_{1})\to \dots \to \sk(\tau_{p})\to Y)\to \sk(\rho))$.
Finally, if $\sigma=\forall X\tau$, then $\sigma^{+}= \forall \C X\sk(\tau)= \forall X(\forall \C X'\sk(\tau))=\forall X\tau^{+}$, where $\C X'=\C X-\{X\}$.
\end{proof}

\begin{lem}\label{nonproper}
For any type $\sigma\in \TT T$,
\begin{enumerate}[label=(\roman*)]
\item if $\sigma\in \forpiudue$, then $\vdash I_{\sigma}:\sigma\to \sigma^{+}$;
\item if $\sigma\in \formenodue$, then $\vdash I_{\sigma}:\sigma^{-}\to \sigma$.
\end{enumerate}
\end{lem}
\begin{proof}
We first prove the following fact: if $\sigma\in \forpiudue$ (resp.\ $\sigma\in \formenodue$), then $\vdash I_{\sigma}:\sigma^{+}\to  \sk(\sigma)$ (resp.\ $\vdash I_{\sigma}: \sk(\sigma)\to \sigma^{-}$). 
If $\sigma\in \forpiudue$, then the claim follows from the fact that $\sigma^{+}$ is of the form $ \forall \C  X \sk(\sigma)$.  
If $\sigma\in \formenodue$, we argue by induction on $\sigma$: if $\sigma=X$ the claim clearly holds; if $\sigma=\tau\to \rho$, then from the induction hypothesis $\vdash I_{\tau}: \tau^{+}\to  \sk(\tau)$ and $\vdash I_{\rho}:  \sk(\rho)\to \rho^{-}$, whence $\vdash \arrow{I_{\tau}}{I_{\rho}}: ( \sk(\tau)\to  \sk(\rho))\to \tau^{+}\to \rho^{-}$ and, since $\sk(\tau\to \rho)=\sk(\tau)\to \sk(\rho)$ and 
$(\tau\to \rho)^{-}=\tau^{+}\to \rho^{-}$, we can conclude by Lemma \ref{lemma:usigmaprop} (i) and subject $\beta$-reduction.

We can now prove claims (i),(ii) by induction on $\sigma$. The case $\sigma=X$ is trivial. Let $\sigma=\tau\to \rho$.
If $\sigma\in \forpiudue$, then $\tau\in \formenodue$ and $\rho\in \forpiudue$.
By what we just proved and the induction hypothesis we have $\vdash I_{\tau}: \sk(\tau) \to \tau^{-}, \vdash I_{\tau}:\tau^{-}\to \tau$ and $\vdash I_{\rho}:\rho^{+}\to  \sk(\rho)$ and $\vdash I_{\rho}:\rho\to \rho^{+}$. By Lemma \ref{lemma:usigmaprop} (ii), $I_{\tau}\circ I_{\tau}\to_{\beta}^{*}I_{\tau}$  and $I_{\rho}\circ I_{\rho}\to_{\beta}^{*}I_{\rho}$, hence  by subject $\beta$-reduction we deduce
$\vdash I_{\tau}: \sk(\tau)\to\tau$ and $\vdash I_{\rho}:\rho\to  \sk(\rho)$. 
We deduce then $\vdash \arrow{I_{\tau}}{I_{\rho}}:\sigma\to \sigma^{+}$ as illustrated in fig. \ref{id1} and we conclude
$\vdash I_{\sigma}:\sigma\to \sigma^{+}$ by Lemma \ref{lemma:usigmaprop} (i) and subject $\beta$-reduction.

\begin{figure}
\begin{center}
\resizebox{0.9\textwidth}{!}{
$$  
\AXC{$\vdash I_{\rho}:\rho\to  \sk(\rho)$}
\AXC{$x:\sigma\vdash x:\tau\to \rho$}
\AXC{$\vdash I_{\tau}:  \sk(\tau) \to \tau$}
\AXC{$y: \sk(\tau)\vdash y: \sk(\tau)$}
\BIC{$y: \sk(\tau)\vdash I_{\tau}y:\tau$}
\BIC{$ x:\sigma, y:\sk(\tau)\vdash x(I_{\tau}y):\rho$}
\BIC{$  x:\sigma, y: \sk(\tau)  \vdash I_{\rho}(x(I_{\tau}y)):  \sk(\rho)$}
\doubleLine
\UIC{$\vdash \lambda x.\lambda y.I_{\rho}(x(I_{\tau}y)): \sigma\to \sigma^{+}$}
\DP
$$
}
\end{center}
\caption{Derivation of $\vdash \arrow{I_{\tau}}{I_{\rho}}:\sigma\to \sigma^{+}$, for $\sigma\in \forpiudue$.}
\label{id1}
\end{figure}

One can argue similarly for $\sigma\in \formenodue$. Finally, if $\sigma=\forall X\tau$, by induction hypothesis $\vdash I_{\tau}: \tau^{+}\to \tau$, from which one easily deduces $\vdash I_{\sigma}:\sigma^{+}\to \sigma$.
\end{proof}

\subsubsection{Proper positive types}

In order to investigate in more detail the relation between positive and $\B \Pi$ types, we must refine our description of quantification. In particular, we must distinguish, as in \cite{Nour1998}, between \emph{proper} and \emph{improper} quantification.

Given a type $\sigma$, for any bound variable $X\in \TT{BV}(\sigma)$ there exists a unique subtype $\sigma_{X}$ of $\sigma$ such that $\forall X\sigma_{X}$ is a subtype of $\sigma$\footnote{We are here exploiting Convention \ref{convention} on the name of bound variables.}.

\begin{defi}[proper quantification]\index{System F!Proper quantification}
Given a type $\sigma$ and a variable $X\in \TT{BV}(\sigma)$, we call $X$ \emph{proper} if $X\in \TT{FV}(\sigma_{X})$ and \emph{improper} if $X\notin \TT{FV}(\sigma_{X})$. 

We let $\TT{PV}(\sigma)$ indicate the set of proper bound variables of $\sigma$ and $\TT{IV}(\sigma)$ indicate the set of improper bound variables of $\sigma$.

We call a type $\sigma$ \emph{proper} if $\TT{BV}(\sigma)=\TT{PV}(\sigma)$ and \emph{strongly improper} if $\TT{BV}(\sigma)=\TT{IV}(\sigma)$.
\end{defi}

For example, the type $\forall Y(\forall X(X\to Y)\to Y)$ is proper and
 the type $\forall Y(\forall X(Z\to Z)\to Z)$ is strongly improper. A type which is both proper and strongly improper must be a simple type.

\begin{defi}[$\forpiupro,\formenopro$ types, \cite{Nour1998}]\label{properforall}\index{System F!Classes of types:!$\forpiupro$ and $\formenopro$}

$\sigma\in\forpiupro$ (resp.\ $\sigma\in\formenopro$) if $\sigma$ is proper and $\sigma\in\forpiudue$ (resp.\ $\sigma$ is proper and $\sigma\in\formenodue$).
\end{defi}

We will see that the distinction between the classes $\forpiudue,\formenodue$ and $\forpiupro,\formenopro$ is related to the failure of the subject $\eta$-reduction property.

 The reverse of Lemma \ref{nonproper} can be established for strongly improper types:

\begin{lem}\label{vain}
For any type $\sigma\in \TT T$,
\begin{enumerate}[label=(\roman*)]
\item if $\sigma\in \forpiudue$ and is strongly improper, then $\vdash I_{\sigma}:\sigma^{+}\to \sigma$;
\item if $\sigma\in \formenodue$ and is strongly improper, then $\vdash I_{\sigma}:\sigma\to \sigma^{-}$.
\end{enumerate}

\end{lem}
\begin{proof}
We first prove the following fact: if $\sigma\in \forpiudue$ (resp.\ $\sigma\in \formenodue$) is strongly improper, then $\vdash I_{\sigma}: \sk(\sigma)\to \sigma^{+}$ (resp.\ $\vdash I_{\sigma}:\sigma^{-}\to  \sk(\sigma)$). 
For $\sigma\in \forpiudue$ the claim follows from the fact that $\sigma^{+}$ is of the form $\forall \C X \sk(\sigma)$ and $\C  X^{\dag}\cap \TT{FV}( \sk(\sigma))=\TT{BV}(\sigma)\cap \TT{FV}(\sk(\sigma))=\emptyset$.  
If $\sigma \in \formenodue$, we argue by induction on $\sigma$: if $\sigma=X$ it is trivial; if $\sigma=\tau\to \rho$, then the claim follows from the induction hypothesis and the definition of $I_{\sigma}$: from $\vdash I_{\tau}: \sk(\tau)\to \tau^{+}$ and $\vdash I_{\rho}:\rho^{-}\to  \sk(\rho)$, it follows $\vdash\arrow{I_{\tau}}{I_{\rho}}: (\tau^{+}\to \rho^{-})\to  \sk(\tau)\to  \sk(\rho)$ and, since 
$\sk(\tau\to \rho)=\sk(\tau)\to \sk(\rho)$ and $(\tau\to \rho)^{-}=\tau^{+}\to \rho^{-}$, we can conclude by Lemma \ref{lemma:usigmaprop} (i) and subject $\beta$-reduction.

We can now prove claims (i),(ii) by induction on $\sigma$. The case $\sigma=X$ is trivial. Let $\sigma=\tau\to \rho$.
If $\sigma\in \forpiudue$, then, by what we just proved and the induction hypothesis, we have $\vdash I_{\tau}:\tau^{-}\to  \sk(\tau), \vdash I_{\tau}:\tau\to \tau^{-}$ and $\vdash I_{\rho}: \sk(\rho)\to \rho^{+}$ and $\vdash I_{\rho}:\rho^{+}\to \rho$. By Lemma \ref{lemma:usigmaprop} (ii), $I_{\tau}\circ I_{\tau}\to_{\beta}^{*}I_{\tau}$ and $I_{\rho}\circ I_{\rho}\to_{\beta}^{*}I_{\rho}$, hence  by subject $\beta$-reduction we deduce
$\vdash I_{\tau}:\tau\to  \sk(\tau)$ and $\vdash I_{\rho}: \sk(\rho)\to \rho$. We deduce then $\arrow{I_{\tau}}{I_{\rho}}:\sigma^{+}\to \sigma$ as illustrated in fig. \ref{id3} and we conclude $\vdash I_{\sigma}:\sigma^{+}\to \sigma$ by Lemma \ref{lemma:usigmaprop} (i) and subject $\beta$-reduction.

\begin{figure}
\begin{center}
\resizebox{0.9\textwidth}{!}{
$$  
\AXC{$\vdash I_{\rho}: \sk(\rho)\to \rho$}
\AXC{$x:\sigma^{+}\vdash x:\sigma^{+}$}
\doubleLine
\UIC{$x:\sigma^{+}\vdash x: \sk(\tau)\to  \sk(\rho)$}
\AXC{$\vdash I_{\tau}: \tau\to  \sk(\tau)$}
\AXC{$y:\tau\vdash y:\tau$}
\BIC{$y:\tau\vdash I_{\tau}y: \sk(\tau)$}
\BIC{$ x:\sigma^{+}, y:\tau\vdash x(I_{\tau}y): \sk(\rho)$}
\BIC{$  x:\sigma^{+}, y:\tau\  \vdash I_{\rho}(x(I_{\tau}y)): \rho$}
\doubleLine
\UIC{$\vdash \lambda x.\lambda y.I_{\rho}(x(I_{\tau}y)): \sigma^{+}\to \sigma$}
\DP
$$
}
\end{center}
\caption{Derivation of $\vdash\arrow{I_{\tau}}{I_{\rho}}:\sigma^{+}\to \sigma$, for $\sigma\in \forpiupro$.}
\label{id3}
\end{figure}

One can argue similarly for the case $\sigma\in \formenodue$. Finally, if $\sigma=\forall X\tau$, by induction hypothesis $\vdash I_{\tau}: \tau\to \tau^{+}$, from which one deduces $\vdash I_{\sigma}:\sigma\to \sigma^{+}$.
\end{proof}

We now prove an important property of proper positive types.

\begin{lem}\label{lemma:chicchi}
Let $M$ be a $\beta$-normal $\lambda$-term, let $\Gamma,\Delta\in \formenopro$ and $\sigma\in \forpiupro$, and let $S\subseteq \TT{PV}(\sigma)\cup \TT{PV}(\Delta)$ such that no variable in $S$ occurs free in $\Gamma$. 
If $\Gamma, \Delta^{S}\vdash M: \sigma^{S}$, then $\Gamma,\Delta\vdash M:\sigma$.

\end{lem}
\begin{proof}
We argue by induction on a derivation $D$ of $\Gamma, \Delta^{S}\vdash M:\sigma^{S}$ which, by Proposition \ref{prop:foralle}, does not use the $\forall$E rule and is such that, for any type declaration $\Gamma'\vdash M':\sigma'$ in $D$, $\Gamma'\in \formenodue$ and $\sigma'\in \forpiudue$. 

If $D$ only consists of an axiom
 $x:\sigma^{S}\vdash x:\sigma^{S}$, then either $\Gamma=\{x:\sigma^{S}\}$ and $\Delta=\emptyset$ or $\Gamma=\emptyset$ and $\Delta=\{x:\sigma\}$. In the first case, since $\TT{FV}(\Gamma)\cap S=\emptyset$, no variable in $S$ can occur in $\sigma^{S}$. But, since
$\sigma$ is proper, this implies that $S=\emptyset$, hence $\sigma=\sigma^{S}$ and the claim trivially holds. 
In the second case, we have $x:\sigma\vdash x:\sigma$.

If $D$ ends by a $\to$I rule, then $\sigma^{S}=\tau\to \rho$. This implies that $\sigma=\sigma_{1}\to \sigma_{2}$ for some $\sigma_{1}\in \formenopro$ and $\sigma_{2}\in \forpiupro$ and $\tau=\sigma_{1}^{S}$, $\rho=\sigma_{2}^{S}$. $D$ is of the form
$$
D\quad =\quad 
\AXC{$D'$}
\noLine
\UIC{$\Gamma, \Delta^{S}, x:\tau^{S}\vdash M': \rho^{S}$}
\RL{$\to$I}
\UIC{$\Gamma, \Delta^{S}\vdash \lambda x.M': \sigma^{S}$}
\DP
$$
By the induction hypothesis we have then $\Gamma, \Delta, x:\tau\vdash M':\rho$, from which we deduce $\Gamma,\Delta\vdash M:\sigma$ by $\to$I.

If $D$ ends by a $\forall$I rule, then $\sigma^{S}=\forall Y\tau$. This implies that $Y\notin S$, that $Y$ does not occur free in $\Gamma$ and $\Delta$ and that $\sigma= \forall Y\sigma_{1}$ for some $\sigma_{1}\in \forpiupro$ such that  $\tau= \sigma_{1}^{S}$.  $D$ is of the form
$$
D\quad =\quad 
\AXC{$D'$}
\noLine
\UIC{$\Gamma, \Delta^{S}\vdash M: \sigma_{1}^{S}$}
\RL{$\forall$I}
\UIC{$\Gamma, \Delta^{S}\vdash M: \sigma^{S}$}
\DP
$$
By the induction hypothesis we have then $\Gamma, \Delta\vdash M: \tau$, from which we deduce $\Gamma,\Delta\vdash M:\sigma$ by $\forall$I.

If $D$ ends by a chain of applications of $\to$E rules, then $D$ is of the form
$$
D\quad =\quad  
\AXC{$D'$}
\noLine
\UIC{$\Gamma, \Delta^{S}\vdash z: \xi_{1}\to \dots \to \xi_{p}\to \sigma^{S}$}
\AXC{$D_{1}$}
\noLine
\UIC{$\Gamma, \Delta^{S}\vdash M_{1}:\xi_{1}$}
\AXC{$\dots$}
\AXC{$D_{p}$}
\noLine
\UIC{$\Gamma, \Delta^{S}\vdash M_{p}:\xi_{p}$}
\doubleLine
\QuaternaryInfC{$\Gamma, \Delta^{S}\vdash M: \sigma^{S}$}
\DP
$$
for some $p\geq 1$ and types $\xi_{1},\dots, \xi_{p}\in \formenopro$.
We must consider two cases:
\begin{enumerate}
\item if $z: \xi_{1}\to \dots \to \xi_{p}\to \sigma^{S}$ occurs in $\Gamma$, then, since $\TT{FV}(\Gamma)\cap S=\emptyset$, no variable in $S$ can occur either in $\sigma^{S}$ nor in $\xi_{1},\dots, \xi_{p}$. But, since $\sigma$ is proper, this implies that $S\subseteq \TT{BV}(\Delta)$, $\sigma=\sigma^{S}$ and  $\xi_{i}=\xi_{i}^{S}$, for $i=1,\dots,p$. From $\Gamma\in\formenopro$ it follows that $\xi_{i}\in \forpiupro$, for $i=1,\dots,p$. Hence $\xi_{i}=\xi_{i}^{S}$, for $i=1,\dots,p$, so we deduce, by the induction hypothesis, that $\Gamma, \Delta\vdash M_{i}:\xi_{i}$, and we can conclude $\Gamma,\Delta\vdash M:\sigma$.

\item if $z: \xi_{1}\to \dots \to \xi_{p}\to \sigma^{S}$ occurs in $\Delta^{S}$, then there exist types $\eta_{1},\dots, \eta_{p}\in\forpiupro$  such that $\xi_{1}=\eta_{1}^{S},\dots, \xi_{p}=\eta_{p}^{S}$ and $z:\eta_{1}\to \dots \to \eta_{p}\to \sigma\in \Delta$. By the induction hypothesis we deduce then
$\Gamma, \Delta\vdash z: \eta_{1}\to \dots \to \eta_{p}\to \sigma$ and 
$\Gamma, \Delta\vdash M_{i}:\eta_{i}$ for $i=1,\dots,p$, from which we can conclude $\Gamma, \Delta\vdash M: \sigma$. \qedhere
\end{enumerate}
\end{proof}

\subsubsection{Comparison of simple and positive types }

We now establish our main result on the comparison of positive and simple types. 

\begin{thm}\label{forpi}
Let $M$ be a closed $\beta$-normal term.
\begin{enumerate}[label=(\roman*)]
\item For all $\sigma\in \forpiupro$, if $\vdash M:\sk(\sigma)$ then $\vdash M:\sigma$;
 
\item For all $\sigma\in \forpiudue$, if $\vdash M:\sk(\sigma)$ then $\vdash M':\sigma$ for some $M'$ such that $M'\to_{\eta}^{*} M$.
 \end{enumerate}
\end{thm}
\begin{proof}
For case (i), let  $\C X$ be such that $\C X^{\dag}=\TT{BV}(\sigma)$; then $\sk(\sigma)= \sigma^{\C X^{\dag}}$. By Lemma \ref{lemma:chicchi} we deduce then $\vdash M: \sigma$.

For case (ii), 
the set $\TT{BV}(\sigma)$ splits as the disjoint union of $\TT{PV}(\sigma)$ and $\TT{IV}(\sigma)$. 
Let $\C X, \C Y$ be such tha $\C X^{\dag}=\TT{PV}(\sigma)$ and $\C Y^{\dag}=\TT{IV}(\sigma)$.
From $\vdash M:\sk(\sigma)$ we deduce $\vdash M: \forall \C Y \sk(\sigma)$. 
Now, since $\forall \C Y \sk(\sigma)= (\sigma^{\C X^{\dag}})^{+}$, by Lemma \ref{vain} we deduce 
$\Gamma\vdash I_{\sigma^{\C X^{\dag}}}M : \sigma^{\C X^{\dag}}$ and by Lemma \ref{lemma:usigma} we deduce that $I_{\sigma^{\C X^{\dag}}}M\to_{\beta}^{*} M'$, for some (obviously closed) $M'$ $\beta$-normal such that $M'\to_{\eta}^{*}M$, whence $\vdash M': \sigma^{\C X^{\dag}}$ holds by subject $\beta$-reduction. We can then conclude by Lemma \ref{lemma:chicchi} that $\vdash M':\sigma$.
\end{proof}

\begin{rem}\label{equivaye}
By putting together Theorem \ref{forpi} and Lemma \ref{beta2} we deduce that, for $\sigma\in\forpiupro$, typability at $\sigma$ is equivalent to typability at $\sk(\sigma)$ for a closed $\beta$-normal term; for $\sigma\in \forpiudue$, typability at $\sigma$ is equivalent to typability at $\sk(\sigma)$ for a closed $\beta$-normal term up to $\eta$-equivalence.
\end{rem}

\begin{exa}\label{etafail}
Theorem \ref{forpi} highlights how the distinction between proper and non proper types is related to $\eta$-expansion. 
Let $\sigma=((\forall YX)\to X)\to (X\to X)\in \forpiudue$. Then $\sigma^{+}=\forall Y((X\to X)\to (X\to X))$. While we have $\vdash \lambda x.x: \sigma^{+}$, it is not true that $\vdash \lambda x.x:\sigma$, but only $\vdash \lambda x.\lambda y.xy:\sigma$. 

\end{exa}

\section{Realizability semantics by closure operators}\label{sec3}

We introduce realizability semantics by means of closure operators over sets of $\lambda$-terms. 
Following \cite{RibaUnion}, we show that, if the closure operator is stable by union, then it generates a realizability semantics which is a topological space. This topological intuition is at the heart of several constructions in the following sections.

We prove a soundness theorem for a class of closure operators (called \emph{adequate $\FF$-closure operators}). This result 
combines two different standard soundness theorems for realizability semantics: the one for the semantics based on sets of $\lambda$-terms (e.g. \emph{$\beta$-stable, $\beta\eta$-stable} and \emph{$\beta$-saturated sets}, \cite{Krivine}) and the one for the semantics based on sets of strongly normalizing $\lambda$-terms (e.g. \emph{reducibility candidates}). The latter is the fundamental ingredient to prove the strong normalization theorem for System F, and is based on the definition of a semantics made of closed sets containing all term variables. We will describe such semantics by means of \emph{regular} closure operators. 

We describe then how the above mentioned semantics fit into this general framework. Moreover, we introduce two variants of Girard's reducibility candidates due to \cite{Cousineau}, which are more adapted to investigate completeness and parametricity.

\subsection{Semantics by closure operators }\label{sec31}

\subsubsection{Closure operators}
We first recall the definition of closure operators. Then we introduce a class of closure operators over sets of $\lambda$-terms well-suited to interpret System $\FF$. 

\begin{defi}[closure operator]\index{Realizability!Closure operator}
A \emph{closure operator} $\Cl$ over a set $L$ is a map $\Cl:\wp({L})\to \wp({L})$ such that, for all $s,t\in \wp({L})$,
\begin{enumerate}
\item $s\subseteq \Cl(s)$;
\item $s\subseteq t \ \To \ \Cl (s) \subseteq \Cl (t)$;
\item $\Cl (\Cl (s))=\Cl (s)$.

\end{enumerate}
\end{defi}

Any closure operator $\Cl$ over $L\subseteq \Lambda$ generates a family of closed subsets of $L$, which are the $s\subseteq L$ such that $\Cl(s)=s$.
For any closure operator, the intersection of closed sets is closed: 

\begin{lem}\label{intersection}
Let $\Cl$ be a closure operator over $L\subseteq \Lambda$. Let $I\neq \emptyset$ and $u_{i}\in \wp(L)$ for all $i\in I$.
If all $u_{i}$ are closed, then
$\bigcap_{i\in I}u_{i}$ is closed.
\end{lem}
\begin{proof}
Let $u=\bigcap_{i\in I}u_{i}$. It suffices to show that $\Cl(u)\subseteq u$. From $u\subseteq u_{i}$, we deduce $\Cl(u)\subseteq \Cl(u_{i})=u_{i}$, for any $i\in I$, whence $\Cl(u)\subseteq \bigcap_{i\in I}u_{i}=u$. 
\end{proof}

By Lemma \ref{intersection}, to interpret System $\FF$ connectives it suffices that closed sets are stable with respect to the following construction:
\begin{defi}
Given sets $s,t\subseteq \Lambda$, we let $s\to t= \{M \mid\forall P \ ( P\in s \ \To \ MP\in t)\}$. 
Moreover, if $L\subseteq \Lambda$, for all $s,t\subseteq \Lambda$, we let $s\to_{L}t= (s\to t) \cap L$.

\end{defi}

This leads to the following definition.

\begin{defi}[$\FF$-closure operator]\index{Realizability!$\FF$-closure operator}
%

A closure operator $\Cl$ over a set $L$ generates a family of closed sets $\C S_{\Cl}=\{ s\in \wp({L})\mid \Cl(s)=s\}\subseteq \wp({L})$.

A closure operator $\Cl$ over $L\subseteq\Lambda$ is a \emph{$\FF$-closure operator} if for all $s,t\in \C S_{\Cl}$, $s\to_{L}t\in \C S_{\Cl}$.
If $\Cl$ is a $\FF$-closure operator over $L$, the family $\C S_{\Cl}$ is called a \emph{semantics over $L$}.

\end{defi}

Two closure operator can be compared as follows:

\begin{defi}
Let $\Cl, \F D$ be closure operators over sets $L,L'\subseteq \Lambda$. We say that $\Cl$ is \emph{finer} than ${\F D}$, noted ${\Cl}\leq {\F D}$, when, for any $s\subseteq L\cap L'$, $\Cl(s)\subseteq \F D(s)$. 

\end{defi}

The following proposition establishes that, when $\Cl,\F D$ are closure operators over the same set $L\subseteq \Lambda$, then the property $\Cl \leq \F D$ corresponds to the inclusion of the associated semantics.
\begin{prop}\label{confront}
Let $\F C, \F D$ be two $\FF$-closure operators over $L, L'\subseteq \Lambda$, respectively. If $L\in \C S_{\F D}$, then $\F C\leq \F D$ iff $\C S_{\F D}\cap L\subseteq \C \C S_{\F C}$, where $\C S_{\F D}\cap L=\{ s\cap L\mid s\in \C S_{\F D}\}$.
\end{prop}
\begin{proof}
First observe that, if $L\in \C S_{\F D}$, then $\C S_{\F D}\cap L\subseteq \C S_{\F D}$, since for any $s\in \C S_{\F D}$, $s\cap L\in \C S_{\F D}$, by Lemma \ref{intersection}. Moreover, since $L\in \C S_{\F D}$, if $s\subseteq L$, then $\F D(s) \subseteq \F D(L)=L$.
Now, for the right-to-left direction, suppose $\C S_{\F D}\cap L\subseteq \C S_{\F C}$ and let $s\subseteq L\cap L'$. Then $\F D(s)\cap L= \F D(s)\in \C S_{\F C}$, and we get $\F C(s)\subseteq \Cl(\F D(s))=\F D(s)$.
For the left-to-right direction, suppose $\Cl(s)\subseteq \F D(s)$ for all $s\subseteq L\cap L'$ and let $t\in \C S_{\F D}\cap L$, i.e. $t= t'\cap L$ for some $t'\in \C S_{\F D}$. Then $t\subseteq L\cap L'$, hence 
$\Cl(t)\subseteq \F D(t)=t$, i.e. $\Cl(t)\subseteq t$; since $t\subseteq \Cl(t)$, we deduce $\Cl(t)=t$, i.e. $t\in \C S_{\F C}$.
\end{proof}

We introduce now an important class of $\FF$-closure operators:

\begin{defi}\index{Realizability!Regular closure operator}
A closure operator $\Cl$ is \emph{regular} when $\TT{TermVar}\subseteq\Cl(\emptyset)$.
\end{defi}

The notion of regular $\FF$-closure operators captures a property which is often used in normalization proofs and that we will exploit to prove soundness (Theorem \ref{soundness}) in the case in which the closure operator is defined over a subset of $\Lambda$ (e.g.\ on $\C{SN}$). In normalization arguments one usually considers a family of closed sets which contain all terms of the form $xP_{1}\dots P_{n}$, where the $P_{i}$ are normalizing terms (see \cite{Gallier, Krivine}).
This construction corresponds to defining a regular closure operator, as we now show.

For any set $L\subseteq \Lambda$, we let $L^{*}$ be the set of all terms of the form $xP_{1}\dots P_{n}$, for some term variable $x$, $n\in \BB N$ and terms $P_{1},\dots, P_{n}\in L$. 

\begin{prop}\label{ellezero1}
Let $L\subseteq \Lambda$. If $\Cl$ is a regular $\FF$-closure operator over $L$, then $L^{*}\subseteq L$ and for all $s\in \C S_{\Cl}$, $L^{*}\subseteq s$.
\end{prop}
\begin{proof}
We show that for all $s\in \C S_{\Cl}$, variable $x$, $n\in \BB N$ and terms $P_{1},\dots, P_{n}\in L$, $xP_{1}\dots P_{n}\in s$. We argue by induction on $n$. For all $s\in \C S_{\Cl}$, from $\emptyset\subseteq s$, it follows $\Cl(\emptyset)\subseteq \Cl(s)=s$. Since $\Cl$ is regular, then $\TT{TermVar}\subseteq \Cl(\emptyset)\subseteq s$. This proves the case $n=0$.
For $n=k+1$, by the induction hypothesis for all $s\in \C S_{\Cl}$, $P_{1},\dots, P_{k}\in L$, $xP_{1}\dots P_{k}\in s$. Let $P\in L$; 
since $\Cl(\{P\})\to_{L}s\in \C S_{\Cl}$, then $xP_{1}\dots P_{k}\in \Cl(\{P\})\to_{L}s$, and this implies that $xP_{1}\dots P_{k}P\in~s$.
\end{proof}

In Subsection \ref{sec32} we will show that the reducibility candidates semantics is generated by a regular $\FF$-closure operator.

\subsubsection{Stability by union}

Stability by union is the property that arbitrary unions of closed sets are closed. This property was investigated in \cite{RibaUnion} in the case of reducibility candidates. Following \cite{RibaUnion}, we show that if a closure operator $\Cl$ is stable by union, then the $\Cl$-closed sets (plus the empty set) form a topology. The topological presentation of a semantics will be exploited throughout the following sections.
In Subsection \ref{sec32} and Appendix \ref{appCR} we will show that all concrete semantics discussed in this paper are stable by union.

\begin{defi}\index{Realizability!Stability by union $\mathsf{SU}$}
Let $L\subseteq \Lambda$ and $\Cl$ be a closure operator over $L$. $\Cl$ satisfies \emph{stability by union} ($\mathsf{SU}$) if 
for all $s\subseteq L$, if $s\neq \emptyset$, $\Cl (s)=\bigcup_{P\in s}\Cl (\{P\})$.
\end{defi}


Given a closure operator $\Cl$ over a set $L$, let $\OV{\Cl}=\{s\subseteq L\mid s\neq \emptyset \ and \ s=\bigcup_{x\in s} \Cl(\{x\})\}\cup \{\Cl(\emptyset)\}$.
We show that $\C S_{\Cl}=\OV{{\Cl}}$ iff $\C S_{\Cl}$ is closed with respect to arbitrary unions (we essentially follow  \cite{RibaUnion}, Prop. 4.1, p. 6).
\begin{prop}\label{union}
For any closure operator $\Cl$, $\OV{{\Cl}}$ is the smallest set such that $\C S_{\Cl}\subseteq \OV\Cl$ and for all non-empty $X\subseteq \OV{\Cl}$, $\bigcup X, \bigcap X\in \OV\Cl$.

\end{prop}
\begin{proof}
For all $s\in \C S_{\Cl}$, if $s\neq \emptyset$, $s=\bigcup_{x\in s}\Cl(\{x\})$. Indeed, $s\subseteq \bigcup_{x\in s}\Cl(\{x\})$ and for all $x\in s$, $\Cl(\{x\})\subseteq \Cl(s)=s$. We can conclude then $\C S_{\Cl}\subseteq \OV\Cl$.

Let now $X\subseteq \OV \Cl$ be a non-empty set. Then for all $s\in X$, $s=\bigcup_{x\in s}\Cl(\{x\})$, hence $\bigcup X=  \bigcup_{x\in \bigcup X}\Cl(\{x\})$, so $\bigcup X\in \OV\Cl$. Moreover, if $x\in \bigcap X$, then $x\in s$, for all $s\in X$, hence $\Cl(\{x\})\subseteq s$, since $s\in \OV\Cl$, and we can conclude then $\Cl(\{x\})\subseteq \bigcap X$. This implies that $\bigcap X= \bigcup_{x\in \bigcap X}\Cl(\{x\})$, whence $\bigcap X\in \OV\Cl$. If $\bigcap X=\emptyset$, observe first that, for all $s\in \OV\Cl$, $\Cl(\emptyset)\subseteq s$. Indeed, either $s=\Cl(\emptyset)$, so the claim is trivial, or $s\neq \emptyset$. In this last case there is some $x\in s$, hence from $\emptyset\subseteq \{x\}$, we deduce $\Cl(\emptyset)\subseteq \Cl(\{x\})\subseteq s$.
Now, since $\Cl(\emptyset)\subseteq s$, for all $s\in \C S_{\Cl}$ and $X$ is non empty, $\Cl(\emptyset)\subseteq \bigcap X$, so it must be $\Cl(\emptyset)=\emptyset$, and we conclude again $\bigcap X\in \OV \Cl$.

Let now $S\subseteq \wp(L)$ be such that $\C S_{\Cl}\subseteq S$ and for all $X\subseteq S$ non empty, $\bigcap X, \bigcup X\in S$. 
If $s\in \OV\Cl$, then either $s=\Cl(\emptyset)$ or $s\neq \emptyset $ and $s=\bigcup_{x\in s}\Cl(\{x\})$. In the first, case, since $\Cl(\emptyset)\in \C S_{\Cl}\subseteq S$, $s\in S$; in the second case, since $X=\{\Cl(\{x\})\mid x\in s\}\subseteq S$, we have $s=\bigcup X\in S$. This shows that $\OV\Cl\subseteq S$.
\end{proof}

\begin{cor}
$\C S_{\Cl}=\OV{{\Cl}}$ iff for all $X\subseteq \C S_{\Cl}$ non-empty, $\bigcup X\in \C S_{\Cl}$. 
\end{cor}
\begin{proof}
If $\C S_{\Cl}=\OV\Cl$, then the claim follows from Proposition \ref{union}. Conversely, if for all $X\subseteq \C S_{\Cl}$ non-empty, $\bigcup X\in \C S_{\Cl}$, since for all $X\subseteq \C S_{\Cl}$ non-empty, $\bigcap X\in \C S_{\Cl}$ by Lemma \ref{intersection}, and trivially $\C S_{\Cl}\subseteq \C S_{\Cl}$, by Proposition \ref{union}, $\OV\Cl\subseteq \C S_{\Cl}$ and $\C S_{\Cl}\subseteq \OV \Cl$, that is, $\C S_{\Cl}=\OV \Cl$.
\end{proof}

$\OV{{\Cl}}\cup\{\emptyset\}$ is the topological closure of $\C S_{\Cl}$. The pair $(L, \OV\Cl\cup\{\emptyset\})$ forms indeed a topological space, as it contains $L, \emptyset$ and is closed by arbitrary unions and intersections. Observe that the addition of $\emptyset$ is necessary to obtain a topology when $\Cl(\emptyset)\neq\emptyset$ (e.g.\ when $\Cl$ is regular) and redundant otherwise. 
  
The following result shows that when $\Cl$ satisfies $\mathsf{SU}$, $(L, \C S_{\Cl}\cup \{\emptyset\})$ is a topological space.

%

\begin{prop}\label{union2}
Let $\Cl$ be a closure operator over $L\subset \Lambda$. Then $\Cl$ satisfies $\mathsf{SU}$ iff $\C S_{\Cl}=\OV{{\Cl}}$.
\end{prop}
\begin{proof}
Suppose $\Cl$ satisfies $\mathsf{SU}$. As $\Cl(s)=\bigcup_{P\in s}\Cl(\{P\})$, $\Cl(s)=s$ iff $s\in \OV{{\Cl}}$. This shows that $\C S_{\Cl}=\OV\Cl$.
Conversely, suppose $\C S_{\Cl}=\OV{\Cl}$. 
Observe that for any $s\subseteq L$, $\Cl(s)$ is the smallest $\Cl$-closed set containing $s$: if $s\subseteq t\in \C S_{\Cl}$, from $\Cl(s)\subseteq \Cl(t)=t$ it follows that $\Cl(s)\subseteq t$. Now, since $\C S_{\Cl}$ is closed by unions, for all $s\subseteq L$, if $s\neq\emptyset$, then $s'=\bigcup_{P\in s}\Cl(\{P\})\in \C S_{\Cl}$ and is the smallest $\Cl$-closed set containing $s$: if $s\subseteq s''\in \C S_{\Cl}$, then for all $P\in s$, $\Cl(\{P\})\subseteq s''$, hence $s'\subseteq s''$. 
We deduce then $s'=\bigcup_{P\in s}\Cl(\{P\})=\Cl(s)$, i.e. $\C S_{\Cl}$ satisfies $\mathsf{SU}$.
\end{proof}

We present a useful application of $\mathsf{SU}$. We show that if a $\FF$-closure operator $\Cl$ over $L$ is stable by union, then there exists a regular $\FF$-closure operator $\F D$ over $L$ such that $\Cl\leq \F D$. 

%
%

\begin{prop}\label{sufirst}
Let $L\subseteq\Lambda$ be such that $L^{*}\subseteq L$ and let $\Cl$ be a $\FF$-closure operator over $L$ satisfying $\mathsf{SU}$. Then the map 
$\Cl^{*}:\wp(L)\to \wp(L)$ given by $\Cl^{*}(s)=\Cl(s)\cup \Cl(L^{*})$ is a regular $\FF$-closure operator over $L$ such that $\Cl\leq \Cl^{*}$. 

\end{prop}
\begin{proof}
We first check that $\Cl^{*}$ is a closure operator. For all $s,t\subseteq L$, it is clear that $s\subseteq \Cl^{*}(s)$ and that if $s\subseteq t$, then $\Cl^{*}(s)\subseteq \Cl^{*}(t)$. Moreover, $\Cl^{*}(\Cl^{*}(s))=\Cl(\Cl^{*}(s))\cup \Cl(L^{*})= \Cl(\Cl(s)\cup \Cl(L^{*}))\cup \Cl(L^{*})=\Cl(s)\cup \Cl(L^{*})=\Cl^{*}(s)$, by $\mathsf{SU}$ and Proposition \ref{union2}.
That $\Cl\leq \Cl^{*}$ is an immediate consequence of the definition of $\Cl^{*}$.

We now show that if $s,t$ are $\Cl^{*}$-closed, then $s\to_{L}t$ is $\Cl^{*}$-closed. First, for all $P\in L^{*}$, and $Q\in s$, $PQ\in L^{*}\subseteq t$. Hence $L^{*}\subseteq s\to_{L}t$. This implies that $\Cl(L^{*})\subseteq \Cl(s\to_{L}t)$. 
 Now we have $\Cl^{*}(s\to_{L}t)= \Cl(s\to_{L}t)\cup \Cl(L^{*})= \Cl(s\to_{L}t)= s\to_{L}t$, where the last passage is a consequence of the fact that $\Cl\leq \Cl^{*}$ and Proposition \ref{confront} (which implies that $s,t\in \C S_{\Cl}$).
\end{proof}

\subsubsection{Soundness for adequate semantics}\label{sec313}

We show that, for any $\FF$-closure operator $\Cl$ over $L\subseteq\Lambda$, any System $\FF $ type $\sigma$ with $n$ free variables can be interpreted as a function $|\sigma|^{\Cl}:\C S_{\Cl}^{n}\to \C S_{\Cl}$.
Then we introduce two conditions under which System $\FF$ typing judgements can be interpreted in the semantics generated by $\Cl$. The first one (called condition $\mathsf K$) is a standard condition for realizability semantics  (see \cite{Krivine}). The second one (condition $\mathsf R$) is only required when $\Cl$ is defined over $L\subsetneq \Lambda$ (e.g.\ $L=\C{SN}$), and underlies a standard construction to prove soundness for Tait's saturated sets semantics and Girard's reducibility candidates semantics.
We prove that, if conditions $\mathsf K$ and $\mathsf R$ hold, then any judgement derivable in System $\FF$ has a sound interpretation in $S_{\Cl}$.

\begin{defi}\label{intedef}\index{Realizability!$\Cl$-interpretations and sets $\vert\sigma\vert^{\Cl}$ and $\|\sigma\|^{\Cl}$}
Let $L\subseteq\Lambda$ and $\Cl$ be a $\FF$-closure operator $\Cl$ over $L$. For any type $\sigma\in \TT T$ and $\C X=\{X_{1},\dots, X_{n}\}\subseteq \TT{TypeVar}$ such that $\TT{FV}(\sigma)\subseteq \C X$ and $\C X\cap \TT{BV}(\sigma)=\emptyset$, we define a map 
$|\sigma|^{\Cl,\C X}:\C S_{\Cl}^{n}\to \C S_{\Cl}$ as follows\begin{equation}
\begin{split}
|X_{i} |^{\Cl ,\C X} (s_{1},\dots, s_{n}) & := s_{i} \\
| \sigma\to \tau |^{\Cl,\C X} (s_{1},\dots, s_{n}) & := |\sigma|^{\Cl,\C X}(s_{1},\dots, s_{n}) \to_{L} |\tau|^{\Cl,\C X}(s_{1},\dots, s_{n}) \\
|\forall X\sigma|^{\Cl,\C X}(s_{1},\dots, s_{n}) & :=  \bigcap_{s\in S}|\sigma|^{\Cl,\{X\}\cup \C X}(s,s_{1},\dots, s_{n})
\end{split}
\end{equation}
We let $|\sigma|^{\Cl}=|\sigma|^{\Cl, \TT{FV}(\sigma)}: \C S_{\Cl}^{card(\TT{FV}(\sigma))}\to \C S_{\Cl}$.

 A \emph{$\Cl$-interpretation} is a map $\C M:\TT{TypeVar}\to \C S_{\Cl}$. 
For any $\Cl$-{interpretation} $\C M$ and type $\sigma\in \TT T$, with $\TT{FV}(\sigma)= \{X_{1},\dots, X_{n}\}$, we let
$|\sigma|^{\Cl}_{\C M}= |\sigma|^{\Cl}(\C M(X_{1}),\dots, \C M(X_{n}))$.

Finally, for all $\sigma\in \TT T$, with
 $\TT{FV}(\sigma)= \{X_{1},\dots, X_{n}\}$, we let 
 $\| \sigma\|^{\Cl}:=\bigcap_{s_{1},\dots, s_{n}\in \C S_{\Cl}}|\sigma|^{\Cl}(s_{1},\dots, s_{n})$. 
A $\lambda$-term $M\in\|\sigma\|^{\Cl}$ will be called \emph{$\Cl$-interpretable at $\sigma$}.
\end{defi}

\begin{lem}\label{emptiset}
Let $L\subseteq \Lambda$ be and $\Cl$ be a $\FF$-closure operator over $L$. Then for all $\Cl$-interpretation $\C M$ and type $\sigma\in \TT T$, 
$\Cl(\emptyset)\subseteq |\sigma|^{\Cl}_{\C M}$.

\end{lem}
\begin{proof}
From $\emptyset\subseteq s$ it follows $\Cl(\emptyset)\subseteq \Cl(s)$, hence $\Cl(\emptyset)$ is a subset of every $\Cl$-closed set.
\end{proof}

From Lemma \ref{emptiset} and Proposition \ref{ellezero1} we deduce the following property, which motivates the use of regular closure operators:

\begin{lem}\label{regular}
Let $L\subseteq \Lambda$ be and $\Cl$ be a regular $\FF$-closure operator over $L$. Then for all $\Cl$-interpretation $\C M$ and type $\sigma\in \TT T$, 
$L^{*}\subseteq|\sigma|^{\Cl}_{\C M}$.

\end{lem}

The following is a standard technical lemma in realizability semantics:

\begin{lem}\label{substitutionlemma}
For any $\FF$-closure operator $\Cl$ over $L\subseteq \Lambda$, types $\sigma,\tau$, where
$\TT{FV}(\sigma)\subseteq \{X,X_{1},\dots, X_{n}\}$ and 
$\TT{FV}(\tau)\subseteq \{X_{1},\dots, X_{n}\}$, $|\sigma|^{\Cl}(|\tau|^{\Cl}(\vec s), \vec s)= |\sigma[\tau/X]|^{\Cl}(\vec s)$.
\end{lem}
\begin{proof}
Induction on $\sigma$. We only prove the case $\sigma=\forall Y\sigma'$. 
By induction hypothesis, $|\sigma'|^{\Cl}(|\tau|^{\Cl}(\vec s), s', \vec s)= |\sigma'[\tau/X]|^{\Cl}(s',\vec s)$. Since we can assume $Y$ not free in $\tau$, we have $|\tau|^{\Cl}(s',\vec s)=|\tau|^{\Cl}(\vec s)$. 


Now, for any term $P\in L$, 
$P\in  |\sigma[\tau/X]|^{\Cl}(\vec s)$ iff for all $s'\in S$, $P\in |\sigma'[\tau/X]|^{\Cl}(s',\vec s)$ iff for all $s'\in S$, 
$P\in |\sigma'|^{\Cl}(|\tau|^{\Cl}(\vec s), s', \vec s)$ iff $P\in |\sigma|^{\Cl}(|\tau|^{\Cl}(\vec s), \vec s)$.
\end{proof}

We can now state the adequacy conditions required to interpret System $\FF$ type judgements.

\begin{defi}[adequacy]\label{adequacy}\index{Realizability!Adequate operator}\index{Realizability!Condition $\mathsf K$}\index{Realizability!Condition $\mathsf R$}
Let $L\subseteq \Lambda$. A $\FF$-closure operator $\Cl$ over $L\subseteq \Lambda$ is \emph{adequate} when it satisfies the conditions below:
\begin{itemize}
\item[($\mathsf K$)]
For all $s,t\in \C S_{\Cl}$ and $M\in L$, if for all $P\in s$, $M[P/x]\in t$, then $\lambda x.M\in s\to_{L} t$;
\item[($\mathsf R$)] either $L= \Lambda$, or exists a regular $\FF$-closure operator $\F D$ over $L$ such that $\Cl\leq \F D$.

\end{itemize}
\end{defi}

If $\Cl$ is adequate and $L\subsetneq \Lambda$, by condition $\mathsf R$, there exists $\F D$ regular such that $\Cl\leq \F D$. By Proposition \ref{ellezero1} $\F D$ generates a family of closed sets containing $L^{*}$. To prove that such a family is a semantics it suffices then to verify that $\F D$ satisfies condition $\mathsf K$.

\begin{lem}\label{use00}
Let $L\subseteq \Lambda $ and $\Cl, \F D$ be $\FF$-closure operators over $L$. If $\Cl$ satisfies condition $\mathsf K$ and $\Cl\leq \F D$, then $\F D$ satisfies condition $\mathsf K$.
\end{lem}
\begin{proof} 
As $\Cl\leq \F D$, by Proposition \ref{confront}, $\C S_{\F D}\subseteq \C S_{\Cl}$. Thus, if $s,t\in \C S_{\F D}$, then $s,t\in \C S_{\Cl}$, hence
$s\to_{L}t\in \C S_{\Cl}$. This implies that if $M\in L$ and for all $P\in s$, $M[P/x]\in t$, then $\lambda x.M\in s\to_{L}t$, since $\Cl$ satisfies $\mathsf K$.
\end{proof}

The following proposition allows one to show that a $\FF$-closure operator is adequate. We will exploit this result later in this section to prove that reducibility candidates are generated by an adequate $\FF$-closure operator.

\begin{prop}\label{sufirst2}
Let $L\subsetneq \Lambda$ and $\Cl$ be a $\FF$-closure operator $\Cl$ over $L$. If $\Cl$ satisfies $\mathsf{SU}$ and $\mathsf{K}$, then $\Cl$ is adequate.
\end{prop}
\begin{proof}
By Proposition \ref{sufirst}, it suffices to show that $\Cl^{*}$ satisfies condition $\mathsf{K}$, if $\Cl$ does. This follows from Lemma \ref{use00}.
\end{proof}

We now prove that adequate $\FF$-closure operators allows to interpret System $\FF$ judgements:

\begin{thm}[soundness]\label{soundness}
For all $L\subseteq\Lambda$ and adequate $\FF$-closure operator $\Cl$ over $L$ the following holds: if $\Gamma\vdash M:\sigma$ is derivable in System $\FF$, where $\Gamma=\{x_{1}:\sigma_{1},\dots, x_{n}:\sigma_{n}\}$ and $\TT{FV}(\Gamma)\cup \TT{FV}(\sigma)\subseteq \{X_{1},\dots, X_{p}\}$, then for any $\Cl$-interpretation $\C M$, for all $P_{1}\in |\sigma_{1}|^{\Cl}_{\C M}, \dots, P_{n}\in |\sigma_{n}|^{\Cl}_{\C M}$, $M[P_{1}/x_{1},\dots, P_{n}/x_{n}]\in |\sigma|^{\Cl}_{\C M}$.

\end{thm}
\begin{proof}
We argue by induction on a typing derivation of $\Gamma\vdash M:\sigma$. 
\begin{itemize}
\item If the derivation is $\Gamma,x:\sigma\vdash x:\sigma$, then if $P\in |\sigma|_{\C M}^{\Cl}$, $P=x[P/x]\in |\sigma|_{\C M}^{\Cl}$.

\item If the derivation ends by $\AXC{$\Gamma, x:\sigma\vdash M:\tau$}\UIC{$\Gamma\vdash \lambda x.M:\sigma\to \tau$}\DP$, we can assume that $x$ does not occur free in $P_{1},\dots, P_{n}$ and is different from $x_{1},\dots, x_{n}$. By the induction hypothesis, for all $P\in |\sigma|_{\C M}^{\Cl}$, $M[P_{1}/x_{1},\dots, P_{n}/x_{n},P/x]\in |\tau|_{\C M}^{\Cl}$ and by the assumptions just made, $M[P_{1}/x_{1},\dots, P_{n}/x_{n},P/x]=M[P_{1}/x_{1},\dots, P_{n}/x_{n}][P/x]$. 

Now, if $L=\Lambda$, then by $\mathsf{K}$ and the fact that $M\in L$, we deduce
$(\lambda x.M)[P_{1}/x_{1},\dots, P_{n}/x_{n}]\in |\sigma|_{\C M}^{\Cl}\to |\tau|_{\C M}^{\Cl}=|\sigma\to \tau|_{\C M}^{\Cl}$. 
If $L\subsetneq \Lambda$, we argue as follows: since $\Cl$ is adequate, there exists a regular $\FF$-closure operator $\F D$ such that $\Cl\leq \F D$. As $\F D$ is regular, by Lemma \ref{regular}, for all $\F D$-interpretation $\C N$, the sets $|\sigma_{i}|^{\F D}_{\C N}$, for $i=1,\dots,n$ and $|\sigma|^{\F D}_{\C N}$ contain $\TT{TermVar}$. Moreover, since $\F D$ is adequate (by Lemma \ref{use00}), by the induction hypothesis, for all $P_{1}\in |\sigma_{1}|^{\F D}_{\C N}$,\dots, $P_{n}\in |\sigma_{n}|^{\F D}_{\C N}$ and $P\in |\sigma|_{\C N}^{\F D}$, $M[P_{1}/x_{1},\dots, P_{n}/x_{n},P/x]\in |\tau|_{\C N}^{\F D}$. By taking $P_{1}=x_{1},\dots, P_{n}=x_{n}, P=x$, we deduce then $M\in |\tau|_{\C N}^{\F D}\subseteq L$. Hence $M\in L$.
Now, since $\Cl$ satisfies condition $\mathsf{K}$ and $M\in L$, we can conclude $(\lambda x.M)[P_{1}/x_{1},\dots, P_{n}/x_{n}]\in |\sigma|_{\C M}^{\Cl}\to_{L} |\tau|_{\C M}^{\Cl}=|\sigma\to \tau|_{\C M}^{\Cl}$. 

\item If the derivation ends by $\AXC{$\Gamma\vdash M_{1}:\tau\to \sigma$}\AXC{$\Gamma\vdash M_{2}:\tau$}\BIC{$\Gamma\vdash M_{1}M_{2}:\sigma$}\DP$, by the induction hypothesis $M_{1}[P_{1}/x_{1},\dots, P_{n}/x_{n}]\in |\tau\to\sigma|_{\C M}^{\Cl}=|\tau|_{\C M}^{\Cl}\to_{L} |\sigma|_{\C M}^{\Cl}$ and $M_{2}[P_{1}/x_{1},\dots, P_{n}/x_{n}]\in |\tau|_{\C M}^{\Cl}$, hence $ M_{1}[P_{1}/x_{1},\dots, P_{n}/x_{n}]M_{2}[P_{1}/x_{1},\dots, P_{n}/x_{n}]= M[P_{1}/x_{1},\dots, P_{n}/x_{n}]\in |\sigma|_{\C M}^{\Cl}$.

\item If the derivation ends by $\AXC{$\Gamma\vdash M:\sigma$}\UIC{$\Gamma\vdash M:\forall X\sigma$}\DP$, where $X\notin \TT{FV}(\Gamma)$, then for all $s\in S_{\Cl}$ and $i\leq n$, $|\sigma_{i}|^{\Cl}(s, \C M(X_{1}),\dots, \C M(X_{p}))=|\sigma_{i}|^{\Cl}_{\C M}$, whence by the induction hypothesis we deduce $M[P_{1}/x_{1},\dots, P_{n}/x_{n}]\in |\sigma|^{\Cl}(s, \C M(X_{1}),\dots, \C M(X_{p}))$, and finally $M[P_{1}/x_{1},\dots, P_{n}/x_{n}]\in |\forall X\sigma|_{\C M}^{\Cl}$.

\item If the derivation ends by $\AXC{$\Gamma\vdash M:\forall X\sigma$}\UIC{$\Gamma\vdash M:\sigma[\tau/x]$}\DP$, then by the induction hypothesis, for all $s\in S_{\Cl}$, $M[P_{1}/x_{1},\dots, P_{n}/x_{n}]\in |\sigma|^{\Cl}(s, \C M(X_{1}), \dots, \C M(X_{p}))$, hence in particular $M[P_{1}/x_{1},\dots, P_{n}/x_{n}]\in |\sigma|^{\Cl}(|\tau|^{\Cl}_{\C M}, \C M(X_{1}),\dots, \C M(X_{n}))$. The claim follows then from Lemma \ref{substitutionlemma}. \qedhere
\end{itemize}
\end{proof}

We conclude this subsection with some results which allow to compare a type $\sigma\in\forpiudue$ and its translation $\sigma^{+}\in \B \Pi$ and that will be used in the next sections.
We suppose given an adequate $\FF$-closure operator $\Cl$ over $L\subseteq \Lambda$.

\begin{lem}\label{fora}
Let $s,u\subseteq L$, $I\neq \emptyset$ and, for all $i\in I$,  $t_{i}\subseteq L$.
\begin{enumerate}[label=(\roman*)]
\item $s\to_{L} \bigcap_{i\in I}t_{i}= \bigcap_{i\in I}(s\to_{L} t_{i})$;
\item $(\bigcap_{i\in I}t_{i}\to_{L} s)\to_{L} u \subseteq \bigcap_{i\in I}((t_{i}\to_{L} s)\to_{L} u)$.
\end{enumerate}

\end{lem}
\begin{proof}
For (i), let $M\in s\to_{L} \bigcap_{i\in I}t_{i}$, $i\in I$ and $P\in s$; then $MP\in t_{i}$, hence $M\in s\to_{L} t_{i}$ for all $i\in I$ and we conclude $M\in \bigcap_{i\in I}(s\to_{L} t_{i})$. Let now $M\in \bigcap_{i\in I}(s\to_{L} t_{i})$ and $P\in s$; for all $i\in I$, $M\in s\to_{L} t_{i}$, hence $MP\in t_{i}$, and we conclude $MP\in \bigcap_{i\in I}t_{i}$, i.e. $M\in s\to_{L} \bigcap_{i}t_{i}$.

For (ii), first observe that for all $i\in I$, $t_{i}\to_{L} s\subseteq \bigcap_{i\in I}t_{i}\to_{L} s$: if $P\in t_{i}\to_{L} s$ and $Q\in \bigcap_{i\in I}t_{i}$, then, since $Q\in t_{i}$, $PQ\in s$. Let now $M\in (\bigcap_{i\in I}t_{i}\to_{L} s)\to_{L} u$, $i\in I$ and $P\in t_{i}\to_{L} s$. Since $P\in \bigcap_{i\in I}t_{i}\to_{L} s$, $MP\in u$, and we conclude $M\in \bigcap_{i\in I}((u_{i}\to_{L} s)\to_{L} u)$. 
\end{proof}

\begin{prop}\label{pifor}
For any type $\sigma\in \TT T$,
\begin{enumerate}[label=(\roman*)]
\item if $\sigma\in \forpiudue$, then $\|\sigma\|^{\Cl}\subseteq\|\sigma^{+}\|^{\Cl}= \| \sk(\sigma)\|^{\Cl}$;

\item if $\sigma\in \formenodue$, then $\|\sigma^{-}\|^{\Cl}\subseteq\|\sigma\|^{\Cl}$.
\end{enumerate}
\end{prop}
\begin{proof}
We will prove that for any interpretation $\C M:\TT{TypeVar}\to \C S_{\Cl}$, $|\sigma|_{\C M}^{\Cl}\subseteq |\sigma^{+}|^{\Cl}_{\C M}$, if $\sigma\in \forpiudue$ and $|\sigma^{-}|^{\Cl}_{\C M}\subseteq |\sigma|_{\C M}^{\Cl}$, if $\sigma\in \formenodue$, by induction on $\sigma$. If $\sigma=X$ then $\sigma=\sigma^{+}=\sigma^{-}$ and both claims are trivial. If $\sigma=\tau\to \rho\in \forpiudue$, where $\tau=\tau_{1}\to \dots \to \tau_{p}\to Y\in \formenodue$ and $\rho\in \forpiudue$, then by induction hypothesis $|\tau|^{\Cl}_{\C M}\supseteq |\tau^{-}|^{\Cl}_{\C M}$ and $|\rho|^{\Cl}_{\C M}\subseteq|\rho^{+}|^{\Cl}_{\C M}$, hence $|\sigma|^{\Cl}_{\C M}\subseteq|\tau^{-}\to \rho^{+}|^{\Cl}_{\C M}$. Now, since $\tau^{-}\to \rho^{+}= (\forall \C X_{1}\tau'_{1}\to \dots \to \forall \C X_{p}\tau'_{p}\to Y)\to \forall \C X\rho'$ and $\sigma^{+}=\forall \C X_{1}\dots \forall \C X_{p}\forall \C X ((\tau'_{1}\to \dots \to \tau'_{p}\to Y)\to \rho')$, by Lemma \ref{fora} we get 
$|\sigma|^{\Cl}_{\C M}\subseteq |\tau^{-}\to \rho^{+}|^{\Cl}_{\C M}=|\forall \C X(\tau^{-}\to \rho')|^{\Cl}_{\C M}\subseteq |\sigma^{+}|$.

If $\sigma=\tau\to \rho\in \formenodue$, where $\tau\in \forpiudue$ and $\rho\in \formenodue$ then, by induction hypothesis, $|\tau|^{\Cl}_{\C M}\subseteq|\tau^{+}|^{\Cl}_{\C M}$ and $|\rho^{-}|^{\Cl}_{\C M}\subseteq |\rho|^{\Cl}_{\C M}$, and we conclude $|\sigma^{-}|^{\Cl}_{\C M}\subseteq |\sigma|^{\Cl}_{\C M}$.

Now, if $\sigma=\forall X\tau$ and $\sigma\in \forpiudue$, then $|\sigma|^{\Cl}_{\C M}=\bigcap_{s\in S}|\tau|^{\Cl}_{\C M[X\mapsto s]}\subseteq \bigcap_{s\in  S}|\tau^{+}|^{\Cl}_{\C M[X\mapsto s]}= |\sigma^{+}|^{\Cl}_{\C M}$. 

It remains to prove $\| \sigma^{+}\|^{\Cl}=\| \sk(\sigma)\|^{\Cl}$: since $\sigma^{+}= \forall \C X\sk(\sigma)$, if we let $FV(\sigma^{+})= \{Y_{1},\dots, Y_{n}\}$ and $\C X=\{X_{1},\dots, X_{n'}\}$, 
we have
\begin{align*}
  \|\sigma^{+} \|^{\Cl}=&~ \bigcap_{s_{1},\dots, s_{n}\in S_{\Cl}}|\sigma^{+}|^{\Cl}(s_{1},\dots, s_{n})  \\
   =&~ \bigcap_{s_{1},\dots, s_{n}\in S_{\Cl}}\left ( \bigcap_{t_{1},\dots, t_{n'}\in S_{\Cl}} | \sk(\sigma)|^{\Cl}(s_{1},\dots, s_{n},t_{1},\dots, t_{n'})\right ) \\ =&~
\bigcap_{s_{1},\dots, s_{n+n'}\in S_{\Cl}}  | \sk(\sigma)|^{\Cl}(s_{1},\dots, s_{n+n'})\\ =&~
\|\sk(\sigma)\|^{\Cl}.\qedhere
  \end{align*}
\end{proof}

\subsection{Some concrete realizability semantics}\label{sec32}

\subsubsection{Stable and saturated sets}

Three well-known semantics over $\Lambda$ are generated by the following closure operators:
\index{Realizability!$\beta$-stable semantics}
\index{Realizability!$\beta\eta$-stable semantics}
\index{Realizability!$\beta$-saturated semantics}

\begin{enumerate}
\item $(\_)^{\beta}: \wp(\Lambda)\to \wp(\Lambda)$ given by
$s^{\beta}=\{M\in\Lambda\mid \exists M'\in s \ M\simeq_{\beta }M'\}$;

\item $(\_)^{\beta\eta}: \wp(\Lambda)\to \wp(\Lambda)$ given by
$s^{\beta\eta}=\{M\in\Lambda\mid \exists M'\in s \ M\simeq_{\beta\eta }M'\}$;

\item $(\_)^{\beta sat}: \wp(\Lambda)\to \wp(\Lambda)$ defined  inductively as follows:
	\begin{itemize}
	\item $s^{\beta sat}_{0}:= s$
	\item $s^{\beta sat}_{n+1}:= \{ (\lambda x.P)QQ_{1}\dots Q_{n} \mid P[Q/x]Q_{1}\dots Q_{n}\in s^{\beta sat}_{n}\}$
	\item $s^{\beta sat}:= \bigcup_{n}s^{\beta sat}_{n}$.
	\end{itemize}

\end{enumerate}

It is easily checked that all three operators above are closure operators over $\Lambda$.
The following is also easily established:
\begin{prop}\label{semas2}
For all $s\subseteq \Lambda$,
\begin{enumerate}
\item $s\in \C S_{\beta}$ iff $s$ is \emph{$\beta$-stable}, i.e. for any term $M,N$, if $M\in s$ and $M\simeq_{\beta} N$, then $N\in s$.
\item $s\in \C S_{\beta\eta}$ iff $s$ is \emph{$\beta\eta$-stable}, i.e. for any term $M,N$, if $M\in s$ and $M\simeq_{\beta\eta} N$, then $N\in s$.

\item $s\in \C S_{\beta sat}$ iff $s$ is \emph{$\beta$-saturated}, i.e. $s$ closed with respect to weak-head expansion: for any $n\in \BB N$ and term $M$ of the form
$(\lambda x.P)QQ_{1}\dots Q_{n}$, if $P[Q/x]Q_{1}\dots Q_{n}\in s$, then $M\in s$. 
\end{enumerate}
\end{prop}

\begin{prop}\label{semasat}
$(\_)^{\beta}, (\_)^{\beta\eta}$ and $(\_)^{\beta sat}$ are $\FF$-closure operators. Moreover,
$(\_)^{\beta sat}\leq (\_)^{\beta}\leq (\_)^{\beta\eta}$.
\end{prop}
\begin{proof}
From Proposition \ref{semas2}, it is clear that $(\_)^{\beta sat}\leq (\_)^{\beta}\leq (\_)^{\beta\eta}$.
We show that $(\_)^{\beta sat}$ is a $\FF$-closure operator. That $(\_)^{\beta},(\_)^{\beta\eta}$ are $\FF$-closure operators can be proved in a similar way.

Let $s,t\in \C S_{\beta sat}$ and suppose $P[Q/x]Q_{1}\dots Q_{n}\in s\to t$. If $s\neq \emptyset$, then for all $R\in s$, 
$P[Q/x]Q_{1}\dots Q_{n}R\in t$ and, since $t\in \C S_{\beta sat}$, $(\lambda x.P)QQ_{1}\dots Q_{n}R\in t$. We conclude that for all 
$R\in s$, $(\lambda x.P)QQ_{1}\dots Q_{n}R\in t$, whence $(\lambda x.P)QQ_{1}\dots Q_{n}\in s\to t$. This shows that $s\to t$ is $\beta$-saturated, hence by Proposition \ref{semas2}, $s\to t\in \C S_{\beta sat}$.
If $s=\emptyset$, then $s\to t= \Lambda$ and clearly $\Lambda^{\beta sat}=\Lambda$, so again $s\to t\in \C S_{\beta sat}$. 
%
%
\end{proof}

The following can be easily verified:
\begin{prop}
$(\_)^{\beta},(\_)^{\beta sat}$ and $(\_)^{\beta\eta}$ satisfy conditions $\mathsf{SU}$ and $\mathsf{K}$.
\end{prop}

Since $(\_)^{\beta},(\_)^{\beta sat}$ and $(\_)^{\beta\eta}$ obviously satisfy condition $\mathsf{R}$, we deduce:
\begin{cor}
$(\_)^{\beta},(\_)^{\beta sat}$ and $(\_)^{\beta\eta}$ are adequate.
\end{cor}



%

\begin{rem}\label{downward}
It is instructive to consider an example of a non adequate semantics: let $(\_)^{\beta\downarrow}:\wp(\Lambda)\to \wp(\Lambda)$ be given by
$$
s^{\beta\downarrow}= \{ P\mid \exists M\in s \ M\to^{*}_{\beta} P\}
$$
It can be verified that $(\_)^{\beta\downarrow}$ is a $\FF$-closure operator. Moreover, $s\in S_{\beta \downarrow}$ iff $s$ is \emph{$\beta$-downward closed}, that is, for any term $M\in s$, if $M\to^{*}_{\beta} M'$, then $M'\in s$. Now let $s=\{ \lambda x.x\}$; then $x[\lambda x.x/x]\in s$, but $\lambda x.x\notin s\to s$, since $(\lambda x.x)\lambda x.x\notin s$.
Hence the semantics $S_{\beta\downarrow}$ does not satisfy condition $\mathsf{K}$ and is therefore not adequate.

\end{rem}

\subsubsection{Reducibility candidates}

We define a closure operator $\CR$ generating Girard's reducibility candidates (in an untyped frame, see \cite{Gallier}) as well as two variants $\CR',\CR''$ which generate variants of reducibility candidates from \cite{Cousineau}. The introduction of these two variants is motivated by the fact that $\CR$ is regular, and regularity is an obstacle to investigate completeness and parametricity (see Subsection \ref{sec41}).

We let $\CR: \wp(\C{SN})\to \wp(\C{SN})$ be the closure operator defined inductively as follows (see \cite{Riba2009}):
\begin{itemize}
\item $\CR_{0}(s)= s^{\beta\downarrow}$ (see Remark \ref{downward});
\item $\CR_{n+1}(s)= \{ M \mid M \in \C N \ and \ \forall M' \ ( M\to_{\beta}M' \To M'\in \CR_{n}(s)) \}$;
\item $\CR(s)= \bigcup_{n}\CR_{n}(s)$.

\end{itemize}

That $\CR(s)\in \C{SN}$ for all $s\in \C{SN}$ is easily checked by induction on $n\in \BB N$.
We recall the original definition of reducibility candidates. 

\begin{defi}[reducibility candidates, \cite{Girard72, Gallier}]\label{redu}\index{Realizability!Reducibility candidates $\CR$}
A set $s$ of untyped $\lambda$-terms is a \emph{reducibility candidate} if it satisfies the following conditions:

\begin{description}
\item[($\CR1$)] if $M\in s$, then $M$ is strongly normalizing;
\item[($\CR2$)] if $M\in s$ and $M\to_{\beta}^{*} M'$, then $M'\in s$;
\item[($\CR3$)] if  $M\in \C N$ and for all $M'$ such that $M\to_{\beta} M'$, $M'\in s$, then $M\in s$.

\end{description}

\end{defi}

\begin{prop}\label{reduprop}
$s\in \C S_{\CR}$ iff $s$ is a reducibility candidate.
\end{prop}
\begin{proof}
If $s$ is a reducibility candidate, it is easily verified by induction that for all $n\in \BB N$, $\CR_{n}(s)\subseteq s$, whence $\CR(s)\subseteq s$ and we conclude $s\in \C S_{\CR}$.
Conversely, suppose $s\subseteq\C{SN}$ and $s=\CR(s)$. By the assumption $s$ satisfies $\CR1$. 
Since $\CR_{0}(s)\subseteq \CR(s)\subseteq s$, it follows that $s$ satisfies $\CR2$. 
Finally, suppose $M\in \C N$ and for all $M'$ such that $M\to_{\beta}M'$, $M'\in s$. Then $M\in \CR_{1}(s)\subseteq \CR(s)\subseteq s$, whence $s$ satisfies $\CR3$.
\end{proof}

It is well-known that if $s,t$ are reducibility candidates, so is $s\to t$ (see \cite{Gallier}). In particular one has that $s\to_{\C{SN}}t= s\to t$. By Proposition \ref{reduprop} this implies that $\CR$ is a $\FF$-closure operator. Moreover, $\CR$ is regular: condition $\CR3$ implies that for all $s\subseteq \C{SN}$, $\TT{TermVar}\subseteq \CR(s)$. From Proposition \ref{ellezero1} we deduce then that $\C{SN}_{0}\in s$, for all reducibility candidate $s$. In other words, for any variable $x$, $n\in \BB N$ and strongly normalizing terms $P_{1},\dots, P_{n}$, $xP_{1}\dots P_{n}\in \CR(s)$, for all $s\subseteq \C{SN}$.


\begin{prop}
$\CR$ is an adequate $\FF$-closure operator. 
\end{prop}
\begin{proof}
Since $\CR$ is regular, by Lemma \ref{sufirst} it suffices to verify that it satisfies condition $\mathsf{K}$. As $\CR$ is regular, all $\CR$-closed sets are non-empty.
Let then $s,t\in \C S_{\CR}$ and $M\in \C{SN}$ be such that for all $P\in s$, $M[P/x]\in t$. We argue by induction on $\SF d(M)+\SF d(P)$ (see Subsection \ref{sec21}) that $(\lambda x.M)P\in t$. If $\SF d(M)+\SF d(P)=0$, then $M,P$ are $\beta$-normal, hence the unique immediate reduct of $(\lambda x.M)P$ is $M[P/x]$, which is in $t$ by assumption, hence $(\lambda x.M)P\in t$ by $\CR3$.
If $\SF d(M)+\SF d(P)>0$, then any immediate reduct of $(\lambda x.M)P$ is either $M[P/x]$, which is in $t$ by assumption, either of the form
$(\lambda x.M')P$, with $M\to_{\beta}M'$, which is in $t$ by the induction hypothesis, or of the form $(\lambda x.M)P'$, with $P\to_{\beta} P'$, which is again in $t$ by the induction hypothesis. Hence all immediate reduct of $(\lambda x.M)P$ is in $t$, so $(\lambda x.M)P\in t$ by $\CR3$. We can now conclude that $\lambda x.M\in s\to t$.
\end{proof}

We now introduce two variants $\CRuno,\CRdue$ of Definition \ref{redu}, corresponding to the variants of reducibility candidates introduced in \cite{Cousineau}. 
We let $\CRuno,\CRdue:\wp(\C{SN})\to \wp(\C{SN})$ be the closure operators below:
\begin{enumerate}
\item $\CRuno(s)$ is defined inductively as follows:
\begin{itemize}
\item $\CRuno_{0}(s)=s^{\beta\downarrow} $;
\item $\CRuno_{n+1}(s)= \{ M \mid M\in \C{N}^* \ and \ \forall M' \ ( M\to_{\beta}M' \To M'\in \CRuno_{n}(s)) \}$;
\item $\CRuno(s)= \bigcup_{n}\CRuno_{n}(s)$.

\end{itemize}

\item $\CRdue(s)$ is defined inductively as follows:\begin{itemize}
\item $\CRdue_{0}(s)= s^{\beta\downarrow} $;
\item $\CRdue_{n+1}(s)= \{ P \mid P=M[F_{1}/x_{1},\dots, F_{n}/x_{n}], F_{1},\dots, F_{n}\in \C{N}^* \ \text{and} \  \forall G_{1},\dots, G_{n}\ (F_{1}\to_{\beta}G_{1},\dots, F_{n}\to_{\beta}G_{n} \ \To \ M[G_{1}/x_{1},\dots, G_{n}/x_{n}]\in \CRdue_{n}(s))\}$;
\item $\CRdue(s)= \bigcup_{n}\CRdue_{n}(s)$.

\end{itemize}

\end{enumerate}

The corresponding variants of reducibility candidates are obtained by modifying condition~$\CR3$.

\begin{defi}[reducibility candidates, first variant, \cite{Cousineau}]\label{redu1}\index{Realizability!Reducibility candidates, first variant $\CR'$}
A set of $\lambda$-terms is a \emph{reducibility candidate} if it satisfies the conditions $\CR1, \CR2$ of Definition \ref{redu} plus the condition $\CR3'$ below:

\begin{description}
\item[($\CR3'$)]  for all $M\in \C{N}^*$, if for all $M'$ such that $M\to_{\beta} M'$, $M'\in s$, then $M\in s$.

\end{description}

\end{defi}

\begin{defi}[reducibility candidates, second variant, \cite{Cousineau}]\label{redu2}
\index{Realizability!Reducibility candidates, second variant $\CR''$}
A set of $\lambda$-terms is a \emph{reducibility candidate} if it satisfies the conditions $\CR1, \CR2$ of Definition \ref{redu} plus the condition $\CR3''$ below:

\begin{description}
\item[($\CR3''$)] for all $n\in \BB N$ and terms $M,F_{1},\dots, F_{n}$, if
\begin{itemize}
\item for all $i\leq n$, $F_{i}\in \C{N}^*$;
\item for all $G_{1},\dots, G_{n}$ such that, for all $i=1,\dots, n$, $F_{i}\to_{\beta} G_{i}$, $M[G_{1}/x_{1},\dots, G_{n}/x_{n}]\in s$,
\end{itemize}
then $M[F_{1}/x_{1},\dots, F_{n}/x_{n}]\in s$.

\end{description}

\end{defi}

The following Proposition is proved similarly to Proposition \ref{reduprop}.
\begin{prop}\label{crcara}\leavevmode
\begin{enumerate}
\item For all $s\subseteq \C{SN}$, $s\in \C S_{\CRuno}$ iff $s$ is a reducibility candidate according to Definition \ref{redu1}.
\item For all $s\subseteq \C{SN}$, $s\in \C S_{\CRdue}$ iff $s$ is a reducibility candidate according to Definition \ref{redu2}.

\end{enumerate}
\end{prop}

\begin{rem}
The two conditions $\CR3'$ and $\CR3''$ correspond to the conditions $\CR3_{aux}$ and $\CR3'$ in \cite{Cousineau}.
We observe that the definition of the two variants of reducibility candidates in \cite{Cousineau} is slightly different as reducibility candidates are there required to be non-empty. This seems problematic for the interpretation of System $\FF$, as the resulting family of sets is not closed with respect to arbitrary non-empty intersections.

\end{rem}

In \cite{RibaUnion} it is proved that Girard's reducibility candidates are stable by union. By Proposition \ref{union2}, this means that $\CR$ satisfies $\mathsf{SU}$. The argument presented there can be extended to $\CR'$ and $\CR''$, as we do in detail in Appendix \ref{appCR}.
\begin{prop}\label{sucousineau}
$\CR, \CR',\CR''$ satisfy $\mathsf{SU}$.
\end{prop}
\begin{proof}
For $\CR$ see \cite{RibaUnion}, for $\CR',\CR''$ the proof is in Appendix \ref{appCR}.
\end{proof}

\begin{prop}
$\CRuno$ and $\CRdue$ are adequate $\FF$-closure operators. 
\end{prop}
\begin{proof}
We only consider the case of $\CRdue$, as the case of $\CRuno$ can be proved similarly to $\CR$.
Let $s,t\in S_{\CR''}$. As we already observed, if $s\neq\emptyset$, then $s\to_{\C{SN}}t=s\to t$. Indeed, for all $P\in s$ and $Q\in s\to_{\C{SN}}t$, $QP\in t\subseteq \C{SN}$, and this forces $Q\in \C{SN}$. 
We now show that for all $n\in \BB N$, $\CR''_{n}(s\to t)\subseteq s\to t$. If $n=0$, then let $P\in s\to t$ and suppose $P\to_{\beta}^{*}P'$. Then for all $Q\in s$, $PQ\to_{\beta}^{*}P'Q$, and since $\CR''(t)\subseteq t$, we deduce $P'Q\in t$. In definitive, $P'\in s\to t$, so $\CR_{0}(s\to t)\subseteq s\to t$. 
Suppose now $\CR_{n}(s\to t)\subseteq s\to t$ and let $P=P_{0}[F_{1}/x_{1},\dots, F_{m}/x_{m}]$ for some $m\in \BB N$ and terms $F_{1},\dots, F_{m}\in \C N^{*}$ such that for all $G_{1},\dots, G_{m}$ such that $F_{i}\to_{\beta}G_{i}$ for $i=1,\dots,n$, 
$P_{0}[G_{1}/x_{1},\dots, G_{n}/x_{m}]\in \CR_{n}(s\to t)\subseteq s\to t$. Then for all $Q\in s$, we can choose $y_{1},\dots, y_{m}$ not free in $Q$ nor in $P_{0}$ so that $P=P_{0}[y_{1}/x_{1},\dots, y_{m}/x_{m}][F_{1}/y_{1},\dots, F_{m}/y_{m}]$; let $P'_{0}=P_{0}[y_{1}/x_{1},\dots, y_{m}/x_{m}]$. Then $PQ= P'_{0}Q[F_{1}/y_{1},\dots, F_{m}/y_{m}]$ and for all $G_{1},\dots, G_{m}$ such that $F_{i}\to_{\beta}G_{i}$ for $i=1,\dots,n$, $P'_{0}Q[G_{1}/y_{1},\dots, G_{m}/y_{m}]=P_{0}[G_{1}/x_{1},\dots, G_{m}/x_{m}]Q\in s\to t$ by the assumption. Hence, by $\CR3''$ we deduce $PQ\in \CR''(t)\subseteq t$. This shows then that $P\in s\to t$.
Finally, if $s=\emptyset$, then $s\to_{\C{SN}}t= \C{SN}= \CR''(\C{SN})$, so again $s\to_{\C{SN}}t$ is $\CR''$-closed.

It remains to show that $\CR''$ is adequate. By Propositions \ref{sufirst2} and \ref{sucousineau}, it suffices to show that $\CR''$ satisfies condition $\mathsf K$. Let $s,t\in S_{\CR''}$. If $s\neq \emptyset$, we can argue similarly to the case of $\CR$.
%
If $s=\emptyset$, we must show that for all $P\in \C{SN}$, $\lambda x.P\in s\to_{\C{SN}}t$, but this follows from the fact that 
 $s\to_{\C{SN}}t=\Lambda \cap \C{SN}=\C{SN}$.
\end{proof}

Observe that we have $\CRuno\leq{\CRdue}$, since condition $\CR3'$ is a particular case of condition $\CR3''$. Moreover, since $\CRdue(s)\subseteq s^{\beta}$, we have ${\CRdue}\leq (\_)^{\beta}$.
In definitive we have $(\_)^{\beta \downarrow}\leq {\CRuno}\leq {\CRdue}\leq (\_)^{\beta}$. In Appendix \ref{appCR} it is also shown that $(\_)^{\beta sat}\leq {\CRuno},{\CRdue}$ (Corollary \ref{wexpa}), that is, that the semantics  $\C S_{\CRuno},\C S_{\CRdue}$ are closed with respect to strongly normalizing weak head expansion.

Finally, one can consider closure operators $\CR_{\eta}, \CRuno_{\eta}, \CRdue_{\eta}$ obtained by replacing $\beta$-reduction with $\beta\eta$-reduction, yielding in a similar way semantics of $\beta\eta$-reducibility candidates (see \cite{Gallier}).

\subsubsection{Properties of closure operators}\label{sec33}

We introduce some technical properties of closure operator that will be used in the next sections.

\begin{defi}\index{Realizability!Stability by application $\mathsf{SA}$}\index{Realizability!Stability by substitution $\mathsf{SS}$}
Let $L\subseteq\Lambda$ and $\Cl$ be a closure operator over $L$. We define the properties of \emph{stability by application} ($\mathsf{SA}$) and \emph{stability by substitution} ($\mathsf{SS}$) as follows:
\begin{description}
\item[($\mathsf{SA}$)] for all $P,P',Q\in L$, if $P'\in \Cl(\{P\})$ and $PQ\in L$, then $P'Q\in \Cl(\{PQ\})$; 
\item[($\mathsf{SS}$)] for all $P,P',Q_{1},\dots, Q_{n}\in L$, if $P'\in \Cl (\{P\})$ and $P[Q_{1}/x_{1},\dots, Q_{n}/x_{n}]\in L$, then \\$P'[Q_{1}/x_{1},\dots, Q_{n}/x_{n}]\in \Cl(\{P[Q_{1}/x_{1},\dots, Q_{n}/x_{n}]\})$.

\end{description}

\end{defi}


$\mathsf{SA}$ is satisfied by all closure operators defined in the previous subsection and will be exploited in Section \ref{sec5} to define closed logical relations.

\begin{prop}
$(\_)^{\beta},(\_)^{\beta\eta}, (\_)^{\beta sat}$ satisfy $\mathsf{SA}$.

\end{prop}
\begin{proof}
If $P'\simeq_{\beta}P$, then $P'Q\simeq_{\beta}PQ$, whence $P'Q\in \{PQ\}^{\beta}$. This shows that $(\_)^{\beta}$ satisfies $\mathsf{SA}$. A similar argument holds for $(\_)^{\beta\eta}$.

If $P'\in \{P\}^{\beta sat}$, then
$P'\to_{wh}P$. Now, for all $Q$, $P'Q\to_{wh}PQ$, whence $P'Q\in \{PQ\}^{\beta sat}$. This shows that $(\_)^{\beta sat}$ satisfies $\mathsf{SA}$.
\end{proof}

\begin{prop}
$\CR,\CRuno,\CRdue$ satisfy $\mathsf{SA}$.

\end{prop}
\begin{proof}
We consider the case of $\CR$ only, as the other two cases are proved in a similar way. 
Let $P,P',Q\in \C{SN}$, where $P'\in \CR(\{P\})$ and suppose $PQ\in \C{SN}$.
We show that $P'Q\in \C{SN}$ and $P'Q\in \CR(\{P\})$ by induction on $ \SF c(P')+\SF d(Q) $, where $\SF c$ is the least $n$ such that $P'\in \CR_{n}(\{P\})$ and $\SF d$ was defined in Subsection \ref{sec21}. 
If $\SF c(P')+\SF d(Q)=0$, then $P\to_{\beta}^{*}P'$. Hence $PQ\to_{\beta}^{*}P'Q$, whence $P'Q\in \CR_{0}(\{PQ\})\subseteq \CR(\{PQ\})$.
 Suppose $\SF c(P')+\SF d(Q)=k+1$. If $\SF c(P')=0$ we can argue as above. Otherwise it must be $P'\in \C N$. Hence any immediate reduct of $P'Q$ is either of the form $P''Q$, where $P''$ is an immediate reduct of $P'$ or of the form $P'Q'$, where $Q'$ is an immediate reduct of $Q$. In both cases, we deduce $P''Q\in\CR(\{P\})$ by the induction hypothesis. Hence all immediate reducts of $P'Q$ are in $\CR(\{PQ\})$ and we conclude $P'Q\in \CR(\{PQ\})$ by $\CR 3$.
\end{proof}

$\mathsf{SS}$ can be easily established for $(\_)^{\beta},(\_)^{\beta sat}$ and $(\_)^{\beta\eta}$.
\begin{prop}
$(\_)^{\beta},(\_)^{\beta sat}$ and $(\_)^{\beta\eta}$ satisfy $\mathsf{SS}$.
\end{prop}
\begin{proof}
 The unique non-trivial case is $(\_)^{\beta sat}$, where $\mathsf{SS}$ follows from the fact that if $P'\to_{wh}^{*}P$, then $P'[Q/x]\to_{wh}^{*}P[Q/x]$.
\end{proof}

Unlike $\mathsf{SU}$ and $\mathsf{SA}$, $\mathsf{SS}$ is not satisfied by reducibility candidates and its variants. 
We show in Appendix \ref{appCR} that $\mathsf{SS}$ fails for ${\CRuno}$ (Corollary \ref{noss}), the argument being easily adapted to $\CR$ and $\CRdue$.


\section{Completeness for positive types}\label{sec4}

The completeness problem for the semantics generated by a closure operator $\Cl$ and a class of types $\C C$ is the one to know  whether for all $\sigma\in \C C$ and closed (normal) term $M$, if $M\in|\sigma|^{\Cl}$, then $M$ has type $\sigma$.
Several completeness results for positive types are known from the literature for the semantics discussed in the previous section. We list the main ones below:
\begin{thm}[\cite{Hindley, LabibSami, Nour1998}]\label{labib}
Let $M$ be a closed $\lambda$-term and $\sigma$ a type. 

\begin{enumerate}
\item if $\sigma\in \TT T_{0}$ and $M\in \|\sigma\|^{\beta}$, then for some $M'\simeq_{\beta} M$, $\vdash M':\sigma$;

\item if $\sigma\in \forpiudue$ and $M\in \|\sigma\|^{\beta\eta}$, then for some $M'\simeq_{\beta\eta} M$, $\vdash M':\sigma$;

\item if $\sigma\in \forpiupro$ and $M\in \|\sigma\|^{\beta sat}$, then for some $M'\simeq_{\beta} M$, $\vdash M':\sigma$.

\end{enumerate}
\end{thm}

\begin{rem}\label{clormal}

Claim $1.$ can be extended to $\forpiupro$ types by means of Theorem \ref{forpi} and Proposition \ref{pifor}:
if $\sigma\in \forpiudue$ and $M\in \|\sigma\|^{\beta}$, then by Proposition \ref{pifor}, $M\in \|\sigma^{+}\|^{\beta}= \|\sk(\sigma)\|^{\beta}$, hence by $1.$, for some $M'\simeq_{\beta} M$, $\vdash M':\sk(\sigma)$; finally, by Theorem \ref{forpi}, we get $\vdash M':\sigma$. 

Claims 1. and 3. can be strengthened in the case of closed $\beta$-normal $\lambda$-terms, as a consequence of subject $\beta$-reduction. For instance, if $\sigma\in \TT T_{0}$ and $M\in \|\sigma\|^{\beta}$ is closed $\beta$-normal, then $\vdash M:\sigma$. Indeed, from claim 1., for some $M'\simeq_{\beta}M$, $\vdash M':\sigma$, and, since it must be $M'\to_{\beta}^{*}M$, we deduce $\vdash M:\sigma$ by subject $\beta$-reduction.

Finally, claim 2. can be strengthened in the case of closed $\beta\eta$-normal $\lambda$-terms, as a consequence of subject $\beta$-reduction. If $\sigma\in \TT T_{0}$ and $M\in \|\sigma\|^{\beta\eta}$ is closed $\beta\eta$-normal, then for some $M''$ such that $M''\to_{\eta}^{*}M$, $\vdash M'':\sigma$. Indeed, from Theorem \ref{labib}, for some $M'\simeq_{\beta\eta}M$, $\vdash M':\sigma$, hence, since it must be $M'\to_{\beta\eta}^{*}M$, by subject $\beta$-reduction we deduce $\vdash M'':\sigma$, where $M''$ is the $\beta$-normal form of $M''$.

\end{rem}

The completeness properties in Theorem \ref{labib} are proved by a similar technique, which consists in constructing an interpretation made of typable terms. 
The aim of this section is to discuss the conditions under which this technique can be applied to prove completeness for an arbitrary semantics generated by a $\FF$-closure operator. 
In particular, we focus on a property, that we call $\FF$-adaptedness (inspired from \cite{Cousineau}), which plays a central role in such arguments.

By relying on this property, and on the syntactic analysis of positive types in Section \ref{sec2}, we reconstruct the completeness arguments for positive types in the more general framework introduced in Section \ref{sec3}.

\subsection{$\FF$-adapted closure operators}\label{sec41}

The proofs of the completeness results mentioned above are based on the construction of a ``term model'', that is, an interpretation made of sets of typable terms. The construction is based on the notion of infinite context that we recall below:

\begin{defi}\label{gammainfinity}\index{Realizability!Context $\Gamma^{\infty}$ and set $\sigma^{\infty}$}
\index{Realizability!Term model $\C M_{\Cl}^{\infty}$}
Let $(\tau_{i})_{i\in \BB N}$ be an enumeration of all types such that for any type $\sigma$, the set $\{\tau_{i}\mid  \tau_{i}=\sigma\}$ is infinite. We let $\Gamma^{\infty}=\{x_{i}:\tau_{i}\mid i\in \BB N\}$ be an infinite set of type declarations. For any $M\in \Lambda$ and $\sigma\in \TT T$, we let $\Gamma^{\infty}\vdash M:\sigma$ indicate the judgement $\Gamma^{\infty}_{M}\vdash M:\sigma$, where $\Gamma^{\infty}_{M}$ is the context $\{x_{i_{1}}:\tau_{i_{1}},\dots, x_{i_{n}}:\tau_{i_{n}}\}$, for $\TT{FV}(M)= \{x_{i_{1}},\dots, x_{i_{n}}\}$. 

For any type $\sigma$, we let $\sigma^{\infty}=\{M\mid \Gamma^{\infty}\vdash M:\sigma\}$.

\end{defi}

Given a closure operator $\Cl$, the sets $\sigma^{\infty}$ need not be $\Cl$-closed. For instance, the sets $\sigma^{\infty}$ are not $\beta$-stable, as typability in System $\FF$ is not closed with respect to $\beta$-equivalence. 
A $\Cl$-interpretation is obtained as follows:

\begin{defi}
Given an adequate $\FF$-closure operator $\Cl$, we let $\C M^{\infty}_{\Cl}$ be the $\Cl$-interpretation given by 
$\C M^{\infty}_{\Cl}(X)=\Cl(X^{\infty})$. 
\end{defi}

To establish completeness one has to show that the sets
$|\sigma|^{\Cl}_{\C M^{\infty}_{\Cl}}$ and $\Cl(\sigma^{\infty})$ coincide when $\sigma$ is a positive type. 
The property below is an essential ingredient to prove this fact.

\begin{defi}[$\FF$-adapted]\label{fada}\index{Realizability!$\FF$-adapted operator}
Let $\Cl$ be a $\FF$-closure operator. $\Cl$ is \emph{$\FF$-adapted} if for all types $\sigma,\tau\in \TT T$, 
\begin{equation}\label{fadapted}
\Cl( (\sigma\to\tau)^{\infty})\ = \  \Cl (\sigma^{\infty})\to \Cl (\tau^{\infty})
\end{equation}
\end{defi}

In other words, a $\FF$-closure operator $\Cl$ is $\FF$-adapted when for all types $\sigma,\tau$ and term $M$, 
the two properties below are equivalent:
\begin{itemize}
\item for all $P$ in the $\Cl$-closure of terms of type $\sigma$, $MP$ is in the $\Cl$-closure of the terms of type $\tau$;
\item $M$ is in the $\Cl$-closure of the terms of type $\sigma\to \tau$.

\end{itemize}

For any $\FF$-adapted $\FF$-closure operator one can easily prove that  $|\sigma|^{\Cl}_{\C M_{\Cl}^{\infty}}=\Cl(\sigma^{\infty})$ holds for simple types:
\begin{prop}\label{simplecom}
Let $\Cl$ be a $\FF$-adapted $\FF$-closure operator. Then for all {simple type} $\sigma\in \TT T_{0}$ $ |\sigma|^{{\Cl}}_{\C M_{\Cl}^{\infty}}= \Cl(\sigma^{\infty})$.
\end{prop}
\begin{proof}
We argue by induction on $\sigma$. If $\sigma=X$, then the claim holds by definition. If $\sigma=\sigma_{1}\to\sigma_{2}$, then by induction hypothesis $|\sigma_{i}|^{\Cl}_{\C M_{\Cl}^{\infty}}= \Cl(\sigma_{i}^{\infty})$, for $i=1,2$. Then
$|\sigma|^{\Cl}_{\C M_{\Cl}^{\infty}}= |\sigma_{1}|^{\Cl}_{\C M^{\infty}}\to |\sigma_{2}|^{\Cl}_{\C M_{\Cl}^{\infty}}=
\Cl(\sigma_{1}^{\infty})\to \Cl(\sigma_{2}^{\infty})= \Cl(\sigma^{\infty})$, since $\Cl$ is $\FF$-adapted.
\end{proof}

Definition \ref{fada} is inspired from some remarks in \cite{Cousineau}. There the introduction of the second variant $\CRdue$ of reducibility candidates is motivated by the need to obtain a semantics satisfying Equation \ref{fadapted} (that is,  $\FF$-adapted). Indeed the following results hold:
\begin{prop}\label{tzerocr}\leavevmode
\begin{enumerate}[label=(\roman*)]
\item$\CRdue$ is $\FF$-adapted.
\item
$\CR$ and $\CRuno$ are not $\FF$-adapted.
\end{enumerate}
\end{prop}
\begin{proof}
Claim (i) is Lemma 4.10, p. 20 in \cite{Cousineau}. For Claim (ii),
let $\tau$ be a simple type and $x$ a variable such that $x:\tau\in \Gamma^{\infty}$. Let $P$ be a closed strongly normalizing term which cannot have type $\tau $ (e.g.\ $P=\lambda z.zz$), and 
let $M=(\lambda y.x) P$. $M$ is neutral and not $\beta$-normal, and $M\to_{\beta} x$. Then $x\in \tau^{\infty}$, whence $M\in \CR(\tau^{\infty})$. Morever, we have $\lambda x'.M\in \CR(\tau^{\infty})\to \CR(\tau^{\infty})$, where $x'$ is a new variable. However, $\lambda x'.M\notin \CR ((\tau\to \tau)^{\infty})$: $\Gamma^{\infty}\vdash\lambda x'.M:\tau\to \tau$ is not derivable as $P$ cannot have type $\tau$, and for all $n\in \BB N$, $\lambda x'.M\notin \CR_{n}((\tau\to\tau)^{\infty})$ as $\lambda x'.M$ is not neutral. 
A similar argument can be made for $\CRuno$.
\end{proof}

We conclude this subsection by proving that $(\_)^{\beta}$ and $(\_)^{\beta\eta}$ are $\FF$-adapted. The arguments below follow the completeness proofs in \cite{Hindley} and \cite{Nour1998}.

\begin{prop}\label{tzerobetaeta}
$(\_)^{\beta\eta}$ is $\FF$-adapted.
\end{prop}
\begin{proof}

Let $\sigma=\tau\to \rho$, $P\in (\tau^{\infty})^{\beta\eta}\to (\rho^{\infty})^{\beta\eta}$ and let $x_{i}$ be a variable not occurring free in $P$ and such that $\tau_{i}= \tau$. From $x_{i}:\tau\vdash x_{i}:\tau$, we deduce $x_{i}\in (\tau^{\infty})^{\beta\eta}$ and we conclude that $Px_{i}\in (\rho^{\infty})^{\beta\eta}$. Hence for some $Q\simeq_{\beta\eta}P x_{i}$, $\Gamma^{\infty}\vdash Q:\rho$ holds, whence $\Gamma^{\infty}\vdash\lambda x_{i}.Q:\sigma$ and from $\lambda x_{i}.Q\simeq_{\beta\eta}\lambda x_{i}.Px_{i}\simeq_{\eta} P$ we deduce $P\in (\sigma^{\infty})^{\beta\eta}$.

%

For the converse direction suppose $P\in (\sigma^{\infty})^{\beta\eta}$, hence for some $P'\simeq_{\beta\eta}P$, $\Gamma^{\infty}\vdash P': \sigma$ holds. Now
 let $Q\in (\tau^{\infty})^{\beta\eta}$, hence for some $Q'\simeq_{\beta\eta}Q$, $\Gamma^{\infty}\vdash Q':\tau$ holds.
 We have then $\Gamma^{\infty}\vdash P'Q':\rho$, whence $PQ\simeq_{\beta\eta}P'Q'\in (\rho^{\infty})^{\beta\eta}$. We conclude that
 $P\in (\tau^{\infty})^{\beta\eta}\to (\rho^{\infty})^{\beta\eta}$.
\end{proof}

\begin{prop}\label{tzerobeta}
$(\_)^{\beta}$ is $\FF$-adapted.
\end{prop}
\begin{proof}
Let $\sigma=\tau\to \rho$, $P\in (\tau^{\infty})^{\beta}\to (\rho^{\infty})^{\beta}$ and let $x_{i}$ be a variable not occurring free in $P$ and such that $\tau_{i}= \tau$. From $x_{i}:\tau\vdash x_{i}:\tau$, we deduce $x_{i}\in (\tau^{\infty})^{\beta}$ and we conclude that $Px_{i}\in (\rho^{\infty})^{\beta}$. 
Hence for some $Q\simeq_{\beta}P x_{i}$, $\Gamma^{\infty}\vdash Q:\rho$ holds. As $Q$ is simply typable, it is strongly normalizable, and from $Px_{i}\simeq_{\beta}Q$ it follows that $Px_{i}$ has a $\beta$-normal form $P'$ (that of $Q$). 
Moreover, as $Q\to^{*}_{\beta} P'$, we have $\Gamma^{\infty}\vdash P':\rho$.
%
%
%
%
%
We consider now two cases:
\begin{itemize}
\item[$1.$] for all terms $Q$ such that $P\to_{\beta}^{*}Q$, $Q$ is neutral. Then $P'$ is neutral and of the form $x_{j}P_{1}\dots P_{n}x_{i}$ where $P\to_{\beta}^{*} x_{j}P_{1}\dots P_{n}$. Now from $\Gamma^{\infty}\vdash P':\rho$ it follows that $\Gamma^{\infty}\vdash x_{j}P_{1}\dots P_{n}: \tau\to \rho$. Since  $P\simeq_{\beta} x_{j}P_{1}\dots P_{n}$, we deduce $P\in (\sigma^{\infty})^{\beta}$.

\item[$2.$] The normal form of $P$ (which exists since $Px_{i}$ has a normal form) is of the form $Q=\lambda y.Q'$. Hence $P'=Q'[x_{i}/y]$ and from $\Gamma^{\infty}\vdash P':\rho$ we deduce $\Gamma^{\infty}\vdash \lambda x_{i}.P':\sigma$. Now $\lambda x_{i}.P'$ and $\lambda y.Q'$ are the same term, so we conclude $P\in (\sigma^{\infty})^{\beta\eta}$.

\end{itemize}
In both cases we showed then that $P\in (\sigma^{\infty})^{\beta}$.

For the converse direction suppose $P\in (\sigma^{\infty})^{\beta}$, hence for some $P'\simeq_{\beta}P$, $\Gamma^{\infty}\vdash P': \sigma$ holds. Now
 let $Q\in (\tau^{\infty})^{\beta}$, hence for some $Q'\simeq_{\beta}Q$, $\Gamma^{\infty}\vdash Q':\tau$ holds.
 We have then $\Gamma^{\infty}\vdash P'Q':\rho$, whence $PQ\simeq_{\beta}P'Q'\in (\rho^{\infty})^{\beta}$. We conclude that
 $P\in (\tau^{\infty})^{\beta}\to (\rho^{\infty})^{\beta}$.
\end{proof}

It is worth stressing that, unlike Proposition \ref{tzerobetaeta}, Proposition \ref{tzerobeta} exploits the fact that $\lambda_{\to}$ is strongly normalizing, as well as the subject $\beta$-reduction property.

\begin{rem}\label{natnat}
When $\sigma$ is not a positive type, the $\Cl$-closed sets $|\sigma|^{\Cl}_{\C M^{\infty}_{\Cl}}$ and $\Cl(\sigma^{\infty})$ can be very different, even for well-behaved closure operators as $\Cl=\beta,\beta\eta$.  
For instance, let $\Cl=(\_)^{\beta}$ and $\sigma= \B{Nat}\to \B{Nat}$, where $\B{Nat}=\forall X((X\to X)\to (X\to X))$. Then $\Cl(\sigma^{\infty})$ contains all terms $\beta$-equivalent to a term $M$ such that $\Gamma^{\infty}\vdash M:\sigma$. It can be easily verified that $|\sigma|^{\Cl}$ contains all terms $M$ coding total unary recursive functions, and from a well-known incompleteness argument (see \cite{Girard1989}), such terms cannot all belong to $\Cl(\sigma^{\infty})$. 
\end{rem}

\subsection{Completeness results}

We reconstruct the completeness arguments for positive types, by exploiting the analysis of the previous subsection of $\FF$-adapted closure operators. Our result allows one to show completeness also for the semantics generated by a non $\FF$-adapted closure operator, as soon as this is finer than $(\_)^{\beta}$ or $(\_)^{\beta\eta}$ (this is the case of $\CRuno$, for instance). 

Our treatment does not cover the semantics generated by a regular $\FF$-closure operator (hence, in particular, Girard's reducibility candidates). It seems that a different method has to be looked for in this case (we shortly discuss this point in the concluding section).

\begin{thm}[$\forpiudue$-completeness for $\Cl\leq (\_)^{\beta\eta}$]\label{pieta}
Let $\Cl$ be an adequate $\FF$-closure operator over $L\subseteq \Lambda$ such that $L$ is $\beta\eta$-closed and
 $\Cl\leq (\_)^{\beta\eta}$. Let $M$ be a closed $\lambda$-term and $\sigma\in \forpiudue$. If $M\in \|\sigma\|^{\Cl}$, then there exists $M'\simeq_{\beta\eta} M$ such that $\vdash M':\sigma$.
\
\end{thm}
\begin{proof}
Let $M$ be a closed term, $\sigma\in \forpiudue$ and suppose $M\in \|\sigma\|^{\Cl}$. From Lemma \ref{pifor} 1. we deduce $M\in \|\sigma^{+}\|^{\Cl}= \|\sk(\sigma)\|^{\Cl}$. 
From Theorem \ref{soundness} it follows that, for all type variable $X_{i}$, if $\Gamma^{\infty} \vdash t:X_{i}$ holds, then 
$t\in |X_{i}|^{\Cl}_{\C M^{\infty}_{\Cl}}= \Cl(X_{i}^{\infty})$. This implies in particular that $X^{\infty}\subseteq \Cl(X_{i}^{\infty})\subseteq L$. Moreover, since $L$ is $\beta\eta$-closed, for all $s\subseteq L$, $s^{\beta\eta}\cap L=s^{\beta\eta}\in S_{\Cl}$ by Proposition \ref{confront}.
Hence, since $\Cl\leq (\_)^{\beta\eta}$, the sets $(X_{i}^{\infty})^{\beta\eta}\cap L= (X_{i}^{\infty})^{\beta\eta}$ are $\Cl$-closed.
Therefore, if we let $s_{i}:=(X_{i}^{\infty})^{\beta\eta}$, $M\in |\sk(\sigma)|^{\Cl}(\vec s)=
 |\sk(\sigma)|^{\Cl}_{\C M^{\infty}_{\beta\eta}}$. By Proposition \ref{tzerobetaeta} $(\_)^{\beta\eta}$ is $\FF$-adapted, hence by Proposition \ref{simplecom}, 
 $M\in (\sk(\sigma)^{\infty})^{\beta\eta}$, that is, there is $M'\simeq_{\beta\eta} M$ such that $\vdash M': \sk(\sigma)$. As $M'$ is closed, by Proposition \ref{forpi}, we conclude that there exists $M''\simeq_{\eta}M'\simeq_{\beta\eta}M$ such that $\vdash M'': \sigma$.
\end{proof}

\begin{thm}[$\forpiupro$-completeness for $\Cl\leq (\_)^{\beta}$]\label{pibeta}
Let $\Cl$ be an adequate $\FF$-closure operator over $L\subseteq \Lambda$ such that $L$ is $\beta$-closed and $\Cl\leq (\_)^{\beta}$. Let $M$ be a closed $\lambda$-term and $\sigma\in \forpiupro$. If $M\in \|\sigma\|^{\Cl}$, then there exists $M'\simeq_{\beta} M$ such that $\vdash M':\sigma$.
\
\end{thm}
\begin{proof}
One can argue similarly to Theorem \ref{pieta}.
\end{proof}

By arguing as in Remark \ref{clormal} we deduce the following:

\begin{cor}\label{forallcr3}
Let $\Cl$ be an adequate $\FF$-closure operator over a $\beta\eta$-closed set $L\subseteq \Lambda$.
\begin{enumerate}[label=(\roman*)]
\item Suppose $\Cl\leq (\_)^{\beta\eta}$. Let $M$ be a $\beta\eta$-normal closed $\lambda$-term and $\sigma\in \forpiudue$. If $M\in \|\sigma\|^{\Cl}$, then there exists $M'\to_{\eta}^{*} M$ such that $\vdash M':\sigma$.

\item Suppose $\Cl\leq (\_)^{\beta}$. Let $M$ be a $\beta$-normal closed $\lambda$-term and $\sigma\in \forpiupro$. If $M\in \|\sigma\|^{\Cl}$, then $\vdash M:\sigma$.

\end{enumerate}
\end{cor}

Corollary \ref{forallcr3} (i) applies to $(\_)^{\beta\eta}$, ${\CRuno_{\beta\eta}}$, ${\CRdue_{\beta\eta}}$, while
Corollary \ref{forallcr3} (ii) applies to $(\_)^{\beta}$,$(\_)^{\beta sat}$, ${\CRuno}$, ${\CRdue}$.
Finally, from Proposition \ref{tzerocr} and some properties of $\CRdue$ proved in Appendix \ref{appCR}, we deduce that $\CRdue$ satisfies a slightly stronger form of completeness:

\begin{thm}[$\forpiudue$-completeness of ${\CRdue}$]\label{picr3}
Let $M$ be a closed $\lambda$-term and $\sigma\in \forpiudue$. If $M\in \|\sigma\|^{{\CRdue}}$, then there exists $M'$ such that $M\in \CRdue(\{M'\})$ and $\vdash M':\sigma$. 
\end{thm}
\begin{proof}
Let $M$ be closed, $\sigma\in \forpiudue$, and suppose $M\in \|\sigma\|^{\CRdue}$. From Lemma \ref{pifor} 1. we deduce $M\in \|\sigma^{+}\|^{\CRdue}= \|\sk(\sigma)\|^{\CRdue}$. Let, for all $i\in \BB N$, $s_{i}:=\CRdue(X_{i}^{\infty})$; then $M\in |\sk(\sigma)|^{\CRdue}(\vec s)=
 |\sk(\sigma)|^{\CRdue}_{\C M^{\infty}}$. By Proposition \ref{tzerocr} and Proposition \ref{simplecom} we deduce
 $M\in \CRdue(\sk(\sigma)^{\infty})$.
Now, since ${\CRdue}$ satisfies $\mathsf{SU}$, $M\in \CRdue(\sk(\sigma)^{\infty})$ implies that for some $M'\in \sk(\sigma)^{\infty}$, $M\in \CRdue(\{M'\})$. 
\end{proof}

\section{Closure operators and logical relations}\label{sec5}

%

We formalise logical relations in the semantics generated by a $\FF$-closure operator. We will suppose that the closure operator satisfies both $\mathsf{SU}$ and $\mathsf{SA}$. When $\mathsf{SU}$ holds, closed sets form a topology and  closed logical relations satisfy then the universal property of the product topology (Proposition \ref{universality}). When $\mathsf{SA}$ holds, closed logical relations can be shown to be closed with respect to System $\FF $ connectives.

We define a notion of parametricity as invariance with respect to $\Cl$-closed logical relations (\emph{$\Cl$-invariance}). We establish within this framework the standard result that typable terms are $\Cl$-invariant.
Since, by the completeness results of last section, for positive types, closed normal $\Cl$-interpretable $\lambda$-terms are typable, we can deduce that such terms are also $\Cl$-invariant.
Moreover, if $\Cl$ is stable by substitutions, we prove that $\Cl$-interpretability implies $\Cl$-invariance at any type (Theorem \ref{param1}). This implies in particular that the closed realizers at any type for the semantics described in the last section are invariant with respect to $\beta$ and $\beta\eta$-stable logical relations. 

Finally, we shortly discuss Reynolds' parametricity: for simple types it coincides with invariance with respect to logical relations, while for second order types it provides a stronger condition which requires an extension of the equational theory of System $\FF$ terms.

\subsection{Logical relations in realizability semantics}\label{sec51}


\subsubsection{Closure operators via the product topology}

If $\Cl$ is stable by union, the extension of a closure operator over sets of $\lambda$-terms into a closure operator over relations between $\lambda$-terms has a natural topological counterpart. Indeed, as mentioned in Section \ref{sec3}, if a closure operator $\Cl$ over $L\subseteq \Lambda$ satisfies $\mathsf{SU}$, the family $\C S_{\Cl}\cup \{\emptyset\}$ is a topology over $L$ having the $\Cl$-closed sets plus $\emptyset$ as open sets. One can define then the \emph{product topology} over $L\times L$, which has as basis 
the sets of the form $s\times t$, for $s,t$ $\Cl$-closed, plus $\emptyset$. 

Observe that if $\Cl$ is stable by union, the topology $\C S_{\Cl}\cup\{\emptyset\}$ has as basis $\C B=\{ \Cl (\{M\})\mid M\in L\}\cup \{\emptyset\}$. Whence the product topology over $L\times L$ has as basis the sets of the form 
$\Cl(\{P\})\times \Cl(\{Q\})$ plus $\emptyset$.

This topological intuition leads to the following definition:

\begin{defi}\label{clodue}\index{Logical relations!Product of closure operators}

Let $\Cl_{1}$ and $\Cl_{2}$ be closure operators over two sets $L_{1},L_{2}$, respectively. Then the map $\Cl_{1}\times \Cl_{2}: \wp(L_{1}\times L_{2})\to\wp(L_{1}\times L_{2})$ is defined as follows:
\begin{itemize}
\item $(\Cl_{1}\times \Cl_{2})(\emptyset)= \Cl_{1}(\emptyset)\times \Cl_{2}(\emptyset)$;
\item if $r\neq \emptyset$, $(\Cl_{1}\times \Cl_{2})(r)= \{(x,y) \mid \exists (x',y')\in r \ \text{s.t.}\ x\in \Cl_{1}(\{x'\}), y\in \Cl_{2}(\{y'\})\}$.
\end{itemize}

\end{defi}

The proposition below shows that $(\Cl_{1}\times \Cl_{2})$ is a closure operator: 

\begin{prop}
If $\Cl_{1},\Cl_{2}$ are closure operators over $L_{1},L_{2}$, then 
$\Cl_{1}\times \Cl_{2}$ is a closure operator over $L_{1}\times L_{2}$, that is for all $r\subseteq L_{1}\times L_{2}$ the following hold:
\begin{enumerate}
\item $r\subseteq (\Cl_{1}\times \Cl_{2})(r)$;
\item $r\subseteq r'\To (\Cl_{1}\times \Cl_{2})(r)\subseteq (\Cl_{1}\times \Cl_{2})(r')$;
\item $(\Cl_{1}\times \Cl_{2})^{2}(r)\subseteq (\Cl_{1}\times \Cl_{2})(r)$.

\end{enumerate}
\end{prop}
\begin{proof}
We first consider the case $r\neq \emptyset$.
Property $1.$ follows from the definition of $\Cl_{1}\times \Cl_{2}$ and the fact that $x\in\Cl_{1}(\{x\}), y\in \Cl_{2}(\{y\})$ for all $(x,y)\in r$. Let now $r\subseteq r'$ and suppose $(x,y)\in(\Cl_{1}\times \Cl_{2}) (r)$. Then there exist $x',y'$ such that $(x',y')\in r$ and $x\in \Cl_{1}(\{x'\}), y\in \Cl_{2}(\{y'\})$. Then by the assumption $(x',y')\in r'$ and by definition $(x,y)\in(\Cl_{1}\times \Cl_{2})(r')$.
Finally, if $(x,y)\in (\Cl_{1}\times \Cl_{2})^{2}(r)$, then there exist $x',y'$ such that $x\in \Cl_{1}(\{x'\})$, $y\in \Cl_{2}(\{y'\})$ and $(x',y')\in 
(\Cl_{1}\times \Cl_{2})(r)$, that is, there exist $x'', y''$ such that $x'\in \Cl_{1}(\{x''\})$, $y'\in \Cl_{2}(\{y''\})$ and $(x'',y'')\in r$.
Since $x'\in \Cl_{1}(\{x''\})$, we have $\Cl_{1}(\{x'\})\subseteq \Cl_{1}( \Cl_{1}(\{x''\}))= \Cl_{1}(\{x''\})$, whence $x\in \Cl_{1}(\{x''\})$. Similarly, $y\in \Cl_{2}(\{y''\})$. We deduce then that $(x,y)\in (\Cl_{1}\times \Cl_{2})(r)$. 

If $r=\emptyset$, then Property 1. is trivial; Property 2. follows from the fact that $\Cl_{1}(\emptyset)\subseteq s$, for all $s\in S_{\Cl_{1}}$ and $\Cl_{2}(\emptyset)\subseteq t$, for all $t\in S_{\Cl_{2}}$. Finally, Property 3. follows from the fact that if $(P,Q)\in (\Cl_{1}\times \Cl_{2})^{2}(\emptyset)$, then $P\in \Cl_{1}(\emptyset), Q\in \Cl_{2}(\emptyset)$, hence $(P,Q)\in (\Cl_{1}\times \Cl_{2})(\emptyset)$.
\end{proof}

Observe that, by definition, if $r\subseteq L_{1}\times L_{2}$ is non-empty, then \[(\Cl_{1}\times \Cl_{2})(r)= \bigcup_{(x,y)\in r}\Cl_{1}(\{x\})\times \Cl_{2}(\{y\}),\] hence $\Cl_{1}\times \Cl_{2}$ is stable by union.

We call a relation $r\subseteq L_{1}\times L_{2}$ \emph{($\Cl_{1}\times\Cl_{2}$)-closed} if $(\Cl_{1}\times \Cl_{2})(r)=r$.
 
\begin{prop}\label{produr}
If $\Cl_{1},\Cl_{2}$ are closure operators over $L_{1},L_{2}$, then for any $(\Cl_{1}\times \Cl_{2})$-closed relation $r\subseteq L_{1}\times L_{2}$, 
$$
r= \bigcup_{(x,y)\in r} \Cl_{1}(\{x\})\times \Cl_{2}(\{y\})
$$

\end{prop}
\begin{proof}
Let $r$ be $(\Cl_{1}\times \Cl_{2})$-closed and let $r^{*}= \bigcup_{(x,y)\in r} \Cl_{1}(\{x\})\times \Cl_{2}(\{r\})$. 
From Definition \ref{clodue} it follows that if $(x,y)\in r$, $x'\in\Cl_{1}(\{x\})$ and $y'\in \Cl_{2}(\{y\})$, then $(x',y')\in r$. This shows that $r^{*}\subseteq r$.
For the converse direction, if $(x,y)\in r$, as $x\in \Cl_{1}(\{x\})$ and $y\in \Cl_{2}(\{y\})$, then $(x,y)\in r^{*}$.
\end{proof}
 
When $\Cl_{1}, \Cl_{2}$ satisfy $\mathsf{SU}$, then the associated topologies are generated by the closure of singletons. Then Proposition \ref{produr} and the fact that $\Cl_{1}\times \Cl_{2}$ always satisfies $\mathsf{SU}$ assures that, in this case, the $(\Cl_{1}\times \Cl_{2})$-closed relations (plus the empty relation) correspond to the product topology.

We recall that the product topology has the following universal property: it is the coarsest topology for which projections are continuous. 
We formulate and prove this property in our framework:
\begin{prop}\label{universality}
Let $L_{1},L_{2}\subseteq \Lambda$, $\Cl_{1},\Cl_{2}$ be closure operators over $L_{1}$ and $L_{2}$, respectively, and let $\pi_{i}:L_{1}\times L_{2}\to L_{i}$, for $i=1,2$ be the projection maps $\pi_{i}((P_{1},P_{2}))=P_{i}$. Then for all $i=1,2$ and $s\in \C S_{\Cl_{i}}$, $\pi_{i}^{-1}(s)$ is $(\Cl_{1}\times \Cl_{2})$-closed.
Moreover, for all closure operators $\F D$ over $L_{1}\times L_{2}$
, if for all $i=1,2$ and $s\in \C S_{\Cl_{i}}$, $\pi_{i}^{-1}(s)$ is $\F D$-closed, then $\F D\leq \Cl_{1}\times \Cl_{2}$.

\end{prop}
\begin{proof}
For all $s\in \C S_{\Cl_{1}}$, it is clear that $\pi_{1}^{-1}(s)= s\times L_{2}$ is $(\Cl_{1}\times \Cl_{2})$-closed, and similarly for $s\in \C S_{\Cl_{2}}$.
Suppose now $\F D$ is a closure operator $L_{1}\times L_{2}$ satisfying $\mathsf{SU}$ and such that, for all $i=1,2$ and $s\in \C S_{\Cl_{i}}$, $\pi_{i}^{-1}(s)$ is $\F D$-closed.
We must prove that for all $r\subseteq L_{1}\times L_{2}$, if $r$ is $(\Cl_{1}\times \Cl_{2})$-closed, then $r$ is $\F D$-closed. If $r\neq \emptyset$ is $(\Cl_{1}\times \Cl_{2})$-closed, then by Proposition \ref{produr}, 
$r=\bigcup_{(P,Q)\in r}\Cl_{1}(\{P\})\times \Cl_{2}(\{Q\})$. Observe that, for all $P\in L_{1},Q\in L_{2}$,
$\Cl_{1}(\{P\})\times \Cl_{2}(\{Q\})= (\Cl_{1}(\{P\})\times L_{2})\cap (L_{1}\times \Cl_{2}(\{Q\}))=
\pi_{1}^{-1}(\Cl_{1}(\{P\}))\cap \pi_{2}^{-1}(\Cl_{2}(\{Q\}))$. 
By the assumption we deduce then that $\Cl_{1}(\{P\})\times \Cl_{2}(\{Q\})$ is $\F D$-closed, and since $\F D$ is stable by union, 
$r=\bigcup_{(P,Q)\in r}\Cl_{1}(\{P\})\times \Cl_{2}(\{Q\})$ is $\F D$-closed.
If $r=\emptyset$ is $(\Cl_{1}\times \Cl_{2})$-closed, then it must be either $\Cl_{1}(\emptyset)=\emptyset$ or $\Cl_{2}(\emptyset)=\emptyset$. Suppose $\Cl_{1}(\emptyset)=\emptyset$, then $\pi_{1}^{-1}(\emptyset)=\emptyset$ is $\F D$-closed. We can argue similarly if $\Cl_{2}(\emptyset)=\emptyset$. 
\end{proof}

\subsubsection{$\Cl$-closed logical relations} 
 
In this subsection we consider given an adequate $\mathsf F$-closure operator $\Cl$ over $L\subseteq \Lambda$ satisfying $\mathsf{SU}$ and $\mathsf{SA}$ (as discussed in Section \ref{sec3} all concrete closure operators considered are of this form).
We introduce the notion of $\Cl$-closed relation: 

\begin{defi}[$\Cl$-closed logical relation]\index{Logical relations!$\Cl$-closed relations}
A \emph{$\Cl$-closed logical relation} is a binary relation $r\subseteq s\times t$, for some $s,t\in \C S_{\Cl}$, such that $r= (\Cl\times \Cl)(r)$.
We let $\C R_{\Cl}(s,t)$ indicate the set of $\Cl$-closed relations over $s,t\in S$ and $\C R_{\Cl}:= \C R_{\Cl}(L,L)= \bigcup_{s,t\in \C S_{\Cl}}\C R_{\Cl}(s,t)$.

\end{defi}

By the results of the previous subsections $(L\times L, \C R_{\Cl}\cup \{\emptyset\})$ is the topological space given by the product topology over $(L, \C S_{\Cl}\cup\{\emptyset\})$. 

The following is easily verified from Definition \ref{clodue}.
\begin{lem}\label{supra}
Let $s,t\in \C S_{\Cl}$. If $r\subseteq s\times t$ is non-empty, then $r\in \C R_{\Cl}(s,t)$ iff for all $(P,Q)\in r$ and $P'\in \Cl(\{P\})$, $Q'\in \Cl(\{Q\})$, $(P',Q')\in r$.
\end{lem}

\begin{prop}
If $s,s',t,t'\in \C S_{\Cl}$, $r\subseteq s\times t$ and $r'\subseteq s'\times t'$ are $\Cl$-closed relations, then the relation $r\to r' \subseteq (s\to_{L} s')\times (t\to_{L} t')$ defined by $P \ (r\to r') \ Q$ if for all $M\in s, N\in t$, if $M \ r \ N$ then $(PM) \ r' \ (QN)$, is a $\Cl$-closed relation.  
\end{prop}
\begin{proof}
We must show that $r''=r\to r'$ is $\Cl$-closed.
Suppose $(F,G)\in r''$ and let $F'\in \Cl(\{F\}), G'\in \Cl(\{G\})$. If $r''$ is non-empty, by Lemma \ref{supra} we must show that $(F',G')\in r''$. 

Suppose $r\neq \emptyset$, so that there exist $P\in s, Q\in t$ such that $(P,Q)\in r$; since $\Cl$ satisfies $\mathsf{SA}$, if $F'\in \Cl(\{F\})$ and $FP\in L$, then $F'P\in \Cl(\{FP\})$. Similarly, if $G'\in \Cl(\{G\})$ and $GQ\in L$, then $G'Q\in \Cl(\{GQ\})$. 
Hence, since $(FP, GQ)\in r'$, it must be $FP,GQ\in L$. Moreover, as
 $r'$ is $\Cl$-closed, we deduce that $(F'P,G'Q)\in r'$, so we conclude that $(F',G')\in r''$.
Suppose $r=\emptyset$: then $r''= (s\to_{L}s')\times( t\to_{L}t')$ so $(F',G')\in r''$.

Let now $r''=\emptyset$. We claim that $r'=\emptyset$: if $(P,Q)\in r'$, then if $x\notin\TT{FV}(P)\cup \TT{FV}(Q)$, $\lambda x.P\in s\to_{L}s'$ and $\lambda x.Q\in t\to_{L}t'$ by condition $\mathsf K$ and $(\lambda x.P, \lambda x.Q)\in r''$, against the assumption. Since $r'$ is $\Cl$-closed, we deduce $\Cl(\emptyset)=\emptyset$, whence $r''$ is $\Cl$-closed.
\end{proof}

\begin{prop}
Let $I\neq \emptyset$ and let, for all $i\in I$, $s_{i},t_{i}\in \C S_{\Cl}$ and $r_{i}\in \C R_{\Cl}(s_{i},t_{i})$. Then $\bigcap_{i\in I}r_{i}\in \C R_{\Cl} (\bigcap_{i\in I}s_{i}, \bigcap_{i\in I}t_{i})$.
\end{prop}
\begin{proof}
Let $r=\bigcap_{i\in I}r_{i}$, $s=\bigcap_{i\in I}s_{i}, t=\bigcap_{i\in I}t_{i}$. 
We claim that $(\Cl\times \Cl)(r)= r$: suppose $P\in s,Q\in t$, $(P,Q)\in r$ and let $P'\in \Cl(\{P\}), Q'\in \Cl(\{Q\})$; we must show that $(P',Q')\in r$. For any $i\in I$, we have then $(P,Q)\in r_{i}$ and, since $r_{i}$ is a $\Cl$-closed relation, $(P',Q')\in r_{i}$. We conclude thus $(P',Q')\in r$. 
\end{proof}

The characterizations of the closure operators considered in Section \ref{sec3} (Propositions \ref{semas2} and \ref{crcara}) induce the following characterization of the associated closed relations:

\begin{prop}
\begin{itemize}
\item $r\subseteq \Lambda\times \Lambda$ is a $(\_)^{\beta}$-closed logical relation iff for all $P,Q\in \Lambda$, if $(P,Q)\in r$, $P\simeq_{\beta} P'$ and $Q\simeq_{\beta} Q'$, then $(P,Q)\in r$;

\item $r\subseteq \Lambda\times \Lambda$ is a $(\_)^{\beta\eta}$-closed logical relation iff for all $P,Q\in \Lambda$, if $(P,Q)\in r$, $P\simeq_{\beta\eta} P'$ and $Q\simeq_{\beta\eta} Q'$, then $(P,Q)\in r$;

\item $r\subseteq \Lambda\times \Lambda$ is a $(\_)^{\beta sat}$-closed logical relation iff for all $P,Q\in \Lambda$, if $(P,Q)\in r$ and if $P=M[P'/x]P_{1}\dots P_{n}$ (resp.\ $Q= N[Q'/y]Q_{1}\dots Q_{m} $), then $((\lambda x.M)PP_{1}\dots P_{n}, Q)\in r$ (resp.\ $(P,(\lambda y.N)QQ_{1}\dots Q_{m})\in r$);

\item  $r\subseteq \C{SN}\times \C{SN}$ is a ${\CR}$-closed logical relation iff the following hold:
\begin{itemize}
\item if $(P,Q)\in r$, $P\to^{*}_{\beta}P', Q\to^{*}_{\beta}Q'$, then $(P',Q), (P,Q')\in r$;
\item if $P\in \C{N}$ (resp.\ $Q\in \C N$) and for all $P'$ (resp.\ $Q'$) such that $P\to_{\beta}P'$ (resp.\ $Q\to_{\beta}Q'$), $(P',Q)\in r$ (resp.\ $(P,Q')\in r$), then $(P,Q)\in r$.

\end{itemize}

\item  $r\subseteq \C{SN}\times \C{SN}$ is a ${\CRuno}$-closed logical relation iff the following hold:
\begin{itemize}
\item if $(P,Q)\in r$ and $P\to^{*}_{\beta}P', Q\to^{*}_{\beta}Q'$, then $(P',Q), (P,Q')\in r$;
\item if $P\in \C{N}^*$ (resp.\ $Q\in \C{N}^*$) and for all $P'$ (resp.\ $Q'$) such that $P\to_{\beta}P'$(resp.\ $Q\to_{\beta}Q'$), $(P',Q)\in r$ (resp.\ $(P,Q')\in r$), then $(P,Q)\in r$.

\end{itemize}

\item  $r\subseteq \C{SN}\times \C{SN}$ is a ${\CRdue}$-closed logical relation iff the following hold:
\begin{itemize}
\item if $(P,Q)\in r  $ and $P\to^{*}_{\beta}P', Q\to^{*}_{\beta}Q'$, then $(P',Q), (P,Q')\in r$;
\item given $P, Q, M\in \C{SN}$ and $F_{1}\dots F_{n}\in \C{N}^*$, if
  $P=M[F_{1}/x_{1},\dots, F_{n}/x_{n}]$ (resp.\ $Q=M[F_{1}/x_{1},\dots,
  F_{n}/x_{n}]$) and for all $G_{1},\dots, G_{n}$ such that
  $F_{i}\to_{\beta}G_{i}$, \\
 $(M[G_{1}/x_{1},\dots, G_{n}/x_{n}] , Q)\in r$ (resp.\ $(P, M[G_{1}/x_{1},\dots, G_{n}/x_{n})\in r$), then $(P,Q)\in r$.\end{itemize}

\end{itemize}
\end{prop}

\subsubsection{Soundness for $\Cl$-closed logical relations} 

We now show that for all System $\FF$ type judgement $\Gamma\vdash M:\sigma$, the interpretation of $M$ in the semantics generated by $\Cl$ preserves all $\Cl$-closed logical relations.

\begin{defi}[$\Cl$-closed relation assignment]\label{reldef}\index{Logical relations!$\Cl$-closed relation assignment and relations $\Rel{}{\sigma}$}
For any type $\sigma\in \TT T$ and $\C X=\{X_{1},\dots, X_{n}\}\subseteq \TT{TypeVar}$ such that $\TT{FV}(\sigma)\subseteq \C X$ and $\C X\cap \TT{BV}(\sigma)=\emptyset$, we define a map 
$\langle\sigma\rangle^{\Cl,\C X}:\C R_{\Cl}^{n}\to \C R_{\Cl}$ as follows:
\begin{equation}
\begin{split}
 \langle{X_{i}}\rangle^{\Cl,\C X}(r_{1},\dots, r_{n})& = r_{i} \\
\langle{\sigma\to \tau}\rangle^{\Cl, \C X}(r_{1},\dots, r_{n}) & = \langle{\sigma}\rangle^{\Cl,\C X}(r_{1},\dots, r_{n})\to \langle{\tau}\rangle^{\Cl,\C X}(r_{1},\dots, r_{n})\\
\langle{\forall X\sigma}\rangle^{\Cl,\C X}(r_{1},\dots, r_{n})& =  \bigcap_{s,t\in S} \left (\bigcap_{r\in \C R_{S}(s,t)}\langle{\sigma}\rangle^{\Cl,\{X\}\cup \C X}(r,r_{1},\dots, r_{n}) \right )
\end{split}
\end{equation}
We let $\Rel{}{\sigma}:=\langle\sigma\rangle^{\Cl,\TT{FV}(\sigma)}:\C R_{\Cl}^{card(\TT{FV}(\sigma))}\to \C R_{\Cl}$.

Let $\C M_{1}, \C M_{2}: \TT{TypeVar}\to S$ be two $\Cl$-interpretations. A \emph{$\Cl$-closed relation assignment $R$ over $\C M_{1}$ and $\C M_{2}$} is a map associating, with any variable $X$, a $\Cl$-closed relation $R(X)\in \C R_{\Cl}( \C M_{1}(X), \C M_{2}(X))$. 

Given a {$\Cl$-closed relation assignment $R$ over $\Cl$-interpretations $\C M_{1}$ and $\C M_{2}$}, for any type $\sigma\in \TT T$, with $\TT{FV}(\sigma)\subseteq\{X_{1},\dots, X_{n}\}$, we
let $\Rel{R}{\sigma}:= \Rel{}{\sigma}(R(X_{1}),\dots, R(X_{n}))$.

\end{defi}

The following is easily verified by induction on types:
\begin{lem}
Given a {$\Cl$-closed relation assignment $R$ over $\Cl$-interpretations $\C M_{1}$ and $\C M_{2}$}, for any type $\sigma\in \TT T$,  $\Rel{R}{\sigma}\in \C R_{\Cl}(|\sigma|^{\Cl}_{\C M_{1}}, |\sigma|^{\Cl}_{\C M_{2}})$.

\end{lem}
Given a $\Cl$-closed relation assignment $R$ over $\C M_{1},\C M_{2}:\TT{TypeVar}\to S$, a type $\sigma$ and terms $P\in |\sigma|^{\Cl}_{\C M_{1}}, Q\in |\sigma|^{\Cl}_{\C M_{2}}$, we will often indicate $(P,Q)\in \Rel{R}{\sigma}$ by $P \ \Rel{R}{\sigma} \ Q$.

The following lemma and proposition extend Lemma \ref{fora} and Proposition \ref{pifor} to the case of $\Cl$-closed relations. 

\begin{lem}\label{forarel}
Let $I\neq \emptyset$, $r,t\subseteq L\times L$ and, for all $i\in I$, $s_{i}\subseteq \Lambda\times \Lambda$. Then
\begin{enumerate}[label=(\roman*)]
\item $r\to \bigcap_{i\in I}s_{i}= \bigcap_{i\in I}(r\to s_{i})$;
\item $(\bigcap_{i\in I}s_{i}\to r)\to t \subseteq \bigcap_{i\in I}((s_{i}\to r)\to t)$.
\end{enumerate}

\end{lem}
\begin{proof}
The argument is similar to that of Lemma \ref{fora}.
\end{proof}

\begin{prop}\label{piforrel}
Let $\C M_{1},\C M_{2}$ be $\Cl$-interpretations and $R$ a $\Cl$-closed relation assignment over them. 
\begin{enumerate}[label=(\roman*)]
\item If $\sigma\in \forpiudue$, then $\Rel{R}{\sigma}\subseteq \Rel{R}{\sigma^{+}}$;

\item if $\sigma\in \formenodue$, then $\Rel{R}{\sigma^{-}}\subseteq \Rel{R}{\sigma}$.
\end{enumerate}
\end{prop}
\begin{proof}
The argument is similar to that of Proposition \ref{pifor}.
\end{proof}

\begin{defi}[$\Cl$-invariance]\label{paramet}\index{Logical relations!$\Cl$-invariance}
Let $\sigma\in \TT T$ and $M\in \|\sigma\|^{\Cl}$. $M$ is \emph{$\Cl$-invariant at $\sigma$} if, for all $\Cl$-interpretations $\C M_{1},\C M_{2}$ and $\Cl$-closed logical relation assignment $R$ over $\C M_{1},\C M_{2}$, $M \  \Rel{R}{\sigma} \ M$.
\end{defi}

%
%
%
%

The following lemma establishes a relational version of condition $\mathsf K$ (Subsection \ref{sec313}).
\begin{lem}\label{adequaterel}
Let $s,s',t,t'\in \C S_{\Cl}$, $r\in \C R_{\Cl} (s, s')$ and $r'\in \C R_{\Cl}( t, t')$. Let $P,Q\in L$ be such that, for all $F\in s, G\in s'$, $P[F/x]\in t$ and $Q[G/y]\in t'$. Then, if for all $F\in s,G\in s'$ such that $(F,G)\in r$, $(P[F/x],Q[G/y])\in r'$, then $(\lambda x.P, \lambda y.Q)\in r\to r'$.

\end{lem}
\begin{proof}
Since $\Cl$ is adequate, from the hypotheses it follows that $\lambda x.P\in s\to_{L} t$ and $\lambda y.Q\in s'\to_{L} t'$. 
Suppose that for all $F\in s,G\in s'$ such that $(F,G)\in r$, $(P[F/x],Q[G/y])\in r$.
Let $F\in s$, $s_{0}=\{F\}$ and $t_{0}=\{P[F'/x]\mid F'\in \Cl(s_{0})\}$; observe that, since $\Cl(s_{0})\subseteq s$ and for all $F\in s$, $P[F/x]\in t$, it follows that $\Cl(t_{0})\subseteq t$. Now, by condition $\mathsf{K}$, $\lambda x.P\in \Cl(s_{0})\to \Cl(t_{0})$, whence $(\lambda x.P)F\in \Cl(t_{0})\subseteq t$. Let $G\in s$, $s'_{0}=\{G\}$ and $t'_{0}=\{Q[G'/y]\mid G'\in \Cl(s'_{0})\}$; by a similar argument we deduce that $(\lambda y.Q)G\in \Cl(t'_{0})\subseteq t'$. 
By $\mathsf{SU}$ there exist then $F'\in \Cl(s_{0}), G'\in \Cl(s'_{0})$ such that $(\lambda x.P)F\in \Cl(\{ P[F'/x]\})$ and $(\lambda y.Q)G\in \Cl(\{Q[G'/y]\})$. 

Suppose now $r\neq\emptyset$ and let $(F,G)\in r$. Since $r$ is a $\Cl$-closed relation, $(F',G')\in r$, whence, by the hypothesis, $(P[F'/x], Q[G'/y])\in r'$ and, since $r'$ is a $\Cl$-closed relation, we deduce $((\lambda x.P)F,(\lambda y.Q)G)\in r'$. We can conclude then $(\lambda x.P, \lambda y.Q)\in r\to r'$.
If $r=\emptyset$, then $r\to r'=(s\to_{L}t)\times (s'\to_{L}t')$, hence $(\lambda x.P, \lambda y.Q)\in r\to r'$.
\end{proof}

\begin{lem}\label{relsubstitutionlemma}
For any $\sigma,\tau\in \TT T$, with $\TT{FV}(\sigma)=\{X,X_{1},\dots, X_{n}\}$ and 
$\TT{FV}(\tau)=\{X_{1},\dots, X_{n}\} $, and for all $r_{1},\dots, r_{n}\in \C R_{\Cl}$, 
 $\Rel{}{\sigma}(\Rel{}{\tau}(\vec r), \vec r)= \Rel{}{\sigma[\tau/X]}(\vec r)$.
 \end{lem}
\begin{proof}
The proof is similar to that of Lemma \ref{substitutionlemma}.
\end{proof}

%


\begin{thm}\label{abstra}
Let $\Gamma\vdash M:\sigma$, where $\Gamma=\{x_{1}:\sigma_{1},\dots, x_{n}:\sigma_{n}\}$ and 
$\TT{FV}(\Gamma)\cup \TT{FV}(\sigma)\subseteq\{X_{1},\dots, X_{p}\}$. Then for any two $\Cl$-interpretations $\C M_{1},\C M_{2}$ and $\Cl$-closed relation assignment $R$ over $\C M_{1},\C M_{2}$, for any $F_{1}\in |\sigma_{1}|^{\Cl}_{\C M_{1}}, G_{1}\in |\sigma_{1}|^{\Cl}_{\C M_{2}},\dots, F_{n}\in |\sigma_{n}|^{\Cl}_{\C M_{1}}, G_{n}\in |\sigma_{n}|^{\Cl}_{\C M_{2}}$, 
if $F_{i} \ \Rel{R}{\sigma_{i}}\ G_{i}$ for $i=1,\dots,n$, then 
$M[F_{1}/x_{1},\dots, F_{n}/x_{n}] \ \Rel{R}{\sigma} \ M[G_{1}/x_{1},\dots, G_{n}/x_{n}]$. 
\end{thm}
\begin{proof}
Induction on a typing derivation of $\Gamma\vdash M:\sigma$. 
\begin{itemize}
\item If the derivation is $\Gamma,x:\sigma\vdash x:\sigma$, then $x[F/x]=F \ \Rel{R}{\sigma} \ G=x[G/x]$.

\item If the derivation ends by $\AXC{$\Gamma, x:\sigma\vdash M:\tau$}\UIC{$\Gamma\vdash \lambda x.M:\sigma\to \tau$}\DP$ then we can assume that $x$ does not occur free in $F_{1},G_{1},\dots, F_{n},G_{n}$ and is different from $x_{1},\dots, x_{n}$. By the induction hypothesis, for all $F\in |\sigma|_{\C M_{1}}^{\Cl}, G\in |\sigma|_{\C M_{2}}^{\Cl}$, $M[F_{1}/x_{1},\dots, F_{n}/x_{n},F/x] \ \Rel{R}{\tau} \ M[G_{1}/x_{1},\dots, G_{n}/x_{n},G/x]$ and by the assumptions made, $M[F_{1}/x_{1},\dots, F_{n}/x_{n},F/x]=M[F_{1}/x_{1},\dots, F_{n}/x_{n}][F/x]$ and $M[G_{1}/x_{1},\dots, G_{n}/x_{n},G/x]=M[G_{1}/x_{1},\dots, G_{n}/x_{n}][G/x]$. Then we can conclude $(\lambda x.M)[F_{1}/x_{1},\dots, F_{n}/x_{n}] \ \Rel{R}{\sigma\to \tau} \ (\lambda x.M)[G_{1}/x_{1},\dots, G_{n}/x_{n}]$ by Lemma \ref{adequaterel}.

\item If the derivation ends by $\AXC{$\Gamma\vdash M_{1}:\tau\to \sigma$}\AXC{$\Gamma\vdash M_{2}:\tau$}\BIC{$\Gamma\vdash M_{1}M_{2}:\sigma$}\DP$, then by the induction hypothesis $M_{1}[F_{1}/x_{1},\dots, F_{n}/x_{n}] \ \Rel{R}{\tau\to \sigma} \ M_{1}[G_{1}/x_{1},\dots, G_{n}/x_{n}]$ and \\ $M_{2}[F_{1}/x_{1},\dots, F_{n}/x_{n}] \ \Rel{R}{\tau} \ M_{2}[G_{1}/x_{1},\dots, G_{n}/x_{n}]$. We conclude that \\
$M_{1}M_{2}[F_{1}/x_{1},\dots, F_{n}/x_{n}] \ \Rel{R}{ \sigma} \ M_{1}M_{2}[G_{1}/x_{1},\dots, G_{n}/x_{n}]$.

\item If the derivation ends by $\AXC{$\Gamma\vdash M:\sigma$}\UIC{$\Gamma\vdash M:\forall X\sigma$}\DP$, where $X\notin \TT{FV}(\Gamma)$, then for all $s,t\in S$, $r\in \C R^{\Cl}(s,t)$ and $i\leq n$, $\Rel{}{\sigma_{i}}(r, R(X_{1}),\dots, R(X_{p}))
=\Rel{R}{\sigma_{i}}$, whence by the induction hypothesis we deduce
$M[F_{1}/x_{1},\dots, F_{n}/x_{n}] \ \Rel{}{\sigma}(r, R(X_{1}),\dots, R(X_{p}))
\ M[G_{1}/x_{1},\dots, G_{n}/x_{n}]$, for all $s,t\in \C S_{\Cl}$ and $r\in \C
R_{\Cl}(s,t)$.

We finally conclude then $M[F_{1}/x_{1},\dots, F_{n}/x_{n}] \ \Rel{R}{\forall X\sigma} \ M[G_{1}/x_{1},\dots, G_{n}/x_{n}]$.

\item If the derivation ends by $\AXC{$\Gamma\vdash M:\forall X\sigma$}\UIC{$\Gamma\vdash M:\sigma[\tau/x]$}\DP$, then by the induction hypothesis, for all $s,t\in S$ and $r\in \C R^{\Cl}(s,t)$, $M[F_{1}/x_{1},\dots, F_{n}/x_{n}] \ \Rel{}{\sigma}(r, R(X_{1}),\dots, R(X_{p})) \ M[G_{1}/x_{1},\dots, G_{n}/x_{n}]$, hence in particular \[M[F_{1}/x_{1},\dots, F_{n}/x_{n}] \ \Rel{}{\sigma}(\Rel{R}{\tau} , R(X_{1}),\dots, R(X_{p})) \ M[G_{1}/x_{1},\dots, G_{n}/x_{n}].\] The claim then follows from Lemma \ref{relsubstitutionlemma}. \qedhere
\end{itemize}
\end{proof}

\subsection{The $\Cl$-invariance theorem}\label{sec52}

We prove that $\Cl$-interpretable terms are $\Cl$-invariant, when $\Cl$ satisfies $\mathsf{SU}$, $\mathsf{SA}$ and $\mathsf{SS}$. Such conditions hold for the semantics $(\_)^{\beta},(\_)^{\beta\eta}, (\_)^{\beta sat}$. As a corollary, we obtain that terms intepretable in the semantics generated by either $\CRuno$ or $\CRdue$ are invariant with respect to $\beta$ and $\beta\eta$-stable logical relations.

The proof of Theorem \ref{param1} is obtained by adapting the ``term model'' of Section \ref{sec4} to the case of $\Cl$-closed logical relations.
Theorem \ref{param1} generalizes a result in \cite{FSCD2017}, showing that the implication between realizability and Reynolds' parametricity in the $\beta\eta$-stable semantics holds for simple types (see Subsection \ref{secRey}).


The infinite context $\Gamma^{\infty}$ (Definition \ref{gammainfinity}) will be here used to define contextual notions of $\Cl$-interpretation and $\Cl$-closed relation assignment. A special interpretation $\C M_{\TT p}$ will be defined such that, for any type $\sigma$ and $\lambda$-term $P$, if $P\in |\sigma|_{\C M_{\TT p}}^{\Cl}$, then for any two $\Cl$-interpretations and $\Cl$-closed relation assignment over them, $P$ is contextually related to $P$ relative to $\sigma$. 

In addition to $\mathsf{SU}$ and $\mathsf{SA}$, which are presupposed by our treatment of $\Cl$-closed logical relations, we use stability by substitution $(\mathsf{SS})$ to show that the contextual typability relations and relation assignments form, respectively, $\Cl$-closed sets and $\Cl$-closed relations.

As in the previous subsection, we will suppose $\Cl$ to be an adequate $\FF$-closure operator over $L\subseteq \Lambda$ satisfying $\mathsf{SU}$ and $\mathsf{SA}$.

\begin{defi}\label{Srela}\index{Logical relations!Contextual interpretation $\vert\Gamma\vdash^{\infty} \tau\vert^{\Cl}_{\C M}$}
\index{Logical relations!Contextual relations $\Rel{R}{\Gamma\vdash^{\infty} \tau}$}
\begin{itemize}
\item Given $\C M:\TT{TypeVar}\to \C S_{\Cl}$, $P\in \Lambda$ and $\tau\in \TT T$, we let the statement $P\in |\Gamma^{\infty}\vdash \tau|_{\C M}^{\Cl}$ hold when, by letting $FV(P)=\{x_{i_{1}},\dots, x_{i_{n}}\}$, for every $Q_{1}\in |\tau_{i_{1}}|^{\Cl}_{\C M}, \dots, Q_{n}\in |\tau_{i_{n}}|^{\Cl}_{\C M}$, 
$P[Q_{1}/x_{i_{1}},\dots, Q_{n}/x_{i_{n}}]\in |\tau|_{\C M}^{\Cl}$;

\item given $\C M_{1},\C M_{2}:\TT{TypeVar}\to \C S_{\Cl}$ and $R$ a $\Cl$-closed relation assignment over $\C M_{1},\C M_{2}$, $P,Q\in \Lambda$ and $\tau\in \TT T$, we let the statement $P \ \Rel{R}{\Gamma^{\infty}\vdash \tau} \ Q$ hold when $P\in |\Gamma^{\infty}\vdash \tau|_{\C M_{1}}^{\Cl}$, $Q\in |\Gamma^{\infty}\vdash \tau|_{\C M_{2}}^{\Cl}$ and, by letting $FV(P)\cup FV(Q)=\{x_{i_{1}},\dots, x_{i_{n}}\}$,
for every $F_{j}\in |\tau_{i_{j}}|_{\C M_{1}}^{\Cl}, G_{j}\in |\tau_{i_{j}}|_{\C M_{2}}^{\Cl}$ such that 
$F_{j} \ \Rel{R}{\tau_{i_{j}}} \ G_{j}$ for $1\leq j\leq n$, we have
\begin{equation}
P[F_{1}/x_{i_{1}},\dots, F_{n}/x_{i_{n}}] \ \Rel{R}{\tau} \ Q[G_{1}/x_{i_{1}},\dots, G_{n}/x_{i_{n}}]
\end{equation}
\end{itemize}

\end{defi}

The following lemmas assure that, if $\Cl$ satisfies $\mathsf{SS}$, Definition \ref{Srela} introduces $\Cl$-closed sets and $\Cl$-closed relations, respectively.

\begin{lem}\label{localrel}
If $\Cl$ satisfies $\mathsf{SS}$, then for all $s,t\in \C S_{\Cl}$, $r\in \C R_{\Cl}(s,t)$,
for all $P,P',Q,Q',F,G\in L$, $P'\in \Cl(\{P\}), Q'\in \Cl(\{Q\})$, if $(P[F_{1}/x_{1},\dots, F_{n}/x_{n}], Q[G_{1}/y_{1},\dots, G_{n}/y_{n}])\in r$, then \\ $(P'[F_{1}/x_{1},\dots, F_{n}/x_{n}], Q'[G_{1}/y_{1},\dots, G_{n}/y_{n}])\in r$;

%
\end{lem}
\begin{proof}
The claim follows from $P'[F_{1}/x_{1},\dots, F_{n}/x_{n}]\in \Cl(\{P[F_{1}/x_{1},\dots, F_{n}/x_{n}]\})$ and \\ $Q'[G_{1}/y_{1},\dots, G_{n}/y_{n}]\in \Cl(\{Q[G_{1}/y_{1},\dots, G_{n}/y_{n}]\})$ by Lemma \ref{supra}. 
\end{proof}

\begin{lem}\label{sclosed}
If $\Cl$ satisfies $\mathsf{SS}$, then for any type $\sigma$, for all $\Cl$-interpretations $\C M_{1}, \C M_{2}$ and $\Cl$-closed relation assignment $R$ over $\C M_{1},\C M_{2}$, $|\Gamma^{\infty}\vdash \sigma|_{\C M_{i}}^{\Cl}$, for $i=1,2$, is $\Cl$-closed and $\Rel{R}{\Gamma^{\infty}\vdash \sigma}$ is a $\Cl$-closed relation.

\end{lem}
\begin{proof}
Given a $\Cl$-interpretation $\C M$, to prove that $|\Gamma^{\infty}\vdash \sigma|_{\C M}^{\Cl}$ is $\Cl$-closed it suffices to show that for all $P\in |\Gamma^{\infty}\vdash \sigma|_{\C M}^{\Cl}$, $\Cl(\{P\})\subseteq |\Gamma^{\infty}\vdash \sigma|_{\C M}^{\Cl}$ (by $\mathsf{SU}$ and Proposition \ref{union2}). $|\Gamma^{\infty}\vdash \sigma|_{\C M}^{\Cl}$ contains all $P$ such that $P[Q_{1}/x_{i_{1}},\dots, Q_{n}/x_{i_{n}}]\in |\sigma|_{\C M}^{\Cl}$, for all $Q_{j}\in|\tau_{i_{j}}|_{\C M}^{\Cl}$, given $\TT{FV}(P)=\{x_{i_{1}},\dots, x_{i_{n}}\}$.
If $\Cl$ is $\mathsf{SS}$, then for all $P'\in \Cl\{ P\}$, $P'[Q_{1}/x_{i_{1}},\dots, Q_{n}/x_{i_{n}}]\in
\Cl\{ P[Q_{1}/x_{i_{1}},\dots, Q_{n}/x_{i_{n}}]\}\subseteq \Cl( |\sigma|^{\Cl}_{\C M})=|\sigma|^{\Cl}_{\C M}$. 
Hence
$P'[Q_{1}/x_{i_{1}},\dots, Q_{n}/x_{i_{n}}]\in |\sigma|_{\C M}^{\Cl}$, from which we deduce $P'\in |\Gamma^{\infty}\vdash\sigma|_{\C M}$. We conclude that $|\Gamma^{\infty}\vdash\sigma|_{\C M}$ is $\Cl$-closed.

Suppose now $P\in |\Gamma^{\infty}\vdash \sigma|_{\C M_{1}}^{\Cl}, Q\in |\Gamma^{\infty}\vdash \sigma|_{\C M_{2}}^{\Cl}$, $P \ \Rel{R}{\Gamma^{\infty}\vdash \sigma} \ Q$ and $P'\in \Cl(\{P\}), Q'\in \Cl(\{Q\})$. Since the sets $|\Gamma^{\infty}\vdash \sigma|_{\C M_{i}}^{\Cl}$ are $\Cl$-closed, $P'\in |\Gamma^{\infty}\vdash\sigma|_{\C M_{1}}^{\Cl}, Q'\in |\Gamma^{\infty}\vdash \sigma|_{\C M_{2}}^{\Cl}$. Let now $\TT{FV}(P)\cup \TT{FV}(Q)=\{x_{i_{1}},\dots, x_{i_{n}}\}$, $F_{j}\in |\tau_{i_{j}}|_{\C M_{1}}^{\Cl}$, $G_{j}\in |\tau_{i_{j}}|_{\C M_{2}}^{\Cl}$ and $F_{j} \ R[\tau_{i_{j}}] \ G_{j}$. Then $P[F_{1}/x_{i_{1}},\dots, F_{n}/x_{i_{n}}] \ \Rel{R}{\sigma} \ Q[G_{1}/x_{i_{1}},\dots, G_{n}/x_{i_{n}}]$ and since $\Rel{R}{\sigma}$ is a $\Cl$-closed relation, by Lemma \ref{localrel}, $P'[F_{1}/x_{i_{1}}, \dots, F_{n}/x_{i_{n}}] \ \Rel{R}{\sigma} \ Q'[G_{1}/x_{i_{1}},\dots, G_{n}/x_{i_{n}}]$, hence $P' \ \Rel{R}{\Gamma^{\infty}\vdash \sigma} \ Q'$, which proves that $\Rel{R}{\Gamma^{\infty}\vdash \sigma}$ is a $\Cl$-closed relation.
\end{proof}



We can now prove the $\Cl$-invariance theorem:
\begin{thm}\label{param1}
Let  $\Cl$ be an adequate $\FF$-closure operator over $L\subseteq \Lambda$ satisfying $\mathsf{SU}$, $\mathsf{SA}$ and $\mathsf{SS}$. Then for any type $\sigma$ and closed term $M$, if $M\in \|\sigma\|^{\Cl}$, then $M$ is $\Cl$-invariant at $\sigma$.
\end{thm}
\begin{proof}
Let $\C M_{\TT p}:\TT{TypeVar}\to S_{\Cl}$ be the assignment such that, for all $X\in \TT{TypeVar}$, $\C M_{\TT p}(X)$ is the set of all terms $M\in  \bigcap_{\C M:\TT{TypeVar}\to S_{\Cl}}|\Gamma^{\infty}\vdash X|_{\C M}^{\Cl}$ such that, for all $\C M_{1}, \C M_{2}:\TT{TypeVar}\to S_{\Cl}$ and $\Cl$-closed relation assignment $R$ over $\C M_{1},\C M_{2}$, $P \ \Rel{R}{\Gamma^{\infty}\vdash X} \ P$. $\C M_{\TT p}(X)$ is a $\Cl$-closed set: $\bigcap_{\C M:\TT{TypeVar}\to S_{\Cl}}|\Gamma^{\infty}\vdash X|_{\C M}^{\Cl}$ is a $\Cl$-closed set, as it is the intersection of a non-empty family of $\Cl$-closed sets (by Lemma \ref{sclosed}); since for all $\Cl$-interpretations $\C M_{1},\C M_{2}$ and $\Cl$-closed relation assignment $R$ over them, $\Rel{R}{ \Gamma^{\infty}\vdash X}$ is a $\Cl$-closed relation (by Lemma \ref{sclosed}), if $P \ \Rel{R}{ \Gamma^{\infty}\vdash X} \ P$ and $P'\in \Cl(\{P\})$, then $P' \ \Rel{R}{ \Gamma^{\infty}\vdash X}\ P'$, hence the set of terms such that $P \ \Rel{R}{ \Gamma^{\infty}\vdash X} \ P$ contains all $\Cl$-closures of its singletons and, as $\Cl$ satisfies $\mathsf{SU}$, is $\Cl$-closed. It follows that $\C M_{\TT p}(X)$ is the intersection of a non-empty family of $\Cl$-closed sets, and is thus $\Cl$-closed.

We claim that, for any type $\sigma$, the following hold:

\begin{enumerate}
\item $P\in |\sigma|_{\C M_{\TT p}}^{\Cl}$ iff, for any $\C M_{1},\C M_{2}:\TT{TypeVar}\to S_{\Cl}$ and $\Cl$-closed relation assignment $R$ over them, $P \ \Rel{R}{\Gamma^{\infty}\vdash \sigma} \ P$;
\item for every variable  $x_{i}$ such that $\tau_{i}=\sigma$, $x_{i}\in | \sigma|_{\C M_{\TT p}}^{\Cl}$.
\end{enumerate}

If for all $\C M_{1},\C M_{2}:\TT{TypeVar}\to S_{\Cl}$ and $\Cl$-closed relation assignment $R$ over them, 
$P\ \Rel{R}{\Gamma^{\infty}\vdash \sigma} \ P$ holds, then, by definition, $P\in |\sigma|_{\C M_{1}}^{\Cl}\cap |\sigma|_{\C M_{2}}^{\Cl}$, hence in definitive $P\in \|\sigma\|^{\Cl}\subseteq |\sigma|_{\C M_{\TT p}}^{\Cl}$. Hence the ``if'' direction of Claim 1. holds. 
We argue for the ``only if'' direction of $1.$ and for $2.$ by a simultaneous induction on $\sigma$. 

If $\sigma=X$, then, by definition of $\C M_{\TT p}$, any $P\in |\sigma|_{\C M_{\TT p}}^{\Cl}$ is such that, for any $\C M_{1},\C M_{2}:\TT{TypeVar}\to S_{\Cl}$ and $\Cl$-closed relation assignment $R$ over them, $P \ \Rel{R}{\Gamma^{\infty}\vdash \sigma} \ P$ holds, so claim $1.$ holds; moreover, if $\tau_{i}=X$, then, for any $\C M_{1}, \C M_{2}:\TT{TypeVar}\to S_{\Cl}$ and $\Cl$-closed relation assignment $R$ over $\C M_{1},\C M_{2}$, $x_{i} \ \Rel{R}{ \Gamma^{\infty}\vdash X} \ x_{i}$: if $F\in \C M_{1}(X), G\in \C M_{2}(X)$ and $F \ R[X] \ G$, then $x_{i}[F/x_{i}]= F \ R[X] \ G=x_{i}[G/x_{i}]$. Hence claim $2.$ holds too.

Let now $\sigma=\sigma_{1}\to \sigma_{2}$. By induction hypothesis, for all $i$, if $\tau_{i}=\sigma_{1}$ then $x_{i}\in |\sigma_{1}|_{\C M_{\TT p}}^{\Cl}$; let then $P\in |\sigma|_{\C M_{\TT p}}^{\Cl}$ and choose an index $i$ such that $\tau_{i}=\sigma_{1}$ and $x_{i}$ does not occur free in $P$; we have that $Px_{i}\in |\sigma_{2}|_{\C M_{\TT p}}^{\Cl}$ hence, by induction hypothesis, for any $\C M_{1},\C M_{2}:\TT{TypeVar}\to S_{\Cl}$ and $\Cl$-closed relation assignment $R$ over them, $(Px_{i}) \ \Rel{R}{\Gamma^{\infty}\vdash \sigma_{2}} \ (Px_{i})$; let then $\C M_{1},\C M_{2}:\TT{TypeVar}\to S_{\Cl}$ and $R$ be a $\Cl$-closed relation assignment over them; by letting $\TT{FV}(P)=\{x_{i_{1}},\dots, x_{i_{n}}\}$, suppose $F_{j}\in |\tau_{i_{j}}|_{\C M_{1}}^{\Cl}, G_{j}\in |\tau_{i_{j}}|_{\C M_{2}}^{\Cl}$ are such that $F_{j} \ \Rel{R}{\tau_{i_{j}}}\ G_{j}$, for $1\leq j\leq n$, and moreover suppose $F\in |\sigma_{1}|_{\C M_{1}}^{\Cl}, G\in |\sigma_{1}|_{\C M_{2}}^{\Cl}$ are such that $F \ \Rel{R}{\sigma_{1}} \ G$; then, since $\TT{FV}(Px_{i})=\{x_{i_{1}},\dots, x_{i_{n}}, x_{i}\}$ (observe that $x_{i}\neq x_{i_{1}},\dots, x_{i_{n}}$ as $x_{i}$ has been chosen not to occur free in $P$), we have
\begin{equation}
\begin{split}
 P[F_{1}/x_{i_{1}},\dots, F_{n}/x_{i_{n}}]F  \qquad \qquad \qquad \quad&  \\
  \ = \ 
(Px_{i})[F_{1}/x_{i_{1}},\dots, F_{n}/x_{i_{n}}, F/x_{i}]
\  & \Rel{R}{\sigma_{2}} \ 
(Px_{i})[ G_{1}/x_{i_{1}},\dots, G_{n}/x_{i_{n}}, G/x_{i}]
\\
& \qquad \qquad \qquad \quad  \ = \ 
P[ G_{1}/x_{i_{1}},\dots, G_{n}/x_{i_{n}}]G
\end{split}
\end{equation}
so we conclude that $P \ \Rel{R}{\Gamma^{\infty}\vdash\sigma} \ P$, and we proved claim $1$.

To prove claim $2.$, suppose $x_{i}$ is a variable such that $\tau_{i}=\sigma$. Let $\sigma=\sigma_{1}\to \sigma_{2}$ and $Q\in |\sigma_{1}|_{\C M_{\TT p}}^{\Cl}$; by induction hypothesis, for all $\C M_{1},\C M_{2}:\TT{TypeVar}\to S_{\Cl}$ and $\Cl$-closed relation assignment $R$ over $\C M_{1},\C M_{2}$, $Q\ \Rel{R}{\Gamma^{\infty}\vdash \sigma_{1}}\ Q$. Let $\C M_{1}, \C M_{2}:\TT{TypeVar}\to S_{\Cl}$, and $R$ be a $\Cl$-closed relation assignment over $\C M_{1},\C M_{2}$; moreover let 
$\TT{FV}(Q)\cup\{x_{i}\}$ be the set $\{x_{i_{1}},\dots, x_{i_{r}}\}$, where $i=i_{p}$, for some $1\leq p\leq r$. Given terms $F_{1},G_{1},\dots, F_{r},G_{r}$ such that $F_{j}\in |\tau_{i_{j}}|^{\Cl}_{\C M_{1}}, G_{j}\in |\tau_{i_{j}}|^{\Cl}_{\C M_{2}}$ and $F_{j}\ \Rel{R}{\tau_{i_{j}}}\ G_{j}$ all hold for $1\leq j\leq r$, we have that
\begin{equation}
(x_{i}Q)[F_{1}/x_{i_{1}},\dots, F_{r}/x_{i_{r}}] \ = \ 
F_{i_{p}}(Q[F_{1}/x_{i_{1}},\dots, F_{r}/x_{i_{r}}])
%
\end{equation}
and
\begin{equation}
 (x_{i}Q) [G_{1}/x_{i_{1}},\dots, G_{r}/x_{i_{r}}]
 \ = \ 
 G_{i_{p}}(Q[G_{1}/x_{i_{1}},\dots, G_{r}/x_{i_{r}}])
%
\end{equation}

Since $Q[F_{1}/x_{i_{1}},\dots, F_{r}/x_{i_{r}}] \ \Rel{R}{\sigma_{1}}\ Q[G_{1}/x_{i_{1}},\dots, G_{r}/x_{i_{r}}]$ holds by induction hypothesis and $F_{i_{p}} \ \Rel{R}{\sigma} \ G_{i_{p}}$, it follows that $x_{i}Q \ \Rel{R}{\Gamma^{\infty}\vdash \sigma_{2}} \ x_{i}Q$. We deduce that $x_{i}Q\in |\sigma_{2}|_{\C M_{\TT p}}^{\Cl}$, that is, $x_{i}\in |\sigma|_{\C M_{\TT p}}^{\Cl}$ and claim $2.$ is proved.

Let now $\sigma=\forall X\tau$. If $P\in |\sigma|_{\C M_{\TT p}}^{\Cl}$, then $P\in |\tau|^{\Cl}(\C M_{\TT p}(X),\C M_{\TT p}(X_{1}),\dots, \C M_{\TT p}(X_{q}))=|\tau|^{\Cl}_{\C M_{\TT p}}$ where $\TT{FV}(\tau)=\{X,X_{1},\dots, X_{q}\}$, hence, by induction hypothesis, for any two assignments $\C M_{1},\C M_{2}:\TT{TypeVar}\to S_{\Cl}$ and $\Cl$-closed relation assignment $R$ over them, $P\ \Rel{R}{\Gamma^{\infty}\vdash \tau} \ P$. 
This implies in particular that, for any two assignments $\C M_{1},\C M_{2}:\TT{TypeVar}\to S_{\Cl}$ and $\Cl$-closed relation assignment $R$ over them, $P\ \Rel{R}{\Gamma^{\infty}\vdash \forall X\tau} \ P$ holds, so we proved claim 1. Now let $x_{i}$ be such that $\tau_{i}=\sigma$ and let $\C M_{1},\C M_{2}:\TT{TypeVar}\to S_{\Cl}$ and $R$ be a relation assignment over them. Let moreover $P\in |\sigma|_{\C M_{1}}^{\Cl}, Q\in |\sigma|_{\C M_{2}}^{\Cl}$ be such that $P \ \Rel{R}{\sigma} \ Q$. Then $x_{i}[P/x_{i}]=P \ \Rel{R}{ \sigma}\ Q=x_{i}[Q/x_{i}]$, so we conclude that for all $\C M_{1},\C M_{2}:\TT{TypeVar}\to S_{\Cl}$ and $\Cl$-closed relation assignment $R$ over then,  $x_{i} \ \Rel{R}{\Gamma^{\infty}\vdash \sigma}\ x_{i}$. From 1. we conclude tgen $x_{i}\in |\sigma|^{\Cl}_{\C M_{\TT p}}$.


Finally, let $\sigma\in \TT T$ and suppose $M$ is closed and $M\in \|\sigma\|^{\Cl}$. Then $M\in |\sigma|_{\C M_{\TT p}}^{\Cl}$, so, for every $\C M_{1}, \C M_{2}:\TT{TypeVar}\to S_{\Cl}$ and $\Cl$-closed relation assignment $R$ over $\C M_{1}, \C M_{2}$,  $M \ \Rel{R}{\Gamma^{\infty}\vdash \sigma} \ M$. As $M$ is closed, this implies $M\ \Rel{R}{\sigma} \ M$.
\end{proof}

Theorem \ref{param1} can now be applied to all non-regular closure operators discussed in Section~\ref{sec3}.

\begin{cor}\label{param2}
\begin{enumerate}[label=(\roman*)]
\item
Let $\Cl$ be any among $(\_)^{\beta}, (\_)^{\beta\eta}, (\_)^{\beta sat}$. Then, for any type $\sigma$ and closed term $M$, if $M\in \|\sigma\|^{\Cl}$, then $M$ is $\Cl$-invariant at $\sigma$.
\item
Let $\Cl$ be either $\CRuno$ or $\CRdue$. Then, for any type $\sigma$ and closed term $M$, if $M\in \|\sigma\|^{\Cl}$, then $M$ is $\beta$-invariant and $\beta\eta$-invariant at $\sigma$.
\end{enumerate}
\end{cor}
\begin{proof}
Claim (i) follows from Theorem \ref{param1} and the fact that $(\_)^{\beta}, (\_)^{\beta\eta}, (\_)^{\beta sat}$ are stable by substitution. Claim (ii) follows from claim (i) and $\CRuno,\CRdue\leq (\_)^{\beta}\leq (\_)^{\beta\eta}$, which implies $\|\sigma\|^{\CRuno},\|\sigma\|^{\CRdue}\subseteq \|\sigma\|^{\beta}$
and $\|\sigma\|^{\CRuno},\|\sigma\|^{\CRdue}\subseteq \|\sigma\|^{\beta\eta}$.
\end{proof}

\subsection{Reynolds' parametricity}\label{secRey}

Reynolds' parametricity \cite{Reynolds1983} is a well-known approach to parametric polymorphism based on the technique of logical relations. 
The general idea is that the interpretation of a universally quantified type $\forall X\sigma$ should contain those polymorphic functions which behave uniformly in all possible instantiations. This is expressed by an invariance condition with respect to logical relations.
We shortly discuss why, in order to extend our approach to logical relations to Reynolds' parametricity, one needs to enrich the structure of $\Cl$-closed sets by equivalence relations extending the $\beta\eta$-theory (e.g. by considering $\mathsf{PER}s$, a standard approach to parametricity, see \cite{Bainbridge1990}).

As for logical relations semantics, in Reynolds' semantics any type receives both an interpretation $\model{\sigma}_{0}$ as a set or more generally as an object of some category, and an interpretation $\model{\sigma}_{1}$ as a binary relation over $\model{\sigma}_{0}$. 
If one restricts to simple types, Reynolds' parametricity coincides with invariance with respect to logical relations. Hence  \ref{param1} shows that $\Cl$-interpretable terms are Reynolds' parametric in the case of simple types.

However, Reynolds' semantics differs from the one defined in the previous pages for the interpretation of the universal quantifier: since the interpretation $\model{\forall X\sigma}_{0}$ of a universally quantified type must contain only those polymorphic functions which are ``parametric in $X$'', the definition of $\model{\forall X\sigma}_{0}$ is given in terms of $\model{\sigma}_{1}$. In other words, while in the semantics of logical relations the two interpretations $\model{\sigma}_{0}$ and $\model{\sigma}_{1}$ are defined independently (Definitions \ref{intedef} and \ref{reldef}), this is not the case for Reynolds' semantics. The dependency  between the two notions of interpretation can be caracterised categorically (see \cite{Dunphy2004, Ghani2015b, Ghani2015}) in terms of fibered functors and fibered natural transformations.

A fundamental ingredient of Reynolds' parametricity is the validity of the so-called \emph{Identity Extension Lemma} ($\mathsf{IEL}$) \cite{Reynolds1983}, which says that for any type $\sigma$, the relation obtained by instantiating the relational interpretation $\model{\sigma}_{1}$ of $\sigma$ on equality relations coincides with the equality relation over $\model{\sigma}_{0}$. 

In our framework, a type $\sigma$ with $n$ free variables is interpreted as a map $|\sigma|^{\Cl}: \C S_{\Cl}^{n}\to \C S_{\Cl}$. For all $s\in \C S_{\Cl}$, let $\mathsf{Eq}_{s}$ be the $\Cl$-closure of the identity relation over $s$.
Given a $\Cl$-interpretation $\C M$, let $\mathsf{Eq}$ be the $\Cl$-closed relation assignment over $\C M,\C M$ such that, for all $X$, $\mathsf{Eq}(X)= \mathsf{Eq}_{\C M(X)}$. Then $\mathsf{IEL}$ can be formulated as the fact that, for any type $\sigma\in \TT T$,
\begin{equation}\label{iel}
\Rel{\mathsf{Eq}}{\sigma} \ = \ \mathsf{Eq}_{|\sigma|^{\Cl}_{\C M}}
\end{equation}
In other words, $\mathsf{IEL}$ expresses the fact that the maps over $\C S_{\Cl}$ corresponding to types must be ``equality-preserving''. 
In more generality, in \cite{Ghani2015b} and \cite{Ghani2015} it is explained in detail how the interpretation of the quantifier in Reynolds' parametric models can be defined in terms of equality-preserving functors.

The proposition below shows that $\mathsf{IEL}$ fails in the case of
$\beta\eta$-relations. 

\begin{prop}\label{noiel}
 Equation \ref{iel} fails for the $\beta\eta$-stable semantics.
 \end{prop}
\begin{proof}
Let $s$ be any non-empty $\beta\eta$-stable set, $\C M$ be the interpretation given by $\C M(X)=s$.
Let $M,N$ be the two $\beta\eta$-distinct terms below:
\begin{equation}\label{MN}
\begin{split}
M & := \lambda f.\lambda x_{1}.\lambda x_{2}. x_{1}(fx_{2}) \\
N &:= \lambda f.\lambda x_{1}. \lambda x_{2}. f(x_{1}x_{2})
\end{split}
\end{equation}
Let $\sigma= \forall Z(Z\to Z)\to (X\to Y)\to (X\to Y)$ and $\C M(X)=\Lambda$ for all $X$.
While $(M,N)\notin \mathsf{Eq}_{|\sigma|^{\beta\eta}_{\C M}}$, we claim that $M \ \Rel{\mathsf{Eq}}{\sigma} \ N$. 
Let $f$ be any term variable and $R^{f}$ be the $\beta\eta$-relation assignment over $\C N$ given by $R^{f}(X)= \{(P,Q)\mid fP\simeq_{\beta\eta}Q\}$, which is clearly $\beta\eta$-closed.
If $F \ \langle \forall Z(Z\to Z)\rangle^{\beta\eta} \ G$, then in particular
$F \ \Rel{R^{f}}{Z\to Z} \ G$. We claim that 
for all $Q\in \Lambda$, $f(FQ)\simeq_{\beta\eta} G(fQ)$: we have $(Q,fQ)\in R^{f}(Z)$, so from $(FQ, G(fQ))\in R^{f}(Z)$ we deduce the claim.
From $f(FQ)\simeq_{\beta\eta} G(fQ)$ it follows then that for all $P,Q, P',Q'$ such that $P\simeq_{\beta\eta }P'$, $Q\simeq_{\beta\eta}Q'$, $MFPQ \simeq_{\beta\eta} NGP'Q' $, which concludes our proof.  
\end{proof}

The proof of Proposition \ref{noiel} suggests that, in order to account Reyolds' parametricity
in terms of sets of $\lambda$-terms, one must enrich the $\beta\eta$-equivalence relation in any closed set (e.g. by endowing the latter with a $\Cl$-closed partial equivalence relation). 

\section{Dinaturality}\label{sec6}

Dinaturality is an approach to parametric polymorphism in which types are interpreted as multivariant functors and terms as transformations between such functors satisfying a particular uniformity condition, which generalizes the usual definition of natural transformations.

In order to compare $\Cl$-invariance and dinaturality, we introduce a syntactic formalisation of the dinaturality condition for pure $\lambda$-terms. We do this in two steps: first, we describe the dinaturality condition over the syntactic category generated by System $\FF$ as given by a family of $\beta\eta$-equations; this corresponds to the definition of syntactic dinaturality in \cite{Delatail2009}. Then, we provide a second syntactic dinaturality condition using indeterminates (i.e. free variables), which has the advantage of being expressed by a single equation, leading to a more manageable syntactic theory. While the first dinaturality condition easily follows from the second one, we do not know if the latter can be deduced from the former. 
%
%

The main result of this section is that the closed terms which are $\Cl$-invariant at a positive type are syntactically dinatural. We show two applications of this result: a second proof that closed simply typable terms are dinatural and a second completeness argument for positive types.

\subsection{Syntactic dinaturality}\label{sec61}

\subsubsection{Functorial polymorphism}

We recall the interpretation of simple types as multivariant functors and simply typable terms as dinatural transformations.

\begin{defi}\index{Dinaturality!Dinatural transformation}

Given a category $\C C$ and functors $F,G: (\C C^{op})^{n}\times \C C^{n}\to \C C$, a \emph{dinatural transformation} between $F$ and $G$ is a family of arrows $\theta_{\vec A}$ satisfying the diagram below:
\begin{equation}\label{esagono1}
\resizebox{0.4\textwidth}{!}{$$
\xymatrix{
 &  F\vec A\vec A \ar[r]^{\theta_{\vec A}} & G\vec A\vec A \ar[rd]^{G \vec A\vec f}&  \\
F\vec B\vec A \ar[rd]_{F\vec B\vec f} \ar[ru]^{F\vec f\vec A}  &   &  &   G\vec A\vec B\\
  & F\vec B\vec B \ar[r]_{\theta_{\vec B}} & G\vec B\vec B  \ar[ru]_{G\vec f\vec B}&
}$$}
\end{equation}
for any choice of objects $A_{i},B_{i}$ in $\C C$ and arrows $f_{i}:A_{i}\to B_{i}$.
\end{defi}
%
%
%

If $\C C$ is a cartesian closed category, $\sigma\in \TT T_{0}$, and $\TT{FV}(\sigma)\subseteq\{X_{1},\dots, X_{n}\}$, then $\sigma$ can be interpreted as a multivariant functor $\llbracket \sigma \rrbracket_{\C C}: (\C C^{op}\times \C C)^{n}\to \C C$ as follows:
\begin{equation}\label{funct0}
\begin{split}
\model{X_{i}}_{\C C}(\vec A, \vec B) \ & := \  B_{i}\\
\model{\sigma\to \tau}_{\C C} (\vec A, \vec B)\ & \ := \model{\tau}_{\C C}(\vec A, \vec B )^{\model{\sigma}_{\C C}(\vec B, \vec A)} 
\end{split}
\end{equation}

 For any $\sigma,\tau\in \TT T_{0}$, a term $M$ such that $x:\sigma\vdash M:\tau$ can be interpreted as a family of arrows $\llbracket M \rrbracket_{\vec A}:\llbracket \sigma\rrbracket_{\C C} \vec A \vec A \to \llbracket \tau \rrbracket_{\C C}\vec A\vec A$. In \cite{Girard1992} it is proved that such families are dinatural.

\begin{thm}[\cite{Girard1992}]\label{dinok}
Let $\C C$ be any cartesian closed category. If $x:\sigma\vdash M:\tau$ in $\lambda_{\to}$, then $\llbracket M \rrbracket_{\vec A}:\llbracket \sigma\rrbracket_{\C C} \vec A \vec A \to \llbracket \tau \rrbracket_{\C C}\vec A\vec A$ is a dinatural transformation.

\end{thm}

\begin{rem}\label{dinaproblem}
The extension of Theorem \ref{dinok} to System $\FF$ is more delicate, and we won't describe it here  (the reader can look at \cite{Bainbridge1990}). There are two obstacles to extend the dinatural interpretation to second order quantifiers. First, the well-known fact that dinatural transformations do not generally compose; second, the fact that the definition of quantifiers requires the existence in the category $\C C$ of certain limits (called \emph{ends}, see \cite{MacLane}).

A dinatural interpretation of System $\FF$ is usually obtained by defining a restricted class of composable dinatural transformations over a suitable cartesian closed category $\C C$ in which ``relativized'' ends (see \cite{Bainbridge1990}) exist.


\end{rem}

\subsubsection{Functorial polymorphism over the syntactic category}

A first syntactic dinaturality condition is obtained by considering the syntactic category generated by typable terms, as we now recall. 

\begin{defi}\index{Dinaturality!Syntactic category $\C F$}
The \emph{syntactic category} $\C F$ has objects the types $\TT T$ and arrows $\C F[\sigma,\tau]$ terms $M$ with exactly one free variable $x$ such that $x:\sigma\vdash M:\tau$, considered up to $\beta\eta$-equivalence.

\end{defi}

Any $\sigma\in \TT T$, with $\TT{FV}(\sigma)\subseteq \{X_{1},\dots, X_{n}\}$, generates a functor
$\widehat{\sigma}: (\C F^{op}\times \C F)^{n}\to \C F$ defined as follows:

\begin{enumerate}

\item given types $\vec{\tau}$ and $\vec{\rho}$, $\widehat \sigma(\vec \tau, \vec \rho)$ is the type obtained by replacing all negative occurrences of $X_{i}$ by $\tau_{i}$ and all positive occurrences of $X_{i}$ by $\rho_{i}$, for $i=1,\dots, n$;

\item given types $\vec{\tau}, \vec{\tau}', \vec{\rho}, \vec \rho'$ and terms
terms $\vec P$, $\vec Q$, where $P_{i}\in \C F(\tau_{i}, \tau'_{i})$ and $Q_{i}\in \C F( \rho_{i}, \rho'_{i})$, we define the terms $\widehat\sigma(\vec P, \vec Q)\in \C F(\widehat \sigma(\vec{\tau'}, \vec{\rho}), \widehat\sigma(\vec \tau, \vec{\rho'}))$ by induction on $\sigma$ as follows:
\begin{equation}\label{funct}
\begin{split}
\widehat{X_{i}}(\vec P, \vec Q) \ & := \  Q_{i}\\
\widehat{\tau\to \rho}(\vec P, \vec Q) \ & := \
 \arrow{ \lambda x.\widehat\tau(\vec Q, \vec P)}{\lambda x.\widehat\rho(\vec P, \vec Q)}x \\
\widehat{\forall X\tau}(\vec P, \vec Q) \ & := \ \widehat\tau((\vec P, x ), (\vec Q, x))
\end{split}
\end{equation}
%

\end{enumerate}

Given types $\vec\tau, \vec\rho,\vec\rho'$, and terms $\vec P, \vec Q$, if $P_{i}=x$ and $Q_{i}:\rho_{i}\to \rho'_{i}$ for $i=1,\dots,n$, we let
$\widehat \sigma(\vec\tau, \vec Q)$ and $\widehat\sigma(\vec Q, \vec\tau)$ be shorthand for 
$\widehat \sigma(\vec P, \vec Q)$ and $\widehat\sigma(\vec Q, \vec P)$, respectively.

If $x:\sigma\vdash M:\tau$ holds in System $\FF$, then $M$ induces a (constant) family of arrows from $\widehat\sigma$ to $\widehat \tau$:  for all types $\vec\rho$, $ x:\widehat\sigma \vec\rho\vec\rho\vdash M: \widehat \tau \vec\rho\vec\rho$ (technically, this follows from Lemma \ref{lemma:extra}).
Since $M$ corresponds to a constant family of arrows, the dinaturality condition for $M$ in $\C F$ is expressed by a family of equations corresponding to diagram \ref{esagono1}:
\begin{equation}\label{dina0}
\widehat\tau(\vec\rho, \vec P) \circ M \circ \widehat\sigma(\vec P, \vec\rho) \ \simeq_{\beta\eta} \ 
\widehat\tau(\vec P, \vec\tau)\circ M \circ \widehat\sigma(\vec\tau, \vec P)
\end{equation}
for all types $\vec\tau, \vec\rho$ and arrows $\vec P$, where $P_{i}\in \C F(\tau_{i},\rho_{i})$.
%

Let System $\mathsf F_{\times}$ be the type system obtained from $F$ by adding a unit type $\B 1$ and binary products $\sigma\times\tau$. Its syntactic category $\C F_{\times}$ is cartesian closed \cite{LambekScott}. Moreover, it is easily verified, that if $\C C$ is the syntactic category $\C F$, 
for all simple type $\sigma\in \TT T_{0}$, $\widehat \sigma= \model{\sigma}_{\C F_{\times}}$.
 From Theorem \ref{dinok}, as remarked in \cite{Girard1992}, the following can then be deduced:
 
 \begin{prop}\label{dina00}
All instances of equations \ref{dina0}, for $\sigma, \tau\in \TT{T}_{0}$, hold in the theory generated by $\beta\eta$-equivalence. 
\end{prop}

\begin{rem}\label{DinaEqua2}
An advantage of describing dinaturality in $\C F$ is that we can easily define the interpretation of all System $\FF$ types (see Remark \ref{dinaproblem}). It is natural to ask then whether Proposition \ref{dina00} can be extended to all System $\FF $ types. In this section we will show that this can be done for positive types. However, syntactic dinaturality fails, as shown in detail in \cite{Delatail2009}, for types containing negative occurrences of quantifiers. We will provide a simple example at Remark \ref{DinaEqua3}.

\end{rem}

\subsubsection{Dinaturality with indeterminates}

We now introduce a more uniform notion of syntactic dinaturality. The basic idea is to allow arrows in the syntactic category to have more than one variable. This corresponds, in the terminology of \cite{LambekScott}, to considering the \emph{polynomial} cartesian closed category $\C F_{\times}[x_{1}^{\sigma_{1}},\dots, x_{n}^{\sigma_{n}}]$ generated by a finite number of indeterminates $x_{i}:\B 1\to \sigma_{i}$:

\begin{defi}\index{Dinaturality!Polynomial category}
Let $\sigma_{1},\dots, \sigma_{n}\in \TT T$ and $x_{1},\dots, x_{n}$ be distinct variables. 
The \emph{polynomial} category $\C F_{\times}[x_{1}^{\sigma_{1}},\dots, x_{n}^{\sigma_{n}}]$ is obtained from
$\C F_{\times}$ by adding the arrows $x_{i}:\B 1\to \sigma_{i}$, for $i=1,\dots, n$. 
\end{defi}

In \cite{LambekScott} it is proved that if a category $\C C$ is cartesian closed, then the polynomial category $\C C[x_{1}^{\sigma_{1}},\dots, x_{n}^{\sigma_{n}}]$ constructed over it is cartesian closed too. We can thus conclude that $\C F_{\times}[x_{1}^{\sigma_{1}},\dots, x_{n}^{\sigma_{n}}]$ is cartesian closed.

As $\C F_{\times}[x_{1}^{\sigma_{1}},\dots, x_{n}^{\sigma_{n}}]$ is cartesian closed, Theorem \ref{dinok} applies to it. This implies that, if $x:\sigma\vdash M: \tau$, where $\sigma,\tau$ are simple types, then all instances of equation \ref{dina0} hold, where now the arrows $P_{i}$ can contain occurrences of the indeterminates $x_{i}$. 

If one considers distinct type variables $X_{i},Y_{i}$ and indeterminates $f_{i}:\B 1\to (X_{i}\to Y_{i})$, all instances of equations \ref{dina0}, which depend on the choice of types $\rho_{i},\rho'_{i}$ and arrows $P_{i}\in \C F[\rho_{i},\rho'_{i}]$, can be deduced, by substitution, from a single equation, by letting $\rho_{i}=X_{i}, \rho'_{i}=Y_{i}$ and $P_{i}=f_{i}$. We now describe this idea in detail.

Let $(\TT{V}_{0},\TT{V}_{1})$ be a partition of $\TT{TypeVar}$ into two disjoint countable sets of type variables $\TT{V}_{0}=\{X_{0},X_{1},X_{2},\dots \}$ and $\TT{V}_{1}=\{Y_{0},Y_{1},Y_{2},\dots\}$. To avoid confusion, we will indicate generic variables in $\TT{TypeVar}$ as $Z_{1},Z_{2},Z_{3},\dots$.


We consider a countable set of indeterminates $f_{1},f_{2},\dots$, where $f_{i}:\B 1\to (X_{i}\to Y_{i})$ for all $i\geq 1$. For each type $\sigma$, we define terms $\TT H_{\sigma},\TT K_{\sigma}$ coding the action of the functors $\widehat\sigma$.
%

We define $\TT H_{\sigma},\TT K_{\sigma}$ by a simultaneous induction over $\sigma$:
\begin{equation}
\begin{split}
\TT H_{Z_{i}} \ & := \  \lambda x.f_{i}x  \\
\TT H_{\tau\to \rho} \ & := \ \arrow{\TT K_{\tau}}{\TT H_{\rho}}   \\
\TT H_{\forall Z_{i}\tau} \ & := \ \TT H_{\tau}[\lambda x.x/f_{i}]  \end{split}
\qquad\qquad
\begin{split}
\TT K_{Z_{i}} \ & := \ \lambda x.x \\
 \TT K_{\tau\to \rho} \ & := \ \arrow{\TT H_{\tau}}{\TT K_{\rho}} \\
  \TT K_{\forall Z_{i}\tau} \ & := \ \TT K_{\tau}[\lambda x.x/f_{i}]  
\end{split}
\end{equation}

For a simple type $\sigma=\sigma_{1}\to \dots \to \sigma_{n}\to Z_{i}$ the definition of the terms $\TT H_{\sigma}$ and $\TT K_{\sigma}$ can be unrolled as follows:
\begin{equation}
\begin{split}
\TT H_{\sigma} \ = & \ \lambda x. \lambda x_{1}.\dots.\lambda x_{n}. f_{i} \big ( x(\TT K_{\sigma_{1}}x_{1})\dots (\TT K_{\sigma_{n}}x_{n})\big )\\
\TT K_{\sigma} \ = & \  \lambda x.\lambda x_{1}.\dots.\lambda x_{n}.  x(\TT H_{\sigma_{1}}x_{1})\dots (\TT H_{\sigma_{n}}x_{n})
\end{split}
\end{equation}

The following proposition shows that, for any type $\sigma$ such that $\TT{FV}(\sigma)=\{Z_{i_{1}},\dots,Z_{i_{m}}\}$ the terms $\TT H_{\sigma},\TT K_{\sigma}$ code the functorial action of $\widehat\sigma$ over the polynomial category $\C F_{\times}[f_{i_{1}},\dots, f_{i_{n}}]$.

\begin{prop}\label{code}
For any type $\sigma$, with $\TT{FV}(\sigma)\subseteq\{Z_{i_1},\dots, Z_{i_n}\}$, types $\vec
 \tau, \vec \rho,\vec\rho'$ and terms $\vec P$, where $P_{i}\in \C F_{\times}(\rho_{i},\rho'_{i})$,
\begin{equation}
\begin{matrix}
\widehat \sigma (\vec\tau ,\vec P)  \ \simeq_{\beta} \ (\TT H_{\sigma}x)[\lambda x.P_{1}/f_{i_{1}},\dots, \lambda x.P_{n}/f_{i_{n}}]  \\
\widehat \sigma (\vec P ,\vec \tau)  \ \simeq_{\beta} \ (\TT K_{\sigma}x)[\lambda x.P_{1}/f_{i_{1}},\dots, \lambda x.P_{n}/f_{i_{n}}]  
\end{matrix}
\end{equation}

\end{prop}
\begin{proof}
Let $\vec f= (f_{1}x,\dots, f_{n}x)$ and $\vec x=(x_{1},\dots, x_{n})$.
By induction on $\sigma$, we prove that $\TT H_{\sigma}x\simeq_{\beta}\widehat \sigma (\vec x,\vec f)$ and 
$\TT K_{\sigma}x\simeq_{\beta} \widehat \sigma( \vec f, \vec x)$, from which the claim follows since 
$\widehat \sigma( \vec \tau,\vec P)\simeq_{\beta} \widehat\sigma(\vec x,\vec f)\theta$ and 
$\widehat \sigma (\vec P,\vec \tau)\simeq_{\beta} \widehat\sigma(\vec f,\vec x)\theta$, where $\theta=[\lambda x.P_{1}/f_{i_{1}},\dots, \lambda x.P_{n}/f_{i_{n}}]$.

If $\sigma=X_{i}$, then $\widehat \sigma (\vec x, \vec f)=f_{i}x\simeq_{\beta}\TT H_{\sigma}x$ and 
$\widehat \sigma (\vec f ,\vec x)= x=\TT K_{\sigma}x$.
If $\sigma= \tau\to \rho$, then observe that 
\begin{equation}
\widehat{\tau\to \rho}(\vec x, \vec f) =
\arrow{ \lambda x.\widehat\tau(\vec f, \vec x)}{\lambda x.\widehat\rho(\vec x, \vec f)}x\simeq_{\beta}\lambda y.\left ( \widehat \rho(\vec x, \vec f) \left [ x\left (\widehat\tau(\vec f, \vec x )[y/x]\right)  / x\right]\right)
\end{equation} 
Now
$\TT H_{\sigma}x\simeq_{\beta}(\TT K_{\tau}\To \TT H_{\rho})x\simeq_{\beta}
\lambda y. \TT H_{\rho} (x(\TT K_{\tau}y)) \simeq_{\beta}
\lambda y. (\widehat\rho( \vec x,\vec f))\left[ x(\widehat\tau(\vec f,\vec x) [y/x])/x\right] \simeq_{\beta}
\widehat\sigma(\vec x,\vec f)$, where in the last step we used the induction hypothesis, which implies that 
$\widehat\tau(\vec f,\vec x) \simeq_{\beta} \TT K_{\tau}x$ and $\widehat\sigma(\vec x,\vec f)\simeq_{\beta }\TT H_{\rho}$.
One can argue similarly for $\widehat \sigma(\vec f,\vec x)$.
Finally, if $\sigma=\forall X_{i}\tau$, then 
$\widehat\sigma(\vec x,\vec f)= \widehat \tau((\vec x,x_{i}), (\vec f,x_{i})) \simeq_{\beta}
\TT H_{\tau}[ \lambda x.x/f_{i}] \simeq_{\beta} \TT H_{\sigma}$, by using the induction hypothesis and the fact that $\widehat \tau(( \vec x,x_{i})(\vec f,x_{i}))\simeq_{\beta} \widehat \tau (\vec x ,\vec f)[ \lambda x.x/f_{i}]$.
One can argue similarly for $\widehat \sigma(\vec f,\vec x)$.
\end{proof}
 
The dinaturality condition for a term $M$ such that $x:\sigma\vdash M:\tau$ can now be expressed by the commutation of a single diagram:
\begin{equation}\label{esagono2}
\resizebox{0.4\textwidth}{!}{$$
\xymatrix{
  &\widehat\sigma \vec X \vec X  \ar[r]^{M}  &   \widehat\tau \vec X \vec X  \ar[dr]^{\TT H_{\tau}} & \\
\widehat\sigma\vec Y\vec X \ar[ur]^{\TT K_{\sigma}}  \ar[dr]_{\TT H_{\sigma}}  & & & \widehat\tau \vec X \vec Y\\
 & \widehat\sigma \vec Y \vec Y \ar[r]_{M}  &  \widehat\tau \vec Y \vec Y \ar[ur]_{\TT K_{\tau}} & \\
}$$}
\end{equation}
i.e. by a single equation 
\begin{equation}
\TT H_{\tau} \circ M \circ \TT K_{\sigma} \ \simeq_{\gamma} \ \TT K_{\tau} \circ M \circ \TT H_{\sigma}
\end{equation}
where $\gamma$ is either $\beta$ or $\beta\eta$,
which can be reformulated as
\begin{equation}\label{dina1}
\TT H_{\sigma\to \tau} M \ \simeq_{\gamma} \ \TT K_{\sigma\to \tau}M
\end{equation}

This leads to the following definition:

\begin{defi}[$\beta$ and $\beta\eta$-dinaturality]\label{dinatu}\index{Dinaturality!$\beta$ and $\beta\eta$-dinaturality}
Let $\sigma$ be a type, $\gamma$ be either $\beta$ or $\beta\eta$ and $M$ be a $\lambda$-term. Then $M$ is \emph{$\gamma$-dinatural at $\sigma$} if $\TT H_{\sigma}M \ \simeq_{\gamma} \ \TT K_{\sigma}M$.
\end{defi}

\begin{exa}\label{ex1}
Let $\sigma$ be the type $\forall Z_{i}((Z_{i}\to Z_{i})\to (Z_{i}\to Z_{i}))$. The $\beta$-dinaturality condition associated with a closed $\lambda$-term $M$ and $\sigma$ is the equation
\begin{equation}\label{nato}
f_{i}\big (M \ \lambda y.x_{1}(f_{i}y) \ x_{2} \big) \ \simeq_{\beta} \ M \ \lambda y.f_{i}(x_{1}y)\  f_{i}x_{2}
\end{equation}
which can be illustrated by the diagram below:
\begin{equation}\label{esagono3}
\resizebox{0.5\textwidth}{!}{
$$\xymatrix{
  &X_{i}\to X_{i}  \ar[r]^{M}  &   X_{i}\to X_{i}  \ar[dr]^{\TT H_{Z_{i}\to Z_{i}}} & \\
Y_{i}\to X_{i} \ar[ur]^{\TT K_{Z_{i}\to Z_{i}}}  \ar[dr]_{\TT H_{Z_{i}\to Z_{i}}}  & & & X_{i}\to Y_{i}\\
 & Y_{i}\to Y_{i} \ar[r]_{M}  &  Y_{i}\to Y_{i} \ar[ur]_{\TT K_{Z_{i}\to Z_{i}}} & \\
}
$$}
\end{equation}
If $M$ is a closed term of type $\sigma$, the $\beta\eta$-normal form of $M$ is of the form $\lambda y.\lambda x.(y)^{n}x$, for some $n\geq 0$. Then we can check that Equation \ref{nato} reduces to the valid equation
\begin{equation}
f_{i}\big ( \underbrace{x_{1}(f_{i} \dots x_{1}(f_{i}}_{n \text{ times}} x_{2})\dots  ) \big ) \ \simeq_{\beta} \ 
\underbrace{f_{i}(x_{1} \dots f_{i}(x_{1}}_{n \text{ times}} (f_{i}x_{2}))\dots ) 
\end{equation}
We can conclude then that $M$ is $\beta$-dinatural at $\sigma$.
\end{exa}

\begin{rem}\label{DinaEqua3}
Dinaturality fails for non positive types. Let $M$ be the term $\lambda x.\lambda y.xy$ and $\sigma=(\forall Y(Y\to Y))\to (X\to X)$. Although $\vdash M:\sigma$ holds, $M$ is not $\beta$-dinatural nor $\beta\eta$-dinatural at $\sigma$, since
\begin{equation}\label{MN2}
\TT H_{\sigma}M \ \simeq_{\beta} \ \lambda x_{1}.\lambda x_{2}.f(x_{1}x_{2})     \ \not\simeq_{\beta\eta} \  
\lambda x_{1}.\lambda x_{2}. x_{1}(fx_{2}) \ \simeq_{\beta} \ \TT K_{\sigma}M
\end{equation}
The investigation of syntactic dinaturality for non positive types requires then to consider equational theories which extend $\beta\eta$-equivalence.
Equation \ref{MN2} was already considered in Subsection \ref{secRey} as a consequence of the $\mathsf{IEL}$ for Reynolds' parametricity. In fact, all equations arising from dinaturality can be deduced from an axiomatization of Reynolds' parametricity, as shown in \cite{Plotkin1993}.

\end{rem}

\subsection{$\Cl$-invariance implies dinaturality}\label{sec62}

We prove that, when $\sigma\in \forpiudue$, the closed terms which are $\Cl$-invariant at $\sigma$ are also $\beta$ and $\beta\eta$-dinatural at $\sigma$. This result generalizes a result contained in \cite{FSCD2017}, which was limited to the $\beta\eta$-stable semantics and to simple types.

We first show that, similarly to the case of completeness, $\gamma$-dinaturality at a $\forpiudue$ type (for $\gamma=\beta,\beta\eta$) can be deduced from $\gamma$-dinaturality at $\B \Pi$ types, by means of the following:

\begin{lem}\label{dinatrivia}
For all $\sigma\in \forpiudue$ and term $M$, if $M$ is $\gamma$-dinatural at $\sk(\sigma)$, then $M$ is $\gamma$-dinatural at $\sigma$, where $\gamma$ is either $\beta$ or $\beta\eta$.
\end{lem}
\begin{proof}
Since $\TT H_{\sigma}=\TT H_{ \sk(\sigma)}[\lambda x.x/f_{i_{1}},\dots, \lambda x.x/f_{i_{p}}]$ and
$\TT K_{\sigma}=\TT K_{ \sk(\sigma)}[\lambda x.x/f_{i_{1}},\dots, \lambda x.x/f_{i_{p}}]$, where
$\TT{BV}(\sigma)=\{Z_{i_{1}},\dots, Z_{i_{p}}\}$, if $M$ satisfies the dinaturality condition at $\sigma$, then, by substitution, $M$ satisfies the dinaturality condition at $\sk(\sigma)$.
\end{proof}

Let now $\gamma$ be either $\beta$ or $\beta\eta$ and $\Cl$ be an adequate $\FF$-closure operator over $L\subseteq \Lambda$ such that $\Cl \leq (\_)^{\gamma}$. 
We introduce two models and a relation assignment which corresponds to $\gamma$-dinaturality.

\begin{defi}\label{newdef}\index{Dinaturality!Relation assignment $R^{f}$}
Let $\C M_{1},\C M_{2}:\TT{TypeVar}\to S$ be given, for any $Z_{i}$, by $\C M_{1}(Z_{i})= \Lambda$ and $\C M_{2}(Z_{i})=f_{i}\Lambda$, where $f_{i}\Lambda$ is the $\gamma$-closure of the set of all $\lambda$-terms of the form $f_{i}Q$, where $Q\in \Lambda$. Let moreover $R^{f}$  be the $\gamma$-closed (hence $\Cl$-closed) relation assignment over $\C M_{1},\C M_{2}$ such that, for any $Z_{i}$, $P\in \Lambda$ and $Q\in f_{i}\Lambda$, $(P,Q)\in R^{f}(Z_{i})$ iff $f_{i}P\simeq_{\gamma} Q$.

Finally, let $\C N:\TT{V}_{1}\cup \TT{V}_{2}\to S$ be the interpretation such that, for all $u\geq 0$, $\C N(X_{i})=\C M_{1}(Z_{i})$ and $\C N(Y_{i})=\C M_{2}(Z_{i})$.
\end{defi}

\begin{prop}\label{HK}
For any type $\tau$, 
\begin{equation}
\begin{matrix}
\TT H_{\tau}\in | \widehat\tau\vec Y\vec X\to\widehat \tau \vec Y \vec Y|_{\C N}^{\Cl}   & \TT K_{\tau}\in | \widehat\tau\vec Y \vec Y\to \widehat\tau\vec X \vec Y |_{\C N}^{\Cl}  \\
\TT K_{\tau}\in |\widehat\tau \vec Y \vec X\to\widehat\tau\vec X \vec X|_{\C N}^{\Cl}       & \TT H_{\tau}\in |\widehat \tau\vec X \vec Y\to \widehat\tau\vec X \vec Y|_{\C N}^{\Cl}
\end{matrix}
\end{equation}

\end{prop}
\begin{proof}
We argue by induction on $\tau$: if $\tau=Z_{i}$, then $\TT H_{\tau}= \lambda x.f_{i}x\in | X_{i}\to Y_{i}|_{\C N}^{\Cl}$ and $\TT K_{\tau}=\lambda x.x\in |X_{i}\to X_{i}|_{\C N}^{\Cl}, |Y_{i}\to Y_{i}|_{\C N}^{\Cl}$. 
If $\tau=\tau_{1}\to \dots \to \tau_{m}\to Z_{i}$ for some $m\geq 1$, then let $E\in | \widehat\tau\vec Y \vec X|_{\C N}^{\Cl}, F\in | \widehat\tau\vec X \vec X|_{\C N}^{\Cl}, G\in |\widehat\tau\vec Y \vec Y|_{\C N}^{\Cl}$ and $E_{i}\in |\widehat\tau_{i}\vec Y \vec X|_{\C N}^{\Cl}, F_{i}\in |\widehat\tau_{i}\vec X \vec X|_{\C N}^{\Cl}, G_{i}\in |\widehat\tau_{i}\vec Y \vec Y|_{\C N}^{\Cl}$, for $1\leq i\leq m$. Then, by induction hypothesis we have
\begin{equation}
\begin{matrix}
\TT H_{\tau_{i}}E_{i}\in | \widehat\tau_{i}\vec Y \vec Y|_{\C N}^{\Cl}  & \TT K_{\tau_{i}}G_{i}\in | \widehat\tau_{i}\vec X \vec Y |_{\C N}^{\Cl}  \\
\TT K_{\tau_{i}}E_{i}\in |\widehat\tau_{i}\vec X \vec X|_{\C N}^{\Cl}       & \TT H_{\tau_{i}}F_{i}\in | \widehat\tau_{i}\vec X \vec Y|_{\C N}^{\Cl}
\end{matrix}
\end{equation}
from which we obtain
\begin{equation}\label{24}
\begin{split}
\TT H_{\tau}E G_{1}\dots G_{n} \ \simeq_{\gamma} \ &f_{i} \big ( E (\TT K_{\tau_{1}}G_{1})\dots (\TT K_{\tau_{n}}G_{n})\big )\in | Y|^{\Cl}_{\C N}  \\
\TT H_{\tau}F E_{1}\dots E_{n}  \ \simeq_\gamma\  &f_{i} \big ( F (\TT K_{\tau_{1}}E_{1})\dots (\TT K_{\tau_{n}}E_{n})\big )\in | Y|_{\C N}^{\Cl}   \\
\TT K_{\tau}E F_{1}\dots F_{n} \  \simeq_\gamma\   &E (\TT H_{\tau_{1}}F_{1})\dots (\TT H_{\tau_{n}}F_{n})\in | X|_{\C N}^{\Cl}  \\
\TT K_{\tau}G E_{1}\dots E_{n}\  \simeq_\gamma\   &G (\TT H_{\tau_{1}}E_{1})\dots (\TT H_{\tau_{n}}E_{n})\in | Y|_{\C N}^{\Cl} 
\end{split}
\end{equation}
where $\gamma$ is either $\beta$ or $\beta\eta$, by exploiting the fact that $\widehat\tau\vec Y \vec X=\widehat\tau_{1}\vec X \vec Y\to \dots \to \widehat\tau_{n}\vec X \vec Y\to X_{i}$.

\end{proof}

For any term variable $x$, $\TT H_{\tau}x\in |\widehat\tau\vec Y \vec Y|_{\C N}^{\Cl}= |\tau|_{\C M_{2}}^{\Cl}$ and $\TT K_{\tau}x\in |\widehat\tau\vec X \vec X|_{\C N}^{\Cl}=|\tau|_{\C M_{1}}^{\Cl}$. Hence we can ask whether $\TT H_{\tau}x$ and $\TT K_{\tau}x$ are related under the $\Cl$-closed relation assignment $R^{f}$.  This is shown by the proposition below in the case of simple types:

\begin{prop}\label{HK2}
Let $\gamma$ be either $\beta$ or $\beta\eta$.
For all $\sigma\in \TT T_{0}$ and variable $x$,
\begin{enumerate}[label=(\roman*)]
\item $(\TT K_{\sigma}x) \ \Rel{R^{f}}{\sigma} \ (\TT H_{\sigma}x)$;
\item if $P\in |\sigma|_{\C M_{1}}, Q\in |\sigma|_{\C M_{2}}$ and $P\ \Rel{R^{f}}{\sigma} \ Q$, then $\TT H_{\sigma}P \simeq_{\gamma} \TT K_{\sigma}Q$.
\end{enumerate}

\end{prop}
\begin{proof}

We prove both claims simultaneously by induction on $\sigma$. If $\sigma=Z_{i}$ then $\TT K_{\sigma}x\simeq_\gamma x, \TT H_{\sigma}x\simeq_\gamma f_{i}x$ and $x\ \Rel{R^f}{\sigma} \ f_{i}x$, from which we deduce $\TT K_{\sigma}x \ \Rel{R^{f}}{\sigma} \ \TT H_{\sigma}x$; moreover, if $P \ \Rel{R^f}{Z_{i}} \ Q$, then  $\TT H_{\sigma}P\simeq_\gamma f_{i}P\simeq_{\gamma} Q\simeq_{\gamma}\TT K_{\sigma}Q$.

If $\sigma=\sigma_{1}\to \dots \to \sigma_{n}\to Z_{i}$ for some $n\geq 1$, then suppose $P_{i}\in|\sigma_{i}|_{\C M_{1}}, Q_{i}\in |\sigma_{i}|_{\C M_{2}}$ and $P_{i} \ \Rel{R^{f}}{\sigma_{i}} \ Q_{i}$, for all $1\leq i\leq n$; then
\begin{equation}
P:= \TT K_{\sigma}xP_{1}\dots P_{n} \ \simeq_\gamma \ x (\TT H_{\sigma_{1}}P_{1})\dots (\TT H_{\sigma_{n}}P_{n}) 
\end{equation}
is related to
\begin{equation}
Q:=\TT H_{\sigma}xQ_{1}\dots Q_{n}\ \simeq_\gamma \ f_{i}\Big (x (\TT K_{\sigma_{1}}Q_{1})\dots (\TT K_{\sigma_{n}}Q_{n}) \Big ) 
\end{equation}
Indeed, by induction hypothesis, $\TT H_{\sigma_{i}}P_{i} \simeq_{\gamma} \TT K_{\sigma_{i}}Q_{i}$, hence $x (\TT K_{\sigma_{1}}P_{1})\dots (\TT K_{\sigma_{n}}P_{n}) \simeq_{\gamma}$ \\ $ x (\TT H_{\sigma_{1}}Q_{1})\dots (\TT H_{\sigma_{n}}Q_{n})$ so $f_{i}P\simeq_{\gamma}Q$. 

Suppose now $P\in |\sigma|_{\C M_{1}}, Q\in |\sigma|_{\C M_{2}}$ and $P\ \Rel{R^f}{\sigma}\ Q$, then let
\begin{equation}
P':=\TT H_{\sigma}Px_{1}\dots x_{n} \simeq_{\beta} f_{i} \Big ( P(\TT K_{\sigma_{1}}x_{1})\dots (\TT K_{\sigma_{n}}x_{n}) \Big )
\end{equation}
and 
\begin{equation}
Q':=\TT K_{\sigma}Qx_{1}\dots x_{n} \simeq_{\beta} Q(\TT H_{\sigma_{1}}x_{1})\dots (\TT H_{\sigma_{n}}x_{n})  
\end{equation}
By induction hypothesis $(\TT K_{\sigma_{i}}x_{i}) \ \Rel{R^f}{\sigma_{i}} \ (\TT H_{\sigma_{i}}x_{i})$ hence, by hypothesis $P(\TT K_{\sigma_{1}}x_{1})\dots (\TT K_{\sigma_{n}}x_{n})$ is related to $ Q'$, that is $P'\simeq_{\gamma}Q'$. We conclude that $\TT H_{\sigma}P  \simeq_{\gamma} \TT K_{\sigma}Q$.
\end{proof}

We can now state and prove that parametric terms are $\beta\eta$-dinatural (resp.\ $\beta$-dinatural) for $\forpiudue$ (resp.\ $\forpiupro$) types:

\begin{thm}[$\Cl$-invariance implies dinaturality]\label{dinatural}
Let $\Cl$ be an adequate $\FF$-closure operator, $M\in \Lambda$ and $\sigma$ a type
\begin{enumerate}[label=(\roman*)]

\item if $\sigma\in \forpiudue$, $ \Cl\leq (\_)^{\beta\eta}$, and $M$ is $\Cl$-invariant at $\sigma$, then $M$ is $\beta\eta$-dinatural at $\sigma$;
\item if $\sigma\in \forpiupro$, $ \Cl\leq (\_)^{\beta}$, and $M$ is $\Cl$-invariant at $\sigma$, then $M$ is $\beta$-dinatural at $\sigma$.

\end{enumerate}
\end{thm}
\begin{proof}
We only prove claim (ii). Let $\sigma\in \forpiudue$ and $M$ be $\Cl$-invariant at $\sigma$. By Proposition \ref{piforrel}, $M$ is $\Cl$-invariant at $\sk(\sigma)$.
Let $\sk(\sigma)=\tau_{1}\to \dots \to \tau_{n}\to Z_{i}$, for some $n\geq0$ and
let $\C M_{1},\C M_{2}$ and $R^{f}$ be as in Definition \ref{newdef}. By Propositions \ref{HK} and \ref{HK2}, $\TT H_{\tau_{i}}g_{i}\in | \tau_{i}|_{\C M_{2}}$, $\TT K_{\tau_{i}}g_{i}\in |\tau_{i}|_{\C M_{1}}$ and $(\TT H_{\tau_{i}}g_{i}) \ \Rel{R^{f}}{\tau_{i}}\ (\TT K_{\tau_{i}}g_{i})$. As $M$ is $\Cl$-invariant, $M \ \Rel{R^{f}}{\tau} \ M$, hence for some $P\simeq_{\beta}M (\TT K_{\tau_{1}}g_{1})\dots (\TT K_{\tau_{n}}g_{n})$ and 
$Q\simeq_{\beta}M(\TT H_{\tau_{1}}g_{1})\dots (\TT H_{\tau_{n}}g_{n})$, $P \ \Rel{R^{f}}{Z_{i}}\ Q$, that is  
$f_{i}\big (M (\TT K_{\tau_{1}}g_{1})\dots (\TT K_{\tau_{n}}g_{n})\big ) \ \simeq_{\beta} \ M(\TT H_{\tau_{1}}g_{1})\dots (\TT H_{\tau_{n}}g_{n})$, whence $M$ is $\beta$-dinatural at $\sk(\sigma)$, hence $\beta$-dinatural at $\sigma$ by Lemma \ref{dinatrivia}. 
\end{proof}

\begin{rem}

Theorem \ref{dinatural} can be compared with the well-known fact that Reynolds' parametricity implies dinaturality, or more precisely, that all instances of the dinaturality equations \ref{dina0} follow from a ``parametricity axiom'', i.e. an axiom stating that any term of a given type $\sigma$ is Reynolds' parametric at $\sigma$ \cite{Plotkin1993}.

 As we discussed in Subsection \ref{secRey}, Reynold's parametricity coincides with invariance with respect to logical relations only for simple types. Moreover, it was seen that Reynolds' parametricity implies the validity of some equations which extend $\beta\eta$-equivalence, as a consequence of $\mathsf{IEL}$. For instance, all equations \ref{dina0} can be deduced from parametricity. 
 On the contrary, since $\Cl$-invariance does not induce new equations between realizers, it
 need not imply dinaturality for types containing negative occurrences of quantifiers (see Remark \ref{DinaEqua3}).%
%

\end{rem}

\subsection{Two applications of syntactic dinaturality}\label{sec63}

We present two applications of Theorem \ref{dinatural}. First, we extend Theorem \ref{dinok}, which states that closed simply typed $\lambda$-terms are dinatural to the cases of $\beta$ and $\beta\eta$-dinaturality and to $\forpiudue$ types.
Then we show a second completeness argument based on a result in \cite{FSCD2017} which shows that $\beta\eta$-dinaturality implies typability for closed normal $\lambda$-terms.
 
By considering $(\_)^{\beta}$ or $(\_)^{\beta\eta}$ as $\FF$-closure operator and composing Theorem \ref{dinatural} with the Theorem \ref{abstra}, we obtain the following:
\begin{thm}\label{dinok2}
Let $\sigma\in \forpiudue$ and $M$ be a closed $\lambda$-term. If $\vdash M:\sigma$, then $M$ is $\beta$-dinatural at $\sigma$.
\end{thm}

This theorem can be compared with Theorem \ref{dinok}. First, in Theorem \ref{dinok2} $\beta\eta$-dinaturality is replaced by $\beta$-dinaturality, so it can be seen as a slight refinement of Theorem \ref{dinok}. Moreover, while Theorem \ref{dinok} is proved by a categorial argument, 
Theorem \ref{dinok2} exploits syntactic dinaturality and the $\Cl$-invariance theorem.

We now deduce a second completeness argument from Theorem \ref{dinatural}. We exploit the following result from \cite{FSCD2017}:

\begin{thm}[dinaturality implies typability]\label{fscd}
 If $\sigma\in \TT T_{0}$ and $M$ is a closed $\beta$-normal (resp.\ $\beta\eta$-normal) term $\beta$-dinatural at $\sigma$ (resp.\ $\beta\eta$-dinatural at $\sigma$), then $\vdash M:\sigma$.
 \end{thm}
The argument in \cite{FSCD2017} is proved only for $\beta\eta$-dinaturality, but the adaptation to $\beta$-dinaturality is straightforward. 
By composing Theorem \ref{param1} and Theorem \ref{dinatural}, we get the following completeness results:

\begin{thm}[completeness by dinaturality]\label{forallcr2}
Let $\Cl$ be an adequate $\FF$-closure operator over $L\subseteq \Lambda$ satisfying $\mathsf{SU}$, $\mathsf{SA}$ and $\mathsf{SS}$. 
\begin{enumerate}[label=(\roman*)]
\item Suppose $\Cl\leq (\_)^{\beta\eta}$. Let $M$ be a closed $\lambda$-term and $\sigma\in \forpiudue$. If $M\in \|\sigma\|^{\Cl}$, then there exists $M'\simeq_{\beta\eta} M$ such that $\vdash M':\sigma$.

\item Suppose $\Cl\leq (\_)^{\beta}$. Let $M$ be a closed $\lambda$-term and $\sigma\in \forpiupro$. If $M\in \|\sigma\|^{\Cl}$, then there exists $M'\simeq_{\beta} M$ such that $\vdash M':\sigma$.

\end{enumerate}
\end{thm}
\begin{proof}
We only prove claim (i), the second being proved similarly.
From $\Cl\leq (\_)^{\beta\eta}$ and $M\in \|\sigma\|^{\Cl}$ we deduce $M\in \| \sigma\|^{\beta\eta}$. By Proposition \ref{pifor} we deduce then $M\in \|\sigma^{+}\|^{\beta\eta}\subseteq \|\sk(\sigma)\|^{\beta\eta}$. Hence, by Theorem \ref{param1}, $M$ is $\beta\eta$-invariant at $\sk(\sigma)$, and by Theorem \ref{dinatural}, $M$ is $\beta\eta$-dinatural at $\sk(\sigma)$. By Theorem \ref{fscd} we have then $\vdash M' :\sk(\sigma)$, for some $M'\simeq_{\beta\eta}M$, and we conclude $\vdash M'':\sigma$, for some $M''\simeq_{\beta\eta}M'\simeq_{\beta\eta}M$ by Theorem \ref{forpi}..
\end{proof}

Theorem \ref{forallcr2} is slighty weaker than Corollary \ref{forallcr3}. In particular, it depends on the hypotheses of Theorem \ref{param1} (i.e. that $\Cl$ satisfies $\mathsf{SU}, \mathsf{SA}$ and $\mathsf{SS}$). 

\section{Conclusions and open questions}\label{sec7}

\subsection{Equivalence of realizability and parametricity for positive types}

By putting together Theorem \ref{forpi}, Theorem \ref{param1}, Theorem \ref{dinatural} and Theorem \ref{fscd} 
we deduce that, in the case of positive types,  realizability, invariance with respect to logical relations, dinaturality and typability are equivalent properties for closed normal $\lambda$-terms:

\begin{thm}\label{general}
Let $\Cl$ be an adequate $\FF$-closure operator satisfying $\mathsf{SU}$, $\mathsf{SA}$ and $\mathsf{SS}$.
\begin{enumerate}[label=(\roman*)]
\item Suppose $\Cl\leq(\_)^{\beta\eta}$. Let $M$ be a closed $\beta\eta$-normal $\lambda$-term and $\sigma\in \forpiudue$. Then the following are equivalent:
\begin{itemize}
\item $M$ is $\Cl$-interpretable at $\sigma$;
\item $M$ is $\beta\eta$-invariant at $\sigma$;
\item $M$ is $\beta\eta$-dinatural at $\sigma$;
\item for some $M'$ such that $M'\to_{\eta}^{*} M$, $\vdash M':\sigma$.

\end{itemize}
\item Suppose $\Cl\leq (\_)^{\beta}$. Let $M$ be a closed $\beta$-normal $\lambda$-term and $\sigma\in \forpiupro$. Then the following are equivalent:
\begin{itemize}
\item $M$ is $\Cl$-interpretable at $\sigma$;
\item $M$ is $\beta$-invariant at $\sigma$;
\item $M$ is $\beta$-dinatural at $\sigma$;
\item $\vdash M:\sigma$.

\end{itemize}
\end{enumerate}
\end{thm}

It seems reasonable to expect that completeness cannot be extended beyond the classes $\forpiupro/\forpiudue$. For instance, by taking the semantics $\C S_{\beta}$, it can be seen that completeness fails for $\mathsf{co}\text{-}\B\Pi$ types as follows: by a result in \cite{Girard1989} there exists a total unary recursive function $f$ such that, for no $\lambda$-term $M_{f}$ computing $f$, $M_{f}$ can be given the $\mathsf{co}\text{-}\B\Pi$ type $\B{Nat}\to \B{Nat}$ (or, equivalently, the $\mathsf{co}\text{-}\B\Pi$ type $\sigma=\forall X(\B{Nat}\to (X\to X)\to (X\to X))$), where $\B{Nat}=\forall X((X\to X)\to (X\to X))$. By letting $M_{f}$ be any closed $\beta$-normal $\lambda$-term which represents ${f}$ (see \cite{Baren1976} or \cite{Bohm} for the construction of $M_{f}$), it can be verified that $M_{f}\in \|\B{Nat}\to \B{Nat}\|^{{\beta}}= \|\sigma\|^{{\beta}}$ (where the last equality is a consequence of Lemma \ref{fora}). Hence there exists a closed $\beta$-normal realizer of a $\mathsf{co}\text{-}\B\Pi$ type $\sigma$ such that $\vdash M:\sigma$ is not derivable in System $\FF$.

Theorem \ref{general} shows that the parametricity condition expressed by realizability, invariance and dinaturality coincide for positive types. As observed in Subsection \ref{secRey} and in Remark \ref{DinaEqua2}, in order to compare realizability with parametricity for types containing negative occurrences of the universal quantifier, one needs to consider a more extensional approach to realizability, including equations between realizers extending $\beta\eta$-equivalence.

\subsection{Open problems and future work}
As discussed in Subsection \ref{sec32}, the completeness argument as well as the approach to logical relations here developed do not apply to regular closure operators, hence to Girard's original reducibility candidates. We
agree with Colin Riba's remark on the ``not trivial and somehow mysterious'' - \cite{RibaUnion}, p.2 - structure of reducibility candidates. 
A possible approach might be to consider, instead of the infinite context $\Gamma^{\infty}$, contexts of the form $\Gamma_{M}=\{x_{1}:\forall XX,\dots, x_{n}:\forall XX\}$, where $\TT{FV}(M)=\{x_{1},\dots, x_{n}\}$. For any type $\sigma$, the set $\sigma^{\C S_{\CR}}$ obtained by $\beta$-closure of the set $\{ M\mid \Gamma_{M}\vdash M: \sigma\}$, is a reducibility candidate. Indeed, for any neutral and normal term $M$ and any type $\sigma$, it can be derived that $\Gamma_{M}\vdash M:\sigma$, so $\sigma^{\C S_{\CR}}$ is closed with respect to condition $\CR3$.

Our treatment of logical relations within realizability semantics uses stability by union in an essential way. Appeal to this property significantly simplifies the structure of the semantics, and was justified by the fact that all semantics considered are stable by union. However, as shown in \cite{RibaUnion}, this property is related to a standardization result for the reduction relations employed (the ``principal reduct property'' recalled in Appendix \ref{appCR}), which might fail for reduction relations and type systems different from those here considered. 

Finally, a natural extension of the approach here presented is by considering Krivine's second order functional arithmetic $\C{AF}2$ \cite{Krivine}, which extends System $\FF$ by means of first-order terms and $n$-ary predicates. Realizability semantics straightforwardly extends to $\C{AF}2$, yielding completeness results analogous to those which hold for System $\FF$ (see \cite{Nour1998}). 
The realizability semantics of $\C{AF}2$ seems a good candidate to relate the results here presented with those in \cite{Wadler2007} and \cite{Bernardy2011} 
on realizability and parametricity treated as formal translations in the second order predicate calculus. 
As recalled in the introduction, these results show that $n$-ary parametricity composed with realizability corresponds to $n+1$-ary parametricity.
This approach highlights the formal similarity between the clauses defining realizability and those defining parametricity. This similarity is implicit in our formalization, as it appears by inspecting the treatment of adequate semantics and the soundness theorem on the one side (Section \ref{sec3}) and of logical relations and its soundness theorem on the other side (Section \ref{sec5}). In $\C{AF}2$ one might try to make this explicit as follows. For any System $\FF$ type $\sigma$ we can define a predicate $x\in \boldsymbol\sigma$ expressing the property ``$x$ is a realizer of $\sigma$''. We can then define a ``realizability translation'' consisting in proving that $M\in \|\sigma\|^{\Cl}$ implies $M\in \| M\in \boldsymbol\sigma\|^{\Cl}$ for $M$ closed. For instance, we have $\B{X\to X}= \forall u(u\in \B X\to xu\in \B X)$ and for $I_{X}=\lambda x.x\in \|X\to X\|^{\Cl}$, we can immediately verify that $I_{X}\in \| I_{X}\in \B{X\to X}\|^{\Cl}$ holds\footnote{Where a $\Cl$-interpretation $\C M$ for $\C{AF}2$ associates term variables $x,y$ with $\beta\eta$-equivalence classes of $\lambda$-terms $x_{\C M}, y_{\C M}$.}.

The generalization of the statement of Theorem \ref{param1} to $\C{AF}2$ might be as follows: for any adequate (and $\mathsf{SU}$, etc.) $\FF$-closure operator and closed $\lambda$-terms $P,Q$, $P \in \|Q\in\boldsymbol\sigma\|^{\Cl}$ iff for all $\Cl$-interpretation $\C M$ and $\Cl$-relation assignment $R$ over $\C M,\C M$,  $P \ \Rel{R}{\sigma} \ Q$. 
On the one hand, this result would generalize Theorem \ref{param1} showing that closed terms $M\in \|\sigma\|^{\Cl}$ are $\Cl$-invariant (passing through $M\in \|M\in \boldsymbol\sigma\|^{\Cl}$). On the other hand, it would show that realizability increases the arity of logical relations, as the realizability translation $M\in\|\sigma\|^{\Cl} \ \mapsto \ M\in \|M\in \boldsymbol\sigma\|^{\Cl}$ would turn out equivalent to the $\Cl$-invariance condition $M \ \Rel{R}{\sigma} \ M$ for closed terms.

\section*{Acknowledgment}
  \noindent The author wishes to thank an anonymous referee for a very detailed and helpful list of
  remarks and improvement suggestions on a previous version of this paper.

\bibliographystyle{alpha}
\bibliography{LMCS.bib}

\appendix

%
\section{Proof of Proposition \ref{prop:foralle}}\label{app:foralle}

Before proving Proposition \ref{prop:foralle}, we establish some useful lemmas.

\begin{lem}\label{lemma:extra}
For all derivation $D$ of $\Gamma\vdash M:\sigma$ and for any type $\rho$, by replacing any occurrence of $X$ by $\rho $ in $D$ one obtains a derivation $D[\rho/X]$ of $\Gamma[\rho/X]\vdash M:\sigma[\rho/X]$.
\end{lem}
\begin{proof}
First observe that we can suppose that the variables appearing bound in $D$ are not among the free variables of $\rho$. Then we can argue by induction on $D$. We only consider the two interesting cases below:
\begin{itemize}
\item if $D$ has conclusion $\Gamma\vdash M:\forall Y\tau$ and ends by a $\forall$I rule whose immediate sub-derivation $D'$ has conclusion $\Gamma\vdash M: \tau$, then by induction hypothesis $D'[\rho/X]$ has conclusion $\Gamma[\rho/X]\vdash M:\tau[\rho/X]$ and by the assumption made, $Y$ does not occur free in $\Gamma[\rho/X]$. Hence by the $\forall$I rule, we deduce $\Gamma[\rho/X]\vdash M:\forall Y(\tau[\rho/X])$. The derivation obtained is of the form $D[\rho/X]$ and we conclude by observing that 
$\forall Y(\tau[\rho/X])=(\forall Y\tau)[\rho/X]$ as $Y$ does not occur free in $\rho$.

\item if $D$ has conclusion $\Gamma\vdash M:\tau[\rho'/X]$ and ends by a $\forall$E rule whose immediate sub-derivation $D'$ has conclusion $\Gamma\vdash M:\forall Y\tau$, then by induction hypothesis $D'[\rho/X]$ has conclusion $\Gamma[\rho/X]\vdash M:(\forall Y\tau)[\rho/X]$. By the assumption made $(\forall Y\tau)[\rho/X]=\forall Y(\tau[\rho/X])$, hence by the $\forall$E rule we deduce $\Gamma[\rho/X]\vdash M: \tau[\rho/X][\rho''/Y]$, where $\rho''=\rho'[\rho/X]$. Now we can conclude since $\tau[\rho/X][\rho''/Y]= \tau[\rho'/Y][\rho/X]$. \qedhere
\end{itemize}
\end{proof}

For any derivation $D$ of a type judgement $\Gamma\vdash M:\sigma$, we let $r(D)$ indicate the number of rules occurring in $D$. We call $D$ \emph{$\forall$E-free} when it contains no occurrence of the $\forall$E rule.

\begin{lem}\label{lemma:foralli}
Suppose $M$ is $\beta$-normal, $\Gamma\in \formenodue$ and $\sigma\in \forpiudue$. If $D$ is a $\forall$E-free derivation of $\Gamma\vdash M:\forall X\sigma$, then $D$ ends by a $\forall$I rule.
%
\end{lem}
\begin{proof}
If $D$ is $\forall$E-free and does not end by a $\forall$I rule, then either $M=x$ and $x:\forall X\sigma\in \Gamma$ or $D$ must end by a $\to$E rule. The first case can be excluded since $\forall X\sigma\notin \formenodue$. Hence, since $M$ is $\beta$-normal, $M=xM_{1}\dots M_{p}$, for some $p\geq 1$, where $\Gamma$ contains a type declaration of the form $x: \sigma_{1}\to \dots \to \sigma_{p}\to \forall X\sigma$ and $\Gamma\vdash M_{i}:\sigma_{i}$ is derivable for $i=1,\dots,p$.
But this is also impossible since $\sigma_{1}\to \dots \to \sigma_{p}\to \forall X\sigma\notin \formenodue$. 
\end{proof}

The following two lemmas are easily established by induction, respectively, on derivations and on types.

\begin{lem}\label{occur2}
Suppose $M$ is $\beta$-normal and $D$ is a $\forall$E-free derivation of $\Gamma\vdash M:\sigma$. Then, for all variable $X\in \TT{FV}(\Gamma)\cup \TT{FV}(\sigma)$, if $X$ only occurs in positive (resp.\ negative) positions in $\sigma$ and in negative (resp.\ positive) positions in the types in $\Gamma$, for all judgement $\Gamma'\vdash M':\sigma'$ in $D$, $X$ only occurs in negative (resp.\ positive) positions in the types in $\Gamma'$ and only occurs in positive (resp.\ negative) positions in $\sigma'$. 
\end{lem}

\begin{lem}\label{occurneg}
$\sigma[\tau/X]\in \forpiudue$ iff $\sigma\in \forpiudue$ and one of the following holds:
\begin{itemize}
\item $X$ only occurs in positive positions in $\sigma$ and $\tau\in \forpiudue$;
\item $X$ only occurs in negative positions in $\sigma$ and $\tau\in \formenodue$;
\item $\tau\in \TT T_{0}$.
\end{itemize}
\end{lem}

%
\begin{proof}[Proof of Proposition \ref{prop:foralle}]
We argue by induction on the construction of $D$. The only non-trivial case is when $D$ ends by a $\forall$E rule. Then $D$ has conclusion $\Gamma\vdash M:\tau[\rho/X]$ and is obtained from $D'$ of conclusion $\Gamma\vdash M:\forall X\tau$. By Lemma \ref{occurneg}, then, $\tau\in \forpiudue$, whence $\forall X\tau\in \forpiudue$. By the induction hypothesis there exists then $D^{*}$ $\forall$E-free of conclusion $\Gamma\vdash M:\forall X\tau$ such that for all judgement $\Gamma'\vdash M':\sigma'$ occurring in $D^{*}$, $\Gamma'\in \formenodue$ and $\sigma'\in \forpiudue$. 
By Lemma \ref{lemma:foralli}, $D^{*}$ must end by a $\forall$I rule. $D^{*}$ is thus obtained from a derivation $D''$ of $\Gamma\vdash M:\tau$ and $X$ does not occur free in $\Gamma$. 
By Lemma \ref{lemma:extra}, then, the $\forall$E-free derivation $D''[\rho/X]$ has conclusion $\Gamma\vdash M: \tau[\rho/X]$.
 
It remains to show that for any judgement $\Gamma'\vdash M':\sigma'$ in $D''[\rho/X]$, $\Gamma'\in \formenodue$ and $\sigma'\in \forpiudue$. Any such judgement is of the form $\Gamma''[\rho/X]\vdash M': \sigma''[\rho/X]$, where $\Gamma''\in \formenodue$ and $\sigma''\in \forpiudue$ (as $D''$ is a subderivation of $D^{*}$).  
According to Lemma \ref{occurneg} we must consider three cases.
First, if $X$ occurs in both positive and negative positions in $\tau$, then $\rho\in \TT T_{0}$, hence we can conclude $\Gamma'\in\formenodue$ and $\sigma'\in \forpiudue$. 
Second, if $X$ only occurs in positive position in $\tau$, then $\rho\in \forpiudue$ and, by Lemma \ref{occur2}, $X$ only occurs in negative positions in types in $\Gamma''$ and in positive positions in $\sigma''$. We can then conclude $\Gamma'\in\formenodue$ and $\sigma'\in \forpiudue$. 
Third, if $X$ only occurs in negative position in $\tau$, then $\rho\in \formenodue$ and, by Lemma \ref{occur2}, $X$ only occurs in positive positions in types in $\Gamma''$ and in negative positions in $\sigma''$. We can then conclude $\Gamma'\in\formenodue$ and $\sigma'\in \forpiudue$. 
\end{proof}

\section{Stability by union for $\CRuno$ and $\CRdue$}\label{appCR}

In \cite{RibaUnion} it is proved that $\C S_{\CR}=\OV{\C S_{\CR}}$, which implies that $\CR$ satisfies $\mathsf{SU}$. We now generalize the argument in \cite{RibaUnion} to ${\CRuno}$ and ${\CRdue}$.


\begin{defi}
For all $P\in \C{SN}$, let $\C V(P)=\{Q\in \C V\mid P\to_{\beta}^{*}Q\}$. We call the elements of $\C V(P)$ the \emph{values of $P$}. If $s\subseteq \C{SN}$, we let $\C V(s)=\bigcup_{P\in s}\C V(P)$. 
\end{defi}

The lemma below relates the closure of a term $P$ with the values of $P$.

\begin{lem}\label{cr11}
Let $P,Q\in \C{SN}$.
\begin{enumerate}[label=(\roman*)]
\item $\C V(P)=\C V( {\CRuno}(\{P\}))$;
\item if $\emptyset\neq \C V(Q)\subseteq \C V(P)$, then $Q\in \CRuno(\{P\})$.
\end{enumerate}
\end{lem}
\begin{proof}
For (i), since the map $s\mapsto  \C V(s)$ is clearly monotone, from $\{P\}\subseteq \CRuno(\{P\})$, we deduce $\C V(P)\subseteq \C V(\CRuno(\{P\}))$. For the converse direction, if $Q\in \C V(\CRuno(\{P\}))$, then $Q\notin \CRuno_{n+1}(\{P\})$, for all $n$, since $Q\notin \C{N}^*$, hence $Q\in \CRuno_{0}(\{P\})$, i.e. $P\to^{*}_{\beta} Q$, from which we deduce $Q\in \C V(P)$. 

 
 For (ii) we argue
 by induction on $\mathsf d(Q)$ (see Subsection \ref{sec21}): if $\SF d(Q)=0$, then $Q$ is $\beta$-normal, hence it must be $Q\in \C V(P)$, so the claim follows from (i). If $\SF d(Q)=N+1$, then either $Q\in \C V(P)$, so again we conclude by (i), or $Q\in \C N^{*}$. In this last case, for all $Q'$ such that $Q\to_{\beta}Q'$, $\SF d(Q')< Q$, hence, by the induction hypothesis $Q'\in \CRuno(\{P\})$. We can thus conclude by $\CRuno3$.
\end{proof}

We define now a preorder relation over strongly normalizing terms which allows to characterize the ${\CRuno}$-closure of a strongly normalizing term.

\begin{defi}\label{sqsub}
Given $P,Q\in \C{SN}$, let $P\sqsubseteq Q$ iff  either $Q\to_{\beta}^{*}P$ or $\emptyset\neq \C V(P)\subseteq \C V(Q)$ or $\C V(P)=\emptyset$ and $P\simeq_{\beta}Q$.  
\end{defi}

When $P$ has a value, then $P\sqsubseteq Q$ implies that all values of $P$ are also values of $Q$. When $P$ has no value, then $P\sqsubseteq Q$ implies that $P\simeq_{\beta}Q$.

\begin{lem}\label{cr12}
For all $P\in \C{SN}$, $\CRuno(\{P\})=\{Q\mid Q\sqsubseteq P\}$.

\end{lem}
\begin{proof}
\begin{description}

\item[($\supseteq$)] Suppose $Q\sqsubseteq P$. If $P\to^{*}_{\beta}Q$, then $Q\in \CRuno(\{P\})$. Suppose then $\emptyset \neq \C V(Q)\subseteq \C V(P)$; if $Q\in \C V$, then $Q\in\C V(P)\subseteq \CRuno(\{P\})$, so we can suppose $Q\in \C{N}^*$, as from $\C V(Q)\neq \emptyset$ it follows that $Q$ cannot be neutral and $\beta$-normal. Now, since $\emptyset \neq \C V(Q)\subseteq \C V(P)$, we can conclude $Q\in \CRuno(\{P\})$ by Lemma \ref{cr11} (ii). Finally, suppose $\C V(Q)=\emptyset$ and $Q\simeq_{\beta} P$. This means that $Q,P$ have the same neutral normal form $Q_{0}$. Since all $\beta$-expansions of $Q_{0}$ are neutral, we can deduce that $\CRuno(\{Q_{0}\})=\{Q_{0}\}^{\beta}$. Now, from $Q_{0}\in \CRuno(\{P\})$, it follows $\CRuno(\{Q_{0}\})\subseteq \CRuno(\{P\})$, whence $Q\in \CRuno(\{P\})$. 

\item[($\subseteq$)] Since $\CRuno(\{P\})=\bigcup_{n}\CRuno_{n}(\{P\})$, we show, by induction on $n$, that $\CRuno_{n}(\{P\})\subseteq \{Q\mid Q\sqsubseteq P\}$, for all $n\in \BB N$. If $Q\in \CRuno_{0}(\{P\})$, then $P\to^{*}_{\beta} Q$, hence $Q\sqsubseteq P$; suppose now $Q\in \CRuno_{n+1}(\{P\})-\CRuno_{n}(\{P\})$. This means that $Q\in \C{N}^*$ and, by induction hypothesis, we can suppose that for any immediate reduct $Q'$ of $Q$, $Q'\sqsubseteq P$, since $Q'\in \CRuno_{n}(\{P\})$. 
Suppose $\emptyset \neq \C V(Q)$; this means that $Q$ reduces to some $Q'\in \C V$. Since all immediate reducts of $Q$ are in $\CRuno(\{P\})$ and the latter is closed by $\to_{\beta}^{*}$-reduction, it follows that $Q'\in \CRuno(\{P\})$. Hence $Q'\in \C V(\CRuno(\{P\}))=\C V(P)$ and we conclude that $\C V(Q)\subseteq \C V(P)$, so $Q\sqsubseteq P$. Suppose now $\C V(Q)=\emptyset$; then any immediate reduct of $Q$ is in $\CRuno_{n}(\{P\})\subseteq \{P\}^{\beta}$, whence $Q\simeq_{\beta}P$, so again $Q\sqsubseteq P$. \qedhere
\end{description}
\end{proof}

From Lemma \ref{cr12} we can deduce that $\sqsubseteq$ is a preorder:
\begin{cor}
$\sqsubseteq$ is reflexive and transitive.
\end{cor}
\begin{proof}
Reflexivity is clear. Transitivity is proved as follows: suppose $P\sqsubseteq Q\sqsubseteq R$. By Lemma \ref{cr12} we have then
$Q\in \CRuno(\{P\})$ and 
$R\in \CRuno(\{Q\})$. From $\CRuno(\{Q\})\subseteq \CRuno(\{P\})$ we deduce then $R\in \CRuno(\{P\}$, so, again by Lemma \ref{cr12}, we conclude $P\sqsubseteq R$.
\end{proof}

Moreover, from Lemma \ref{cr12} we can also deduce that $\CRuno$ is not stable by substitution:
\begin{cor}\label{noss}
$\CRuno$ does not satisfy $\mathsf{SS}$.
\end{cor}
\begin{proof}
The term $P'=x((\lambda x.x)y)$ is in the $\CRuno$-closure of $P=xy$, as $P'\in \C{N}^*$ and $P$ is its sole immediate reduct; however, given $Q=\lambda u.\lambda v.u$, the term $P'[Q/x]$ is not in the $\CRuno$-closure of $P[Q/x]$: $P'[Q/x]$ reduces in one step to the value $\lambda v.(\lambda x.x)y$ which is not among the values of $P[Q/x]=(\lambda u.\lambda v.u)y$. Hence $\emptyset\neq \C V(P'[Q/x])\not\subseteq \C V(P[Q/x])$ and this implies $P'[Q/x]\not\sqsubseteq P[Q/x]$, so by Lemma \ref{cr12}, we deduce $P'[Q/x]\notin \CRuno(\{P[Q/x]\})$.
\end{proof}

For any $P\in \C{SN}$, we let $P^{\downarrow}=\{Q\mid \exists Q' \ P\to_{\beta}Q'\to_{\beta}^{*}Q\}$.

\begin{lem}\label{cr13}
For all $P\in \C{N}^*$ there exists $Q\in P^{\downarrow}$ such that $P\sqsubseteq Q$. 

\end{lem}
\begin{proof}
If $P$ is not in weak head normal form, then if $Q$ is the weak head reduct of $P$, $P\sqsubseteq Q$: if $\C V(P)\neq \emptyset$, it must be $\C V(P)\subseteq \C V(Q)$; if $\C V(P)=\emptyset$, then $\C V(Q)=\emptyset $ and $P\simeq_{\beta}Q$. 
If $P$ is in weak head normal form, then for any immediate reduct $Q$ of $P$, $P\sqsubseteq Q$. Indeed, $\C V(P)=\emptyset$ and $P\simeq_{\beta}Q$. 
\end{proof}
Similarly to \cite{RibaUnion} (def. 4.11, p. 8), we call a term $Q\in P^{\downarrow}$, for $P\in\C{N}^*$, such that $P\sqsubseteq Q$, a \emph{principal reduct} of $P$.
A corollary of lemma is that $(\_)^{\beta sat}\leq {\CRuno},{\CRdue}$, i.e. that the sets in $\C S_{\CRuno}$ and $\C S_{\CRdue}$ are closed with respect to strongly normalizing weak-head expansion.

\begin{cor}\label{wexpa}
For all $s\in \C S_{\CRuno}$ (resp.\ $s\in \C S_{\CRdue}$), if $P\in s$, $Q\in \C{SN}$ and $Q\to_{wh} P$, then $Q\in s$.
\end{cor}
\begin{proof}
Let $s\in \C S_{\CRuno}$, $P\in s$ and $Q\in \C{SN}$ be such that $Q\to_{wh}P$. Then $Q\in \C{N}^*$ and $P$ is a principal reduct of $Q$, i.e. $Q\sqsubseteq P$. Hence, by Lemma \ref{cr12}, $Q\in \CRuno(\{P\})\subseteq s$.
From $\C S_{\CRuno}\leq \C S_{\CRdue}$ it follows that the same holds for $\C S_{\CRdue}$.
\end{proof}

We can now prove that $\C S_{\CRuno}$ is stable by union. 

\begin{prop}\label{decocr1}
${\CRuno}$ satisfies $\mathsf{SU}$.
\end{prop}
\begin{proof}

Let $\C P =\{ s\subseteq \C{SN}\mid s\neq \emptyset \ \text{and} \ s=\bigcup\{\CRuno(\{P\})\mid P\in s\}\}$ and $\C O= \{ s\subseteq \C{SN}\mid s\neq \emptyset \ \text{and} \ \forall P,Q \ (P\in s, Q\sqsubseteq P \To Q\in s)\}$ be the set of non-empty, $\sqsubseteq$-downward closed sets of strongly-normalizing $\lambda$-terms. 

We claim that $\C P=\C O$. Indeed, $s\in\C P$ iff $s\neq \emptyset $ and $s=\bigcup\{\CRuno(\{P\})\mid P\in s\}=\{ Q \mid \exists P\in s \ Q\sqsubseteq P\}$ (by Lemma \ref{cr12}) i.e. iff $s\in \C O$.

We show now that $\C S_{\CRuno}= \C O\cup \{\emptyset\}$, from which we can conclude that $\C S_{\CRuno}=\C P\cup\{\emptyset\}=\OV\CRuno$, i.e. that $\C S_{\CRuno}$ is $\mathsf{SU}$ (by Proposition \ref{union2}).
\begin{description}
\item[($\C S_{\CRuno}\subseteq \C O\cup \{\emptyset\}$)] Let $s\in \C S_{\CRuno}$ be non-empty, $P\in s$ and $Q\sqsubseteq P$. Then, by Lemma \ref{cr12}, $Q\in \CRuno(\{P\})\subseteq s$, hence $Q\in s$.

\item[($\C O\cup \{\emptyset\}\subseteq \C S_{\CRuno}$)] First, $\CRuno(\emptyset)=\emptyset$, so $\emptyset\in \C S_{\CRuno}$. Let now $s\in \C O$. Since $s^{\beta\downarrow}\subseteq s$, it suffices to verify that if $P\in \C{N}^*$ and $P^{\downarrow}\subseteq s$, then $P\in s$. By Lemma \ref{cr13}, if $P\in \C{N}^*$, there exists $Q\in P^{\downarrow}$ such that $P\sqsubseteq Q$. As $Q\in s$ and $s$ is $\sqsubseteq$-downward closed, $P\in s$. \qedhere
\end{description}
\end{proof}

We now adapt the argument above to ${\CRdue}$.
Given terms $M,F_{1},\dots, F_{n},G_{1},\dots, G_{n}$ and variables $x_{1},\dots, x_{n}$, we let $M[F_{i}/x_{i}]$ (resp.\ $M[G_{i}/x_{i}]$) be shorthand for $M[F_{1}/x_{1},\dots, F_{n}/x_{n}]$ (resp.\ $M[G_{1}/x_{1},\dots, G_{n}/x_{n}]$).

\begin{lem}\label{cr21}
Let $P=M[F_{i}/x_{i}],Q=M[G_{i}/x_{i}]\in \C{SN}$, for some $i\leq n$, where $F_{i}\in \C{N}^*$.
If for any $i$, either $\emptyset\neq \C V(F_{i})\subseteq \C V(G_{i})$ or $\C V(G_{i})=\emptyset$ and $F_{i}\simeq_{\beta} G_{i}$, then  $P\in \CRdue(\{Q\})$.

%
%
\end{lem}
\begin{proof}



We argue by induction on $N=\sum_{i}^{n}\SF d(F_{i})$ (see Subsection \ref{sec21}). By $\CRdue3$, it suffices to show that for all choice of terms $F'_{i}$ be such that $F_{i}\to_{\beta} F'_{i}$, $M[F'_{i}/x_{i}]\in \CRdue(\{Q\})$. 
Let then  $F'_{i}$ be such that $F_{i}\to_{\beta} F'_{i}$. If, for all $i=1,\dots, n$, $F'_{i}\in \C V$, then $F'_{i}\in \C V(G_{i})$, for all $i=1,\dots, n$. 
Hence $Q\to_{\beta}^{*}M[F'_{i}/x_{i}]$, whence $M[F'_{i}/x_{i}] \in \CRdue(\{Q\})$. 
Otherwise, there is some $i$ such that $F'_{i}\in \C N$. If $F'_{i}$ is $\beta$-normal, then $\C V(F'_{i})=\emptyset$, so by the hypothesis it must be $\C V(G_{i})=\emptyset$ and $G_{i}\to_{\beta}^{*}F'_{i}$. Hence, if for all $i$ such that $F'_{i}\in \C N$, $F'_{i}$ is $\beta$-normal, then for all $i=1,\dots, n$, $G_{i}\to_{\beta}^{*}F'_{i}$ (indeed, if $F'_{i}\notin \C N$, then $F'_{i}\in \C V$, hence $F'_{i}\in \C V(G_{i})$). Therefore $Q\to_{\beta}^{*}M[F'_{i}/x_{i}]$, whence  $M[F'_{i}/x_{i}] \in \CRdue(\{Q\})$. Otherwise, for some $i$, $F'_{i}\in \C N^{*}$. By the induction hypothesis, then 
$Q'=M[F_{1}/x_{1},\dots, F'_{i}/x_{i},\dots, F_{n}/x_{n}]\in \CRdue(\{Q\})$, and since
$Q'\to_{\beta}^{*}M[F'_{i}/x_{i}]$, we deduce $M[F'_{i}/x_{i}] \in \CRdue(\{Q\})$.
\end{proof}

We now define, similarly to Definition \ref{sqsub}, a preorder relation $\trianglelefteq^{*}$ over strongly normalizing terms which allows to characterize the ${\CRdue}$-closure of a strongly normalizing term.

\begin{defi}\label{B3}
Given, $P,Q\in \C{SN}$, $P\trianglelefteq Q$ iff either $Q\to_{\beta}^{*}P$ or $P=M[F_{i}/x_{i}]$, for some $i\leq n$, where $F_{i}\in \C{N}^*$, $Q=M[G_{i}/x_{i}]$ and for all $i\leq n$, either $\emptyset\neq \C V(F_{i})\subseteq \C V(G_{i})$ or $\C V(F_{i})=\emptyset$ and $F_{i}\simeq_{\beta}G_{i}$. 
We let $\trianglelefteq^{*}$ indicate the transitive closure of $\trianglelefteq$. 

\end{defi}

We can prove a property similar to the principal reduct property for $\trianglelefteq$:
\begin{lem}\label{cr23}
For all $P=M[F_{i}/x_{i}]\in \C{SN}$, with $F_{i}\in \C{N}^*$, there exist terms $G_{i}\in F_{i}^{\downarrow}$ such that $P\trianglelefteq Q=M[G_{i}/x_{i}]$. 

\end{lem}
\begin{proof}

If $P=M[F_{1}/x_{1},\dots, F_{n}/x_{n}]$ with $F_{i}\in \C{N}^*$, by Lemma \ref{cr13}, for any $F_{i}$ there exists an immediate reduct $G_{i}$ of $F_{i}$ such that $F_{i}\sqsubseteq G_{i}$, and in particular (since $F_{i}\to_{\beta}G_{i}$) such that either $\emptyset \neq \C V(F_{i})\subseteq \C V(G_{i})$ or $\C V(F_{i})=\emptyset$ and $F_{i}\simeq_{\beta} G'_{i}$. This implies that $Q\trianglelefteq Q''=M[G_{i}/x_{i}]$. \end{proof}

\begin{lem}\label{cr22}
For all $P\in \C{SN}$, $\CRdue(\{P\})=\{Q\mid Q\trianglelefteq^{*} P\}$.

\end{lem}
\begin{proof}
\begin{description}

\item[($\supseteq$)] Suppose $Q\trianglelefteq^{*} P$. We argue by induction on the length $k+1$ of a chain $Q=Q_{0}\trianglelefteq Q_{1}\trianglelefteq \dots \trianglelefteq Q_{k}=P$.  Suppose $k=0$, i.e. $Q\trianglelefteq P$. If $P\to^{*}_{\beta}Q$, then $Q\in \CRdue(\{P\})$. Suppose then $
P=M[F_{i}/x_{i}]$, $Q=M[G_{i}/x_{i}]$, where $F_{i}\in \C{N}^*$ and for all $i\leq n$, either $\emptyset\neq \C V(F_{i})\subseteq \C V(G_{i})$ or $\C V(F_{i})=\emptyset$ and $F_{i}\simeq_{\beta}G_{i}$. By Lemma \ref{cr21}
we conclude that $P\in \CRdue(\{Q\})$.
%
%
Suppose now $k> 1$, i.e. $Q\trianglelefteq Q'\trianglelefteq^{*}P$. By induction hypothesis $Q'\in \CRdue(\{P\})$ and by the argument above $Q\in \CRdue(\{Q'\})\subseteq \CRdue(\{P\})$.

\item[($\subseteq$)] Since $\CRdue(\{P\})=\bigcup_{n}\CRdue_{n}(\{P\})$, we show, by induction on $n$, that $\CRdue_{n}(\{P\})\subseteq \{Q\mid Q\trianglelefteq^{*}P\}$, for all $n\in \BB N$. If $Q\in \CRdue_{0}(\{P\})$, then $P\to^{*}_{\beta} Q$, hence $Q\trianglelefteq P$; assume now $Q\in \CRdue_{n+1}(\{P\})-\CRdue_{n}(\{P\})$. This means that $Q=M[F_{i}/x_{i}]$, where $F_{i}\in \C{N}^*$ and for all $Q'=M[G_{i}/x_{i}]$ such that $F_{i}\to_{\beta}G_{i}$, $Q'\in \CRdue_{n}(\{P\})$. By Lemma \ref{cr23}, there exist immediate reducts $G'_{i}$ of $F_{i}$ such that $Q\trianglelefteq Q''=M[G'_{i}/x_{i}]$. Moreover, by the assumption, $Q''\in \CRdue_{0}(\{P\})$ and, by the induction hypothesis, this implies $Q'' \trianglelefteq^{*}P$. We conclude then $Q\trianglelefteq^{*}P$. \qedhere
\end{description}
\end{proof}

\begin{prop}\label{decocr2}
${\CRdue}$ satisfies $\mathsf{SU}$.
\end{prop}
\begin{proof}

Let $\C P=\{ s\subseteq \C{SN}\mid s\neq \emptyset \ \text{and} \ s=\bigcup\{\CRdue(\{P\})\mid P\in s\}\}$ and $\C O= \{ s\subseteq \C{SN}\mid s\neq \emptyset \ \text{and} \ \forall P,Q \ P\in s, Q\trianglelefteq P \To Q\in s\}$ be the set of non-empty, $\trianglelefteq$-downward closed sets of strongly-normalizing $\lambda$-terms. 

We claim that $\C P=\C O$. Indeed, $s\in \C P$ iff $s\neq \emptyset $ and $s=\bigcup\{\CRuno(\{P\})\mid P\in s\}\}=\{ Q \mid \exists P\in s \ Q\trianglelefteq^{*} P\}$ (by Lemma \ref{cr22}) i.e. iff $s\in \C O$.

We show now that $\C S_{\CRdue}= \C O\cup\{\emptyset\}$, from which we can conclude that $\C S_{\CRdue}=\C P\cup\{\emptyset\}=\OV{{\CRdue}}$, i.e. that $\C S_{\CRdue}$ is $\mathsf{SU}$ (by Proposition \ref{union2}).
\begin{description}
\item[($\C S_{\CRdue}\subseteq \C O\cup\{\emptyset\}$)] Let $s\in \C S_{\CRdue}$ be non-empty, $P\in s$ and $Q\trianglelefteq P$. Then, by Lemma \ref{cr22}, $Q\in CR(\{P\})\subseteq s$, hence $Q\in s$.

\item[($\C O\cup\{\emptyset\}\subseteq\C S_{ \CRdue}$)] First, $\CRdue(\emptyset)=\emptyset$, so $\emptyset\in \C S_{\CRdue}$. Let now $s\in \C O$. Since $s^{\beta \downarrow}\subseteq s$, it suffices to verify that if $P=M[F_{i}/x_{i}]$, with $F_{i}\in \C{N}^*$ and for all $G_{i}\in F_{i}^{\downarrow}$, $M[G_{i}/x_{i}]\in s$, then $P\in s$. Suppose then $P=M[F_{i}/x_{i}]$, with $F_{i}\in \C{N}^*$, and for all $G_{i}\in F_{i}^{\downarrow}$, $M[G_{i}/x_{i}]\in s$. By Lemma \ref{cr23}, there exist terms $G_{i}\in F_{i}^{\downarrow}$ such that $P\trianglelefteq Q= M[G_{i}/x_{i}]$. Since $Q\in s$ and $s$ is $\trianglelefteq$-downward closed, we conclude that $P\in s$. 
\end{description}
\end{proof}

\printindex


\end{document}